%% file: amore_tdr.tex
\documentclass[11pt,a4paper]{report}
\pdfoutput=1
\usepackage{array}
\usepackage{epsfig}
\usepackage{epstopdf}
\usepackage{amsmath}
\usepackage{graphicx}
\usepackage{watermark}
\usepackage{comment}
\usepackage{multirow}
\usepackage{pdflscape}
\usepackage{dcolumn}
\usepackage{bm}
\usepackage{setspace}
\includecomment{printrobustnotes}
\includecomment{printallnotes}
\usepackage{xspace}
\usepackage{color}
\usepackage{xparse}
\usepackage{enumerate}
\usepackage[a4paper, margin=1.3in]{geometry} 
\usepackage[affil-it]{authblk}
\usepackage{authblk}

\begin{document}

\newenvironment{cfigure}[1][tbp]{\begin{figure}[#1]\centering}{\end{figure}}
\newenvironment{cfigure1c}[1][tbp]{\begin{figure*}[#1]\centering}{\end{figure*}}


\newcommand\geniso[2]{\ensuremath{\rm ^{#2}#1}\xspace}

\newcommand\newiso[3][]{
\ifx&#1&
  \expandafter\newcommand\csname #2\endcsname[1][#3]{\geniso{#2}{##1}}
\else
  \expandafter\newcommand\csname #1\endcsname[1][#3]{\geniso{#2}{##1}}
\fi
}

%
%

\newiso{Cs}{137}
\newiso{Rb}{87}
\newiso{K}{40}
\newiso{Ar}{40}
\newiso{U}{238}
\newiso{Th}{232}
\newiso{Rn}{}
\newiso{Ra}{}
\newiso{Na}{22}
\newiso{I}{125}
\newiso{Te}{}
\newiso{Pb}{210}
\newiso{Bi}{}
\newiso{Ac}{}
\newiso{Tl}{208}
\newiso{Po}{}
\newiso{Am}{241}
\newiso{Fe}{55}
\newiso{Mo}{100}
\newiso{Zr}{}
\newiso{Co}{60}
\newiso{Nd}{}
\newiso{Tc}{}
\newiso{Ba}{}
\newiso{Ca}{}
\newiso{Y}{}


\newcommand\ppm{\ensuremath{\mu}g/g\xspace}
\newcommand\ppb{{ng/g}\xspace}
\newcommand\ppt{{pg/g}\xspace}
\newcommand\uBqkg{\ensuremath{\rm \mu Bq/kg}\xspace}
\newcommand\csonethirtyseven{\ensuremath{^{137}Cs}\xspace}
\newcommand\csonethirtyfour{\ensuremath{^{134}Cs}\xspace}
\newcommand\rbeightyseven{\ensuremath{^{87}Rb}\xspace}
\newcommand\kforty{\ensuremath{\rm ^{40}K}\xspace}
\newcommand\arforty{\ensuremath{\rm ^{40}Ar}\xspace}
\newcommand\utwothirtyeight{\ensuremath{\rm ^{238}U}\xspace}
\newcommand\thtwothirtytwo{\ensuremath{\rm ^{232}Th}\xspace}
\newcommand\rntwotwotwo{\ensuremath{\rm ^{222}Rn}\xspace}
\newcommand\natwentytwo{\ensuremath{\rm ^{22}Na}\xspace}
\newcommand\natwentythree{\ensuremath{\rm ^{23}Na}\xspace}
\newcommand\ionetwentyfive{\ensuremath{\rm ^{125}I}\xspace}
\newcommand\ionetwentynine{\ensuremath{\rm ^{129}I}\xspace}
\newcommand\teonetwentyfive{\ensuremath{\rm ^{125}Te}\xspace}
\newcommand\teonetwentyseven{\ensuremath{\rm ^{127}Te}\xspace}
\newcommand\pbtwoten{\ensuremath{\rm ^{210}Pb}\xspace}
\newcommand\potwoten{\ensuremath{\rm ^{210}Po}\xspace}
\newcommand\bitwoforteen{\ensuremath{\rm ^{214}Bi}\xspace}
\newcommand\thtwotwentyeight{\ensuremath{\rm ^{228}Th}\xspace}
\newcommand\actwotwentyeight{\ensuremath{\rm ^{228}Ac}\xspace}
\newcommand\ratwotwentysix{\ensuremath{\rm ^{226}Ra}\xspace}
\newcommand\rntwotwenty{\ensuremath{\rm ^{220}Rn}\xspace}
\newcommand\tltwooeight{\ensuremath{\rm ^{208}Tl}\xspace}
\newcommand\potwoforteen{\ensuremath{\rm ^{214}Po}\xspace}
\newcommand\potwosixteen{\ensuremath{\rm ^{216}Po}\xspace}
\newcommand\bitwotwelve{\ensuremath{\rm ^{212}Bi}\xspace}
\newcommand\potwotwelve{\ensuremath{\rm ^{212}Po}\xspace}
\newcommand\amtwofortyone{\ensuremath{\rm ^{241}Am}\xspace}
\newcommand\fefiftyfive{\ensuremath{\rm ^{55}Fe}\xspace}
\newcommand\geseventysix{\ensuremath{\rm ^{76}Ge~}\xspace}
\newcommand\seeightytwo{\ensuremath{\rm ^{82}Se}\xspace}
\newcommand\cdonesixteen{\ensuremath{\rm ^{116}Cd}\xspace}
\newcommand\teonetwentyeight{\ensuremath{\rm ^{128}Te}\xspace}
\newcommand\teonethirty{\ensuremath{\rm ^{130}Te}\xspace}
\newcommand\zrnintysix{\ensuremath{\rm ^{96}Zr\xspace}}
\newcommand\xeonethirtysix{\ensuremath{\rm ^{136}Xe}\xspace}
\newcommand\ndonefifty{\ensuremath{\rm ^{150}Nd}\xspace}
\newcommand\baonethirty{\ensuremath{\rm ^{130}Ba}\xspace}
\newcommand\moninetytwo{\ensuremath{\rm ^{92}Mo}\xspace}
\newcommand\zrninetytwo{\ensuremath{\rm ^{92}Zr}\xspace}
\newcommand\mohundred{\ensuremath{\rm ^{100}Mo}\xspace}

\newcommand\cs{\ensuremath{^{137}Cs}\xspace}
\newcommand\ces{\ensuremath{^{134}Cs}\xspace}
\newcommand\rb{\ensuremath{^{87}Rb}\xspace}
\newcommand\cof{\ensuremath{^{57}Co}\xspace}
\newcommand\fe{\ensuremath{^{55}Fe}\xspace}
\newcommand\mn{\ensuremath{^{54}Mn}\xspace}


\newcommand\tm{\ensuremath{\times}\xspace} 
\newcommand\sm{\ensuremath{\sim}\xspace}
\newcommand\camobf{\ensuremath{\bf CaMoO_{4}}\xspace}
\newcommand\camo{\ensuremath{\rm CaMoO_{4}}\xspace}
\newcommand\cmo{\ensuremath{\rm CaMoO_{4}}\xspace}
\newcommand\CMO{\ensuremath{\rm CaMoO_{4}}\xspace}
\newcommand\caco{\ensuremath{\rm CaCO_{3}}\xspace}
\newcommand\moothree{\ensuremath{\rm MoO_{3}}\xspace}
\newcommand\doublecamo{\ensuremath{\rm ^{48depl}Ca^{100}MoO_{4}}\xspace}
\newcommand\cawo{\ensuremath{\rm CaWO_{4}}\xspace}
\newcommand\pbmo{\ensuremath{\rm PbMoO_{4}}\xspace}
\newcommand\cafortyeight{\ensuremath{\rm ^{48}Ca}\xspace}

\newcommand\bpec{\ensuremath{\rm \beta^{+}/EC}\xspace}
\newcommand\bp{\ensuremath{\beta^{+}}\xspace}
\newcommand\bpbp{\ensuremath{\beta^{+}\beta^{+}}\xspace}
\newcommand\bmbm{\ensuremath{\beta^{-}\beta^{-}}\xspace}
\newcommand\ep{\ensuremath{e^{+}}\xspace}
\newcommand\ms{\ensuremath{\mus}\xspace}
\newcommand\mm{\ensuremath{\mum}\xspace}
\newcommand\zeronu{\ensuremath{\rm 0\nu}\xspace}
\newcommand\zeronuecec{\ensuremath{\rm 0\nuEC/EC}\xspace}
\newcommand\zeronubpec{\ensuremath{\rm 0\nu\beta^{+}/EC}\xspace}
\newcommand\twonu{\ensuremath{\rm 2\nu}\xspace}
\newcommand\bb{\ensuremath{\beta\beta}\xspace}
\newcommand\multi{\ensuremath{\times}\xspace}
\newcommand\twonubb{2\ensuremath{\nu\beta\beta}\xspace}
\newcommand\zeronubb{0\ensuremath{\nu\beta\beta}\xspace}
\newcommand\meff{\ensuremath{\langle m_{\beta\beta}\rangle}\xspace}

\newcommand\plusminus{\ensuremath{\pm}\xspace}
\newcommand\halft{\ensuremath{T_{1/2}}\xspace}
\newcommand\cpd{\ensuremath{counts/(keV \cdot kg \cdot day)}\xspace}
\newcommand\naitl{\ensuremath{\textrm{NaI(T}\mathrm{\ell}\textrm{)}}\xspace}
\newcommand\csitl{\ensuremath{\textrm{CsI(T}\mathrm{\ell}\textrm{)}}\xspace}


\newcommand\PREP{Phys. Rep.}

\def\Journal#1#2#3#4{{#1} {\bf #2} (#3) #4}

\newcommand\NCA{Nuovo Cimento}
\newcommand\NIM{Nucl. Instrum. Methods}
\newcommand\NIMA{Nucl. Instrum. Methods A}
\newcommand\NPB{Nucl. Phys. B}
\newcommand\PLB{Phys. Lett.  B}
\newcommand\PRL{Phys. Rev. Lett.}
\newcommand\PRD{Phys. Rev. D}
\newcommand\ZPC{Z. Phys. C}
\newcommand\ASP{Astropart. Phys.}
\newcommand\JKPS{J. Kor. Phys. Soc.}
\newcommand\etal{{\it et al.}}

\newcommand{\gevcc}[1]{$#1~\mathrm{GeV}/c^{2}$}

\title{\vspace{-5.0cm}Technical Design Report for 
  the AMoRE $0\nu\beta\beta$ Decay Search Experiment}


\newcounter{authindex}
\stepcounter{authindex}\edef\tempf{\alph{authindex}}
\stepcounter{authindex}\edef\tempi{\alph{authindex}}
\stepcounter{authindex}\edef\tempm{\alph{authindex}}
\stepcounter{authindex}\edef\templ{\alph{authindex}}
\stepcounter{authindex}\edef\tempk{\alph{authindex}}
\stepcounter{authindex}\edef\tempq{\alph{authindex}}
\stepcounter{authindex}\edef\tempp{\alph{authindex}}
\stepcounter{authindex}\edef\tempr{\alph{authindex}}
\stepcounter{authindex}\edef\tempg{\alph{authindex}}
\stepcounter{authindex}\edef\tempb{\alph{authindex}}
\stepcounter{authindex}\edef\tempe{\alph{authindex}}
\stepcounter{authindex}\edef\tempc{\alph{authindex}}
\stepcounter{authindex}\edef\tempo{\alph{authindex}}
\stepcounter{authindex}\edef\temps{\alph{authindex}}
\stepcounter{authindex}\edef\tempt{\alph{authindex}}
\stepcounter{authindex}\edef\temph{\alph{authindex}}
\stepcounter{authindex}\edef\tempu{\alph{authindex}}
\stepcounter{authindex}\edef\tempn{\alph{authindex}}
\stepcounter{authindex}\edef\tempa{\alph{authindex}}
\stepcounter{authindex}\edef\tempd{\alph{authindex}}
\stepcounter{authindex}\edef\tempj{\alph{authindex}}

\author[\tempf]{V.~Alenkov}
\author[\tempi]{P.~Aryal} 
\author[\tempm]{ J.~Beyer}
\author[\templ]{R.S.~Boiko} 
\author[\tempk]{K.~Boonin} 
\author[\tempf]{ O.~Buzanov} 
\author[\tempk]{ N.~Chanthima} 
\author[\tempq]{M.K.~Cheoun} 
\author[\templ]{ D.M.~Chernyak} 
\author[\tempp]{ J.~Choi} 
\author[\tempp]{S.~Choi} 
\author[\templ]{ F.A.~Danevich} 
\author[\tempr]{M.~Djamal} 
\author[\tempm]{D.~Drung} 
\author[\tempg]{C.~Enss} 
\author[\tempg]{ A.~Fleischmann} 
\author[\tempb]{ A.M.~Gangapshev} 
\author[\tempg]{ L.~Gastaldo} 
\author[\tempb]{Yu.M.~Gavriljuk } 
\author[\tempb]{ A.M.~Gezhaev } 
\author[\tempb]{V.I.~Gurentsov} 
\author[\tempi]{ D.H.~Ha} 
\author[\tempe]{I.S.~Hahn}  
\author[\tempi]{ J.H.~Jang} 
\author[\tempc]{E.J.~Jeon} 
\author[\tempc]{ H.S.~Jo} 
\author[\tempp]{ H.~Joo} 
\author[\tempk]{ J.~Kaewkhao} 
\author[\tempc]{ C.S.~Kang} 
\author[\tempo]{S.J.~Kang}
\author[\tempc]{ W.G.~Kang} 
\author[\tempi]{ S.~Karki} 
\author[\tempb]{ V.V.~Kazalov} 
\author[\temps,\tempt]{N.~Khanbekov} 
\author[\tempc]{ G.B.~Kim} 
\author[\tempi]{ H.J.~Kim} 
\author[\tempi]{ H.L.~Kim}
\author[\tempc]{ H.O.~Kim}
\author[\tempp]{ I.~Kim} 
\author[\temph]{J.H.~Kim} 
\author[\tempp]{ K.~Kim} 
\author[\tempp]{ S.K.~Kim}
\author[\tempc]{ S.R.~Kim} 
\author[\tempc]{ Y.D.~Kim} 
\author[\tempc]{Y.H.~Kim} 
\author[\tempk]{ K.~Kirdsiri} 
\author[\templ]{ V.V.~Kobychev} 
\author[\temps]{ V.~Kornoukhov}
\author[\tempb]{ V.V.~ Kuzminov} 
\author[\tempc]{ H.J.~Lee} 
\author[\tempc]{ H.S.~Lee} 
\author[\temph]{ J.H.~Lee} 
\author[\temph]{ J.M.~Lee} 
\author[\tempi]{ J.Y.~Lee} 
\author[\temph]{ K.B.~Lee} 
\author[\tempc]{ M.H.~Lee}  
\author[\temph]{ M.K.~Lee} 
\author[\tempc]{ D.S.~Leonard} 
\author[\tempc]{ J.~Li} 
\author[\tempu]{J.~Li} 
\author[\tempu]{ Y.J.~Li} 
\author[\tempk]{ P.~Limkitjaroenporn} 
\author[\tempn]{K.J.~Ma}
\author[\tempb]{ O.V.~Mineev} 
\author[\templ]{ V.M.~Mokina}
\author[\tempc]{ S.L.~Olsen} 
\author[\tempb]{ S.I.~Panasenko} 
\author[\tempi]{ I.~Pandey} 
\author[\tempc]{ H.K.~Park} 
\author[\temph]{ H.S.~Park}
\author[\tempc]{ K.S.~Park}
\author[\templ]{ D.V.~Poda} 
\author[\templ]{O.G.~Polischuk} 
\author[\temps]{ P.~Polozov} 
\author[\tempr]{ H.~Prihtiadi}
\author[\tempc]{ S.J.~Ra} 
\author[\tempb]{ S.S.~Ratkevich} 
\author[\tempa]{G.~Rooh}
\author[\tempd]{ K.~Siyeon} 
\author[\tempk]{ N.~Srisittipokakun}
\author[\tempc]{ J.H.~So} 
\author[\tempi]{ J.K.~Son}
\author[\tempb]{ J.A.~Tekueva} 
\author[\templ]{ V.I.~Tretyak}
\author[\tempb]{ A.V.~Veresnikova} 
\author[\tempj]{R.~Wirawan}
\author[\tempb]{ S.P.~Yakimenko} 
\author[\tempb]{ N.V.~Yershov} 
\author[\tempc]{ W.S.~Yoon}
\author[\tempc]{ Y.S.~Yoon} 
\author[\tempu]{ Q.~Yue}

\affil[\tempf]{JSC FOMOS-Materials, Moscow 107023, Russia}
\affil[\tempi]{Department of Physics, Kyungpook National University, Daegu 41566, Korea}
\affil[\tempm]{Physikalisch-Technische Bundesanstalt (PTB), D-38116 Braunschweig, Germany}
\affil[\templ]{Institute for Nuclear Research, MSP 03680 Kyiv, Ukraine}
\affil[\tempk]{Nakhon Pathom Rajabhat University, Nakhon Pathom 73000, Thailand}
\affil[\tempq]{Department of Physics, Soongsil University, Seoul 06978, Korea}
\affil[\tempp]{Department of Physics, Seoul National University, Seoul 08826, Korea}
\affil[\tempr]{Institut Teknologi Bandung, Jawa Barat 40132, Indonesia}
\affil[\tempg]{Kirchhoff-Institute for Physics, D-69120 Heidelberg, Germany}
\affil[\tempb]{Baksan Neutrino Observatory of INR RAS, Kabardino-Balkaria 361609, Russia}
\affil[\tempe]{Ehwa Womans University, Seoul 03760, Korea}
\affil[\tempc]{Center for Underground Physics, Institute of Basic Science, Daejeon 34047, Korea}
\affil[\tempo]{Semyung University, Jecheon 27136, Korea}
\affil[\temps]{Institute of Theoretical and Experimental Physics, Moscow 117218, Russia}
\affil[\tempt]{National Research Nuclear University MEPhI, Moscow, 115409, Russia}
\affil[\temph]{Korea Research Institute for Standard Science, Daejeon 34113, Korea}
\affil[\tempu]{Tsinghua University, 100084 Beijing, China}
\affil[\tempn]{Department of Physics, Sejong University, Seoul 05000, Korea}
\affil[\tempa]{Department of Physics, Abdul Wali Khan University, Mardan 23200, Pakistan}
\affil[\tempd]{Department of Physics, Chung-Ang University, Seoul 06911, Korea}
\affil[\tempj]{University of Mataram, Nusa Tenggara Bar. 83121, Indonesia}

\maketitle

\newgeometry{margin=1.5in}

\begin{abstract}
\input{tex/abstract}
\end{abstract}

\addcontentsline{}{chapter}{}
\tableofcontents
\newpage

\input{tex/introduction}
\input{tex/science_goal}
\input{tex/CaMoO4_crystal}

\input{tex/cryogenic_particle_det}
\input{tex/experimental_design}

\input{tex/simulation_tools}

\input{tex/enriched_materials}
\input{tex/offline_software}
\input{tex/time_schedule}

\bibliographystyle{elsarticle-num}
\bibliography{amore_tdr}

\end{document}

%% file: tex/abstract.tex
The AMoRE (Advanced Mo-based Rare process Experiment) project is a series of experiments that
use advanced cryogenic techniques to search for the neutrinoless double-beta decay of \mohundred.
The work is being carried out by an international collaboration of researchers from eight countries.
These searches involve high precision measurements of radiation-induced temperature changes
and scintillation light produced in ultra-pure \Mo[100]-enriched and \Ca[48]-depleted
calcium molybdate ($\mathrm{^{48depl}Ca^{100}MoO_4}$) crystals that are located in a deep underground laboratory in Korea.
The \mohundred  nuclide was chosen for this \zeronubb decay search because of its high
$Q$-value and favorable nuclear matrix element. Tests have demonstrated that \camo crystals produce
the brightest scintillation light among all of the molybdate crystals, both at room and at cryogenic
temperatures.  $\mathrm{^{48depl}Ca^{100}MoO_4}$ crystals are being operated at milli-Kelvin temperatures
and read out via specially developed metallic-magnetic-calorimeter (MMC) temperature sensors that have
excellent energy resolution and relatively fast response times.  The excellent energy resolution
provides good discrimination of signal from backgrounds, and the fast response time is important
for minimizing the irreducible background caused by random coincidence of two-neutrino double-beta
decay events of \mohundred nuclei. Comparisons of the scintillating-light and phonon yields and
pulse shape discrimination of the phonon signals will be used to provide redundant rejection of
alpha-ray-induced backgrounds. An effective Majorana neutrino mass sensitivity that reaches the
expected range of the inverted neutrino mass hierarchy, i.e., 20-50 meV, could be achieved with
a 200~kg array of $\mathrm{^{48depl}Ca^{100}MoO_4}$ crystals operating for three years.   

%% file: tex/introduction.tex
\chapter{Introduction}
Even though we now know that neutrinos have mass, their absolute masses and their
fundamental nature still remain a mystery~\cite{Mohapatra07, Giunti09}. The Standard Model
of three-neutrino mixing has been firmly established by a number of neutrino oscillation
measurements, including the recent determination of $\theta_{13}$ by the Daya Bay, Double-Chooz
and RENO experiments. However, mixing measurements do not discriminate between Majorana- and
Dirac-type neutrinos and only provide information on mass differences, not on mass values
themselves.  At present, five fundamental questions about neutrinos remain:
\begin{enumerate}[(i)]

\item Are they Majorana-type or Dirac-type?
\item What is their absolute mass scale?
\item What is the mass hierarchy between the three neutrinos?
\item Is lepton number conserved?
\item Does neutrino mixing violate CP symmetry? 
\end{enumerate}
Among these, the first question about the nature of neutrinos is most fundamental in that it
remains the major unknown aspect of the Standard Model.  Moreover it is an essential element
for any theoretical model of neutrino masses.

The universe is comprised of matter and not antimatter; the cause of this matter-antimatter
asymmetry is not understood.  Since the now well established $CP$-symmetry violation in the quark
sector is not sufficient to generate the observed matter excess, particle physicists suspect
that the neutrino sector may be responsible for the current matter-antimatter asymmetry
(by a process called leptogenesis). If this turns out to be the case, this will solve one of the
most interesting and fundamental puzzles about the development of the  universe. However, even if
$CP$ is observed to be violated in neutrino oscillation experiments, the theory for a leptogenesis-induced
matter-antimatter asymmetry depends sensitively on whether or not neutrinos are Majorana particles.
Investigation of neutrinoless double-beta decays ($\zeronubb$) is the only practical way to
determine the nature of the neutrino (Majorana or Dirac particle), check lepton-number conservation,
and determine the absolute scale and the neutrino mass
hierarchy~\cite{Giunti09, Vergados02, Fogli07, Avignone08, Rodejohann11, Vergados12, Barea12}.   

The half-life of $\zeronubb$ decay, $T_{1/2}(\zeronubb)$, is related to the effective Majorana
neutrino mass ($\meff$) and nuclear matrix element ($M_{0\nu}$), as follows:
\begin{equation}
 [T^{0\nu}_{1/2}]^{-1}=G_{0\nu}|M_{0\nu}|^{2}\left(\frac{\langle m_{\beta\beta}\rangle}{m_e}\right)^2 ,
\end{equation}
where $G_{0\nu}$ is the phase-space factor and \meff is given by
\begin{equation}
\langle  m_{\beta\beta}\rangle=\Sigma m_iU_{ei}^2 \sim 
\frac{1}{2} \left|m_1+m_2 e^{2i\beta}+m_3e^{2i(\gamma-\delta)} \right|.
\end{equation}
Here $U_{ei}$ is the Pontecorvo-Maki-Nakagawa-Sakata (PMNS)
neutrino mixing matrix, $m_i$ are the light Majorana neutrino eigenstate
masses, and $\beta$, $\gamma$ and $\delta$ are CP-violating phases.

\begin{figure}\centering
  \includegraphics[width=0.8\textwidth]{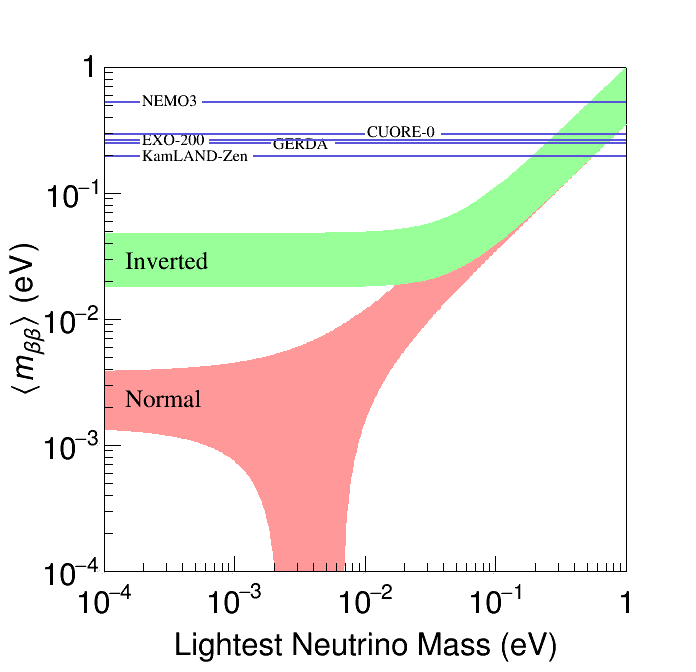}
\caption{Effective Majorana neutrino mass (\meff ) as a function of the mass of the lightest neutrino
together with current limits on the mass from a selection of the most sensitive experiments~\cite{Alfonso15,Auger12,AgostiniGERDA15,Gando13,BarabashNEMO11}.  To convert the experimental half life limits to \meff limits, the products of  $G_{0\nu}|M_{0\nu}|^{2}$ were chosen as the central values from the ranges given in~\cite{Faessler12}. (See Sec.~2.5 for the sensitivity of AMoRE-II experiment after five years of data taking). }  
\label{fig:neutrino_mass}
\end{figure}

Neutrino oscillation experiments give the mass differences:
$\Delta m_{23}^2\sim 2.43\times 10^{-3}\  \mathrm{eV}^2$,
$\Delta m_{12}^2\sim 8.0\times 10^{-5} \ \mathrm{eV}^2$,
and the mixing angles. 
The expected value of \meff is shown as a function of the smallest neutrino mass
in Fig.~\ref{fig:neutrino_mass}. For high \meff
values (and correspondingly lower \zeronubb half-lives), the neutrino mass scale is larger
than the mass differences and the neutrinos are nearly degenerate. If the value of \meff
approaches that of the mass differences, \meff lies in one of the horizontal bands, depending
on the neutrino mass hierarchy;  for the inverted hierarchy, the value of \meff is in
the 0.02 -- 0.05~eV range. This interval of neutrino masses could be accessed with a zero
background $\zeronubb$ detector with a total mass of several hundred kilograms. On the other
hand, orders-of-magnitude larger-scale experiments would be needed to access the normal neutrino
mass hierarchy.

It should be stressed that \zeronubb decay-like signals could result from  the influence of
hypothetical, beyond the Standard Model particles and/or interactions. This might involve,
for example, an admixture of right-handed currents in weak interactions, the emission of massless
(or very light) Nambu-Goldstone bosons (so-called Majorons), or a variety of other processes that
have been proposed in extensions of the SM ~\cite{Rodejohann11, Vergados12, Bilenky12, Deppisch12}.
Therefore, investigations of \zeronubb decay processes are powerful methods for searching
for beyond-the-SM effects.

The AMoRE experiment's aim is to search for $\zeronubb$ decay of \mohundred nuclei using \camo scintillating
crystals operating at milli-Kelvin temperatures. The ultimate goal of the experiment is to achieve
a sensitivity that covers the entire \meff range that is allowed by the inverted neutrino mass
hierarchy.  This will require advancing the current states-of-the-art
in background rejection, radio-pure crystal growing, and cryogenic
radiation detector techniques.  To accomplish these ambitious goals, we have formed an international
team of researchers that includes experts in each of these areas and plan on using a phased approach
that starts with modest experiments with current state-of-the-art
technologies and advance from there.  This report describes the status of our R\&D efforts and
accomplishments in radio-pure crystal growing, cryogenic detection, background simulation and suppression, and the
current status of, and plans for, our step-wise approach to a full experiment with a 200~kg
detector.  

%% file: tex/science_goal.tex
\chapter{Scientific goals}
\section{Brief summary of  $\mathrm{\beta\beta}$ experiments}
Many experimental techniques for double-beta-decay searches have been developed, starting with the
first experiment in 1948 that used Geiger counters, but did not see \bb any signal~\cite{Fireman48}. The SM-allowed,
second-order weak two-neutrino double-beta (\twonubb ) decay mode was first observed in tellurium and
selenium by means of geochemical techniques (for a review, see, e.g.,~\cite{Manuel91}). The first
observation of \twonubb decay in a direct counting experiment was accomplished in 1987 with an apparatus that consisted
of a time-projection chamber that surrounded a thin \seeightytwo film source~\cite{Elliott87}. At present, \twonubb
decay modes have been detected for eleven nuclides: \cafortyeight, \geseventysix, \seeightytwo, \zrnintysix,
\mohundred, \cdonesixteen, \teonetwentyeight, \teonethirty, \xeonethirtysix, \ndonefifty and \U (for reviews, 
see refs.~\cite{Tretyak02, Barabash10, Saakyan13} and references therein).  Indications for
two-neutrino double-electron capture in $\rm ^{78}Kr$~\cite{Gavrilyuk13} and \baonethirty \cite{Meshik01, Pujol09}
have also been reported.

In contrast, despite more than sixty years of experimental effort, no unambiguous examples of \zeronubb decays have
yet to be observed. The best half-life limits are at levels of $T_{1/2} = 10^{23} \sim 10^{25}$~yrs, depending on the
nuclide (see~\cite{Tretyak02, Elliott12, Giuliani12, Cremonesi14} and results of recent
experiments~\cite{Albert14, Agostini13, Asakura14, Arnold14, Alfonso15}). These half-life limits have been used to
restrict (using currently available theoretical calculations of the nuclear matrix elements) the effective Majorana
neutrino mass to be below the level of \meff$\sim (0.2 - 2)$ eV. The best limits on the half-lives for the most
studied nuclei and the effective Majorana neutrino mass are summarized in Table~\ref{tab:dbb_limits}.

There is also a claim by Klapdor-Kleingrothaus {\it et al.} for the detection of \zeronubb decays of
\geseventysix with a half-life of $T_{1/2}\sim 2\times 10^{25}$~yrs (\meff$>\sim 0.3$~eV)~\cite{Klapdor06}.
This was derived from the data of the
Heidelberg-Moscow experiment~\cite{Klapdor01} that utilized an 11~kg array of $\mathrm{^{76}Ge}$-enriched high-purity
germanium (HPGe) detectors~\cite{Kla04, Kla04b}. In the context of the SM, this half-life implies
nearly degenerate neutrino masses.  This claimed observation was criticized in
refs.~\cite{Aalseth02a, Feruglio02, Zdesenko02} and was recently challenged by the GERDA experiment that
used a similar HPGe spectrometry technique~\cite{Agostini13} and found no evidence for $\zeronubb$ decay of 
\geseventysix and set a lower limit at the level of $\lim T_{1/2} = 2.1\times 10^{25}$~yr. Unfortunately, the GERDA
experiment exposure (21.6~kg$\times$yr) was sufficient to permit only a marginal exclusion of the Klapdor claim.

\begin{table}
  \caption{Half-life and Majorana-neutrino-mass limits from the most sensitive
    neutrinoless double-beta decay experiments.}
  \input{tables/table2-1.tex}
  \label{tab:dbb_limits}
\end{table}

\begin{figure}\centering
  \includegraphics[width=0.8\textwidth]{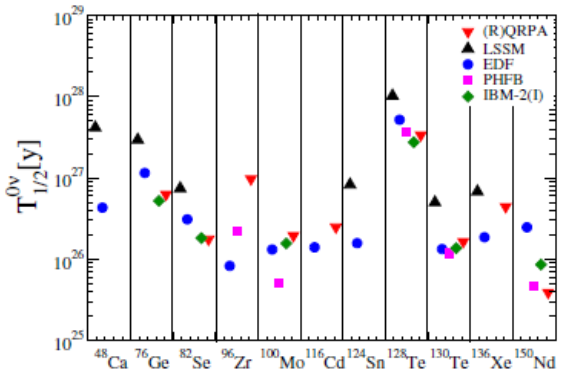}
\caption{The $0\nu\beta\beta$-decay half-lives calculated for an effective Majorana neutrino mass \meff$=0.05$~eV using different
theoretical methods to calculate the nuclear matrix element (Figure 24 from ref.~\cite{Vergados12}).}
\label{fig:0nubbhalf}
\end{figure}

However, as one can see in Fig.~\ref{fig:neutrino_mass}, even the most sensitive current experiments only explore
the degenerate region of the neutrino mass pattern.  Several new experiments have been proposed to reach the minimum
\meff region allowed for the inverted neutrino mass hierarchy, i.e., \meff$\leq 0.05$~eV (see reviews in 
refs.~\cite{Avignone08, Elliott12,  Giuliani12, Cremonesi14, Giuliani10, Barabash12, Gomez12, Schwingenheuer13,Bilenky15} and
original proposals~\cite{Arnaboldi04, Arnold10, Beeman12a, Alvarez12, Gomez14, Graham14, Ishidoshiro14, Heisel14,
Abgrall14, Bongrand13, Jones14}). To cover this region, the half-life sensitivity of the next generation experiments
should be at the level of $T_{1/2} \sim 10^{26}-10^{27}$~yrs (see Fig.~\ref{fig:0nubbhalf}).

To accomplish these ambitious goals, detectors are required to: sample a sufficiently large number of $\beta\beta$-active
nuclei  (i.e., $10^{27} \sim 10^{28}$ nuclei, which corresponds to $10^3 \sim 10^4$ moles of the isotope of interest); 
have as low as possible (ideally zero) radioactive background and as high as possible (ideally 100\%) detection efficiency;
have the ability to distinguish a \zeronubb signal from different background components; and, especially, have excellent
energy resolution. 

The choice of the candidate nuclide is determined by the scale of the experiment, the extreme background requirements, 
the possibility of using calorimetric  (``source=detector'') techniques to achieve high detection efficiency, and the energy
resolution.  Among the theoretically most promising nuclides (see Fig.~\ref{fig:0nubbhalf}), the mass production
of enriched isotopes is available for  $^{76}$Ge, $^{82}$Se, $^{100}$Mo, $^{116}$Cd, $^{130}$Te and $^{136}$Xe~\cite{Giuliani10, Barabash12}.
Suppression of background is easier for $\beta\beta$ nuclides that have a $Q_{\beta\beta}$-value that is above the 2615~keV gamma
line from $^{208}$Tl (a daughter nuclide in the $^{232}$Th chain). This limits the most promising candidate nuclides to
$^{82}$Se, $^{100}$Mo and $^{116}$Cd.

\section{Choice of $\mathrm{^{100}}$Mo for $\mathrm{0\nu\beta\beta}$ experiments}
\begin{figure}\centering
  \includegraphics[width=0.8\textwidth]{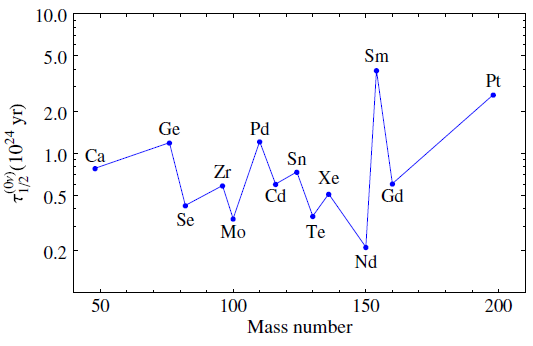}
  \caption{The computed neutrinoless double-beta decay half-lives for various nuclei for \meff$= 1$~eV
(Figure 2 from ~\cite{Barea12}).}
  \label{fig:fig2-2}
\end{figure}

Among the double-beta-decaying nuclides, we selected \mohundred for the AMoRE experiment because of its high transition
energy ($Q_{\beta\beta} = 3034.40(17)$~keV~\cite{Rahaman08}), relatively large natural isotopic abundance
($\delta = 9.82(31)\%$~\cite{Berglund12}), and the encouraging theoretical predictions for the nuclear
matrix-element~\cite{Rodin06a, Rodin06b, Kortelainen07, Simkovic08, Faessler08, Barea09, Kotila10, Rath10, Rodriguez10}.
As shown in Figs.~\ref{fig:0nubbhalf} and~\ref{fig:fig2-2}, the \zeronubb half-life for \mohundred is expected to be
relatively shorter than those for other candidate nuclei. It should be noted that hundreds of kilograms of the \mohundred isotope
can be enriched at a reasonable price by centrifugation methods~\cite{Barabash14}.

The most precise measurement to-date of the \mohundred \twonubb half-life was reported by the NEMO-3 experiment:
$T_{1/2} = (7.11 \pm 0.54) \times 10^{18}$ years~\cite{Arnold05}. The best 90\% confidence-level lower limits on the half-life
for the \zeronubb mode of \mohundred are $T_{1/2}(0\nu\beta\beta) > 5.5 \times 10^{22}$~yrs from the ELEGANT~V
experiment~\cite{Ejiri01}, which used a 171~g sample of 94.5\%-enriched \mohundred, and
$T_{1/2}(0\nu\beta\beta) > 1.1 \times 10^{24}$~yrs from NEMO-3, which used a 6.9~kg \mohundred sample~\cite{Arnold14}. Both
experiments used a spectroscopic (passive source) technique with tracking devices exposed to thin Mo foils. In both cases,
the energy resolution was moderate (8\% $\simeq$12\% FWHM) and the detection efficiency (11\%$\simeq$19\%) was relatively low.
AMoRE is using a calorimetric (source=detector) technique with active scintillating crystals that contain \mohundred.

Several molybdate crystal scintillators that would allow for the realization of a high-detection-efficiency experiment have been
identified. The most promising of these are ZnMoO$_4$~\cite{Beeman12a, Barabash14, Gironi10, Beeman12b, Beeman12c, Beeman12d, Chernyak13,
Berge14}, $\mathrm{CaMoO_4}$~\cite{Belogurov05, Annenkov08, KRISS_CMO, So13}, $\mathrm{CdMoO_4}$~\cite{Mikhailik06},
$\mathrm{PbMoO_4}$~\cite{Danevich10} and $\mathrm{Li_2MoO_4}$~\cite{Barinova10, Cardani13, Bekker14}. Among these,
$\mathrm{CaMoO_4}$  crystal scintillators have the advantages of a high mass-fraction of molybdenum (48\%), the highest
scintillation efficiency (it is the only molybdate that scintillates at room temperature), and excellent low-temperature
bolometric properties.

\section{Development of \camo crystal scintillators}
\begin{table}
  \input{tables/table2-3.tex}
\end{table}

Among the inorganic scintillating materials that contain Mo in the structural form of Scheelite or
Wolframite, \CMO produces the largest amount of scintillation light at room temperature.
We recognized the potential of \CMO crystals for a \mohundred \zeronubb decay search and started developing
\CMO crystals in 2002, when the first \CMO crystal was grown with the Czochralski (CZ) technique in Korea.
Discussions with V.~Kornoukhov (ITEP, Russia) at the 4th International Conference on Non-accelerator New Physics (2003) led to a collaborative R\&D program to study the application of $^{100}$Mo-enriched \CMO
crystals for a $0\nu\beta\beta$-decay search experiment. Kornoukhov subsequently sent Russian-grown \CMO
crystals to Korea for evaluation.  This R\&D program confirmed that \CMO is a promising material for a \zeronubb
search experiment and first results were presented in 2004~\cite{Kim04}. As part of these R\&D activities, other molybdate crystals and cryogenic readout techniques were considered~\cite{Belogurov05}. The INR (Ukraine) group led by F.~Danevich was invited to join the \CMO collaboration in 2006. This group, which had already performed a sensitive experimental search using a $\mathrm{CdWO_4}$ scintillating crystal~\cite{Danevich03,Polischuk15} that demonstrated the applicability of scintillation crystal techniques for \zeronubb decay searches, added valuable operational experience to our team. Some \CMO crystals were produced at the Institute of Materials~(IM), Lviv, Ukraine, and results from characterization and background studies of these crystals were published in Ref.~\cite{Annenkov08}.

These R\&D efforts gained support from the International Science and Technology Center (ISTC), an
intergovernmental organization connecting scientists from Russia and other countries of the Commonwealth of
Independent States (CIS) with their peers in other countries, including Korea. Our first ISTC Project (\#3293)
was carried out between 2005 and 2007. A large, high-quality crystal (200~mm$\times$30~mm$\times$30~mm) was grown~\cite{Korzhik08}
 and characterized~\cite{IEEE572010}. Figures~\ref{fig:fig3-3},~\ref{fig:fig3-4} and~\ref{fig:fig3-5} show photographs
of $\mathrm{CaMoO_4}$ crystals produced at different places. The initial $\mathrm{CaMoO_4}$ crystals were produced by the
Bank at Pusan National University, Korea.  Subsequently, better quality and larger $\mathrm{CaMoO_4}$ crystals
were produced by the Innovation Center of the Moscow Steel and Alloy Institute (ICMSAI) in Moscow, Russia, while others
were produced at the Institute of Materials (IM) in Lviv, Ukraine.

In 2007, a KRISS (Korea Research Institute for Standards and Science) group specialized in advanced detector development
joined our collaboration and initiated the development of cryogenic techniques for reading out \CMO crystals; results
from their initial studies were published in Ref.~\cite{KRISS_CMO}.  A second ISTC Project started in
2008 with the primary goal of growing a \doublecamo crystal with 3~kg of $^{100}$Mo-enriched and $^{48}$Ca-depleted,
radio-pure powders.  Characterization and background studies of these crystals were reported in Ref.~\cite{IEEE592012}.

The AMoRE collaboration, with members from nine different institutions from five countries was officially formed in 2009.
The collaboration has subsequently expanded to $\sim$90~researchers from 17~institutions from eight countries.  
Recently, the Russian members of the AMoRE collaboration were awarded a \$8.7M (US) grant to support further
development of large-scale quantities of ultra-low-background \doublecamo crystals for the 200~kg phase of AMoRE.

\section{AMoRE project - Parameters}
The basic parameters of the AMoRE experiment are summarized as follows:
\begin{itemize}
\item	\doublecamo cryogenic scintillating detectors enriched in
  \mohundred and depleted in \cafortyeight
\item	$^{100}$Mo enrichment  $> 95\%$
\item	Operating temperature: 10-30 mK
\item	Energy resolution: 5 keV @ 3 MeV
\item	Individual detector-element mass: 0.5~kg
\item	Location: Y2L (Yangyang underground laboratory) 700 meter depth
         (AMoRE-pilot, AMoRE Phase I) \& a new deeper underground
         laboratory (for AMoRE Phase II) 
\item	Collaboration: 8 countries, 17 institutes, $\sim$90 researchers
\item	Phases\\[10pt]
\input{tables/table2-4.tex}

\end{itemize}

\section{Sensitivity of AMoRE}
\begin{figure}\centering
  \includegraphics[width=0.8\textwidth]{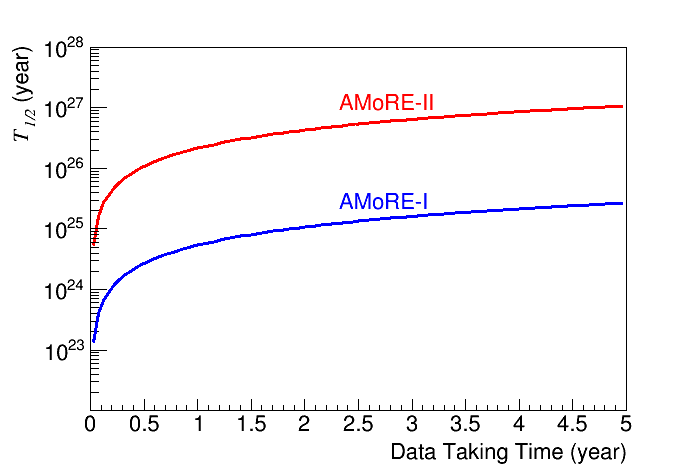}
\caption{Expected half-life sensitivity for AMoRE-I and AMoRE-II as a function of running time.}
  \label{fig:fig2-3}
\end{figure}

\begin{figure}\centering
  \includegraphics[width=0.8\textwidth]{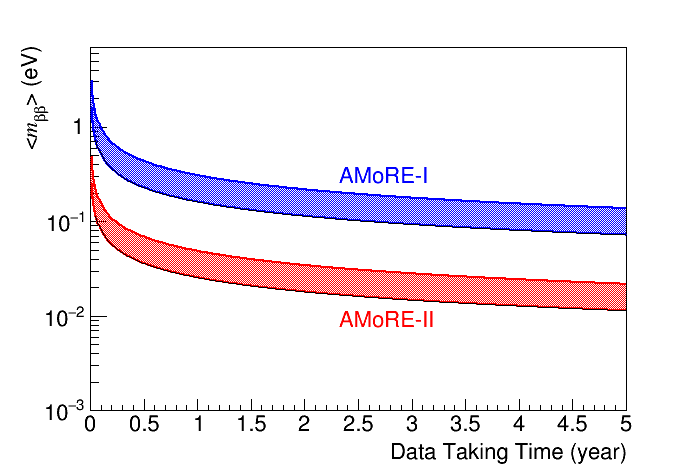}
\caption{Effective Majorana neutrino mass (\meff ) sensitivity vs. data taking time for AMoRE-I and AMoRE-II. The widths of the bands correspond to a range of nuclear matrix element calculations from Refs.~\cite{Barea12,Faessler12,Simkovic09}. }
  \label{fig:fig2-4}
\end{figure}

The sensitivity for a $0\nu\beta\beta$ experiment is usually defined as the half-life level,
$T_{1/2}^{0\nu}$, at which the expected signal would have the same strength as a $1\sigma$ fluctuation
of the background level. For a source=detector arrangement, this is given by
\begin{equation}
T_{1/2}^{0\nu}=\ln{2}\times N_A \frac{a}{A}\varepsilon\sqrt{\frac{M\cdot t}{b\cdot \Delta E}}\times 10^3,
\end{equation}
where $N_A$ is Avogadro's number, $a$ is the istopic abundance, $\varepsilon$ is the detection efficiency,
$A$ is the (dimensionless) atomic mass number of the decaying nuclide, $M$ is the total detector mass of the nuclide of
interest (in kilograms), $t$ is the exposure time (in years), $b$ is the background level in the signal region
(in counts/keV/kg/yr) and $\Delta E$ is the energy window where the signal is expected, in keV (comparable to FWHM energy resolution).  For the AMoRE experiment, where we aim for
``zero-background\footnote{Less than 0.1 events in the experimental region of interest},'' the experimental
half-life sensitivity can be expressed as:
\begin{equation}
T_{1/2}^{0\nu}=\ln{2}\times  \frac{a\ \varepsilon \ N_A}{A} \frac{M\cdot t}{n_{\rm CL}}\times 10^{3} ,
\end{equation}
where $n_{\rm CL}=2.4$ corresponds to a 90\% confidence level.  This expression gives a sensitivity that is,
at least in principle, independent of both the background level and the detector resolution. In the following,
the zero-background relation is used for the evaluation of the experiments, including those in the current proposal.
Figure~\ref{fig:fig2-3} shows the resulting $n_{CL}=2.4$, 90\% confidence level \zeronubb half-life sensitivities
\emph{versus} time for 10~kg and 200~kg arrays of enriched \CMO. The effective Majorana neutrino mass (\meff ) sensitivity as a function of data taking time
is shown in Fig.~\ref{fig:fig2-4}.

Our strategy for achieving a large mass ``zero-background'' experiment includes the following considerations:
\begin{itemize}
\item Since the highest energy single-$\beta/\gamma$ quanta from naturally occurring radio-nuclides is the
2615~keV $\gamma$-line from $\mathrm{^{208}Tl}$, we have focused on candidate \zeronubb nuclides with $Q_{\beta\beta}$ values
above this level. Our currently preferred nuclide is $\mathrm{^{100}Mo}$, which has $Q=3034$~keV.
\item Since most naturally occurring $\alpha$ emitters have $Q_{\beta\beta}$-values above 3034~keV, we need powerful and
reliable methods to distinguish $\alpha$-induced signals from $\beta /\gamma$-induced signals. We have demonstrated
that in \CMO crystals, the amount of scintillation light and the pulse-shape of the temperature signals provide two
independent $\beta /\gamma$-$\alpha$ discriminators, each with a more than $5\sigma$ discrimination power.
\item The effects of accidental time overlap of two low-energy signals combining to form one of higher
energy (i.e., pileup) will be controlled by segmenting the detector array into $\sim$400~g independent elements and by application of pulse-shape discrimination of randomly coinciding events. 
\item In the case of \cmo , background from \twonubb decay of $\mathrm{^{48}Ca}$ will be reduced by using
crystals produced from calcium depleted in $^{48}$Ca.
\item We will continue and expand our aggressive R\&D program on reducing radioactive contaminations
in the crystal powder preparation and growing procedures that was started over ten years ago.
\item The levels of radioactive contaminants in all materials used in the detector, including the cryostat elements and the external-radioactivity shields, will be measured and their influence on the experiment will be carefully simulated.
\item The experiment will be done in a series of phases where at each phase we will exploit what is learned from
the previous one. 
\item The cosmic muon flux will be minimized by the underground location of the experiments and by installation of an active muon veto system.
\end{itemize}

\begin{figure}\centering
  \includegraphics[width=0.8\textwidth]{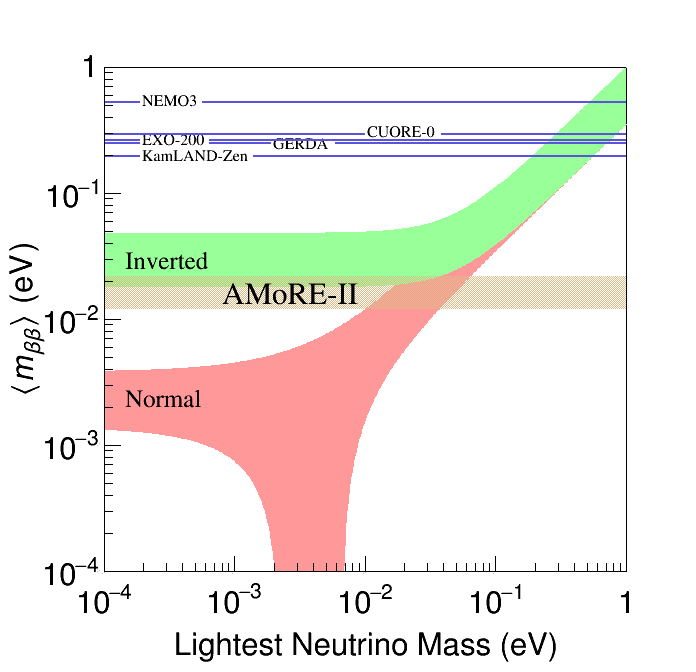}
\caption{Effective Majorana neutrino mass (\meff ) as a function of the mass of the lightest neutrino
together with current limits on the mass from a selection of the most sensitive experiments~\cite{Alfonso15,Auger12,AgostiniGERDA15,Gando13,BarabashNEMO11}, as compared with expectations for AMoRE-II with five years of data taking.
To convert the experimental half life limits to \meff limits, the products of  $G_{0\nu}|M_{0\nu}|^{2}$ were chosen as the central values from the ranges given in~\cite{Faessler12}.  For comparison with the inverted hiearchy region, the  AMoRE-II band is shown using the full range of values.} 
\label{fig:hierarchy-constraint}
\end{figure}

As mentioned above, we plan to perform $\mathrm{CaMoO_4}$ experiments in two phases. We are currently commissioning an
array of five $\mathrm{^{100}Mo}$-enriched \doublecamo crystals, with total mass of~1.5~kg in a cryostat located
in the A5 area of Y2L, as an initial ``pilot'' experiment.  We will start data-taking with this setup later
this year (2015) and continue to take data for about a year. Using measured background rates in
these detectors (as discussed below), we estimate a background rate in the \zeronubb signal region
of $\sim$0.01~counts/keV/kg/yr and a half-life sensitivity of $T^{0\nu}_{1/2}>1.1\times 10^{24}$~yrs, which is
comparable to the current world-best value from NEMO 3~\cite{Arnold14}.   While this pilot experiment is running,
we will exploit the experience we gain from implementing, commissioning and operating it to prepare and assemble
``AMoRE-I,'' a $\sim 5$~kg array of \CMO crystals that will, when ready, replace the pilot array. Our goal
for AMoRE-I is a background level on the order of $10^{-3}$~counts/keV/kg/yr and a $T^{0\nu}_{1/2}$ sensitvity for a
two-year exposure of $\sim 8\times 10^{24}$~yrs, which would correspond to an effective Majorana neutrino
mass in the range of 0.07 eV -- 0.14 eV (see Fig.~\ref{fig:fig2-4}), and could confirm or rule out the controversial
$\zeronubb$ evidence reported by Klapdor.   The next phase will be ``AMoRE-II,'' which will start with a 70~kg array
of  \doublecamo crystals with a background level that is an order-of-magnitude below that for AMoRE-I
is planned.  As background conditions permit, we will increase the detector mass to as much as $\sim 200$~kg.  The
projected sensitivity of a five-year exposure with AMoRE-II would be $T_{1/2} \approx 1\times 10^{27}$ years
(\meff$\approx 20$~meV).  Figure~\ref{fig:hierarchy-constraint} compares the projected sensitivity with the inverted neutrino hierarchy region and existing limits. Such a sensitivity will require exquisite energy resolution and an order-of-magnitude
improvement in background rejection power.  Since the available space at the existing Y2L underground laboratory will
not be sufficient to accommodate an experiment of this scale, we are proposing the development of a larger and deeper
underground laboratory in Korea. 

%% file: tables/table2-1.tex
\begin{tabular}{|p{0.9cm}cp{1.3cm}ccp{1.8cm}p{1.cm}p{1cm}|}																	
\hline																	
	Nucleus	&$Q_{\beta\beta}{\rm \cite{Wang12}}	$\newline $\rm	[keV]	$&	Natural \newline abundance ~\cite{Berglund12}	&$	T_{1/2}	$\newline $\rm	[years]	$&$\rm	\langle m_{\beta\beta} \rangle	$\newline $\rm	[eV]	$&	Experiment\newline Technique	&	Enrich-ment	\newline [\%]	&	Ref.	\tabularnewline	
$\rm		$&&&&&		&	&		\tabularnewline	\hline\hline
$\rm	^{48}Ca	$&$\rm	4267.0(4)	$&$\rm	0.187(21)	$&$\rm	> 5.8\times10^{22}	$&$\rm	< 3.5\text{~--~}22 	$&	Elegant VI\newline $\rm CaF_2(Eu)$ scintillator	&	Nat.	&	~\cite{Umehara08}	\tabularnewline	\hline
$\rm	^{76}Ge	$&$\rm	2039.06(1)	$&$\rm	7.73(12)	$&$\rm	> 1.9\times10^{25}	$&$\rm	< 0.35	$&	H-M,\newline HPGe	&	86	&	~\cite{Klapdor01}	\tabularnewline	
$\rm		$&$\rm		$&$\rm		$&$\rm	> 1.57 \times10^{25}  	$&$\rm	< 0.33\text{~--~}1.35	$&	IGEX,\newline HPGe	&	86	&	~\cite{Aalseth02b}	\tabularnewline	
$\rm		$&$\rm		$&$\rm		$&$\rm	2.23^{+0.44}_{-0.31}	$&$\rm	0.32\pm0.03	$&	HPGe	&	86	&	~\cite{Klapdor06}	\tabularnewline	
$\rm		$&$\rm		$&$\rm		$&$\rm	> 2.1\times10^{25}	$&$\rm	< 0.2\text{~--~}0.4	$&	GERDA,\newline HPGe	&	86	&	~\cite{Agostini13}	\tabularnewline	\hline
$\rm	^{82}Se	$&$\rm	2996.4(15)	$&$\rm	8.73(22)	$&$\rm	> 3.6\times10^{23}	$&$\rm	< 0.89\text{~--~}2.43	$&	NEMO-3,\newline tracking	&	97	&	~\cite{BarabashNEMO11}	\tabularnewline	\hline
$\rm	^{100}Mo	$&$\rm	3034.37(40)	$&$\rm	9.82(31)	$&$\rm	> 1.1\times10^{24}	$&$\rm	< 0.3\text{~--~}0.9	$&	NEMO-3\newline tracking	&	95-99	&	~\cite{Arnold14}	\tabularnewline	\hline
$\rm	^{116}Cd	$&$\rm	2813.50(13)	$&$\rm	7.49(18)	$&$\rm	> 1.7\times10^{23}	$&$\rm	< 1.5\text{~--~}1.7	$&	Solotvina,\newline $\rm^{116}CdWO_4$ scintillator	&	83	&	~\cite{Danevich03}	\tabularnewline	\hline
$\rm	^{128}Te	$&$\rm	866.5(9)	$&$\rm	31.74(8)	$&$\rm	> 8.1\times10^{24}	$&$\rm	< 1.1\text{~--~}1.5	$&	Geochem.	&		&	~\cite{Bernatow93}	\tabularnewline	\hline
$\rm	^{130}Te	$&$\rm	2527.51(1)	$&$\rm	34.08(62)	$&$\rm	> 4.0\times10^{24}	$&$\rm	< 0.27\text{~--~}0.76	$&	Cuoricino, CUORE-0\newline Cryogenic TeO$_2$ bolometer	&	Nat.	&	~\cite{Alfonso15}	\tabularnewline	\hline
$\rm	^{136}Xe	$&$\rm	2457.99(27)	$&$\rm	8.8573(44)	$&$\rm	> 1.1\times10^{25}	$&$\rm	< 0.19\text{~--~}0.45	$&	EXO-200,\newline TPC	&	80.6	&	~\cite{Albert14}	\tabularnewline	
$\rm		$&$\rm		$&$\rm		$&$\rm	> 2.6\times10^{25}	$&$\rm	< 0.14\text{~--~}0.28	$&	KamLAND-Zen,\newline Liquid scintillator	&	90	&	~\cite{Asakura14}	\tabularnewline	\hline
$\rm	^{150}Nd	$&$\rm	3371.38(20)	$&$\rm	5.638(28)	$&$\rm	> 1.8\times10^{22}	$&$\rm	< 4.0\text{~--~}6.3	$&	NEMO-3\newline tracking	&	91	&	~\cite{Argyriades09}	\tabularnewline	\hline
\end{tabular}																	

%% file: tables/table2-3.tex
\begin{tabular}{|p{2cm}p{12cm}|}						
\hline						
	Year	&	Summary	\tabularnewline		\hline
	2002	&	\CMO idea for DBD, First \CMO crystal was grown in Korea	\tabularnewline		
	2003	&	ITEP(Russia)-Korea collaboration on \CMO R\&D	\tabularnewline		
	2004	&	First conference presentation	\tabularnewline		
	2005-2007	&	$\rm 1^{st}$ ISTC project for large \CMO growing 	\tabularnewline		
	2006	&	INR (Ukraine) joined the collaboration, \CMO grown in Ukraine	\tabularnewline		
	2007	&	R\&D on cryogenic readout techniques for \CMO started	\tabularnewline		
	2008	&	$\rm2^{nd}$ ISTC project for enriched \doublecamo growing	\tabularnewline		
	2009	&	AMoRE collaboration formed with institutions from 5 countries	\tabularnewline		
	2010-2011  	& \doublecamo crystal characterization and internal background studies	\tabularnewline		
	2012	&	ITEP group awarded 8.7M\$ for \doublecamo production line R\&D	\tabularnewline		
\hline						
\end{tabular}

%% file: tables/table2-4.tex
\begin{tabular}{|p{5cm}p{2.5cm}p{2.5cm}p{2.5cm}|} \hline
		&	AMoRE-pilot	&	Phase I	&	Phase II	\tabularnewline		\hline
Mass of \doublecamo	&	1.5 kg	&	5~kg	&	200~kg	\tabularnewline		
Background $\rm[counts/(keV\cdot kg \cdot year)]$ & $10^{-2}$ &
$10^{-3}$ & $10^{-4}$  \tabularnewline
$T_{1/2}$ sensitivity [years]	&  $3.2 \times 10^{24}$ & $2.7 \times 10^{25}$	& $1.1 \times 10^{27}$	\tabularnewline
\meff sensitivity [meV] & 210--400 &	70--140	& 12--22	\tabularnewline		
	Schedule	&	2015--2016	&	2016--2018	&	2018--2022	\tabularnewline		\hline
\end{tabular}										

%% file: tex/CaMoO4_crystal.tex
\chapter{\camobf crystal scintillators}
\section{Production of high-purity crystals}
\begin{figure}\centering
  \includegraphics[width=0.4\textwidth]{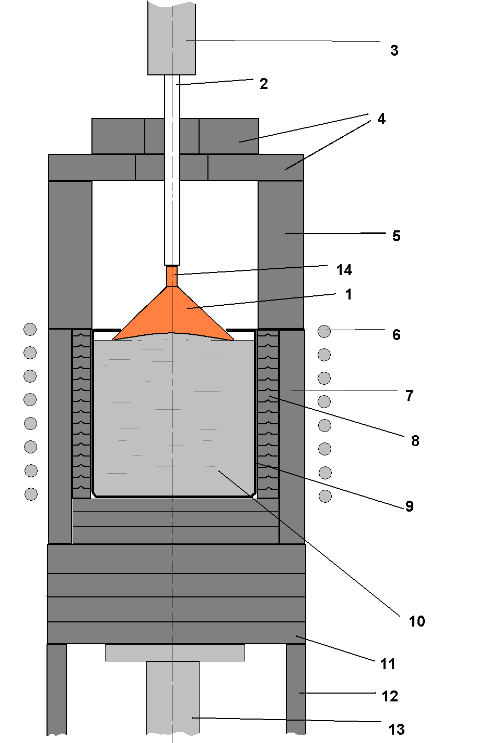}
\caption{a sketch of a Czochralski crystal-growing system. 1-- crystal; 2--seedholder; 3--main shaft;
4,5--heat insulation screens; 6--induction coil; 7,8,11-- heat insulation ceramics;
9--crucible; 10--melt; 12--bottom support; 13-- bottom shaft; 14--seed crystal}
\label{fig:fig3-1}
\end{figure}

\begin{figure}\centering
  \includegraphics[width=0.4\textwidth]{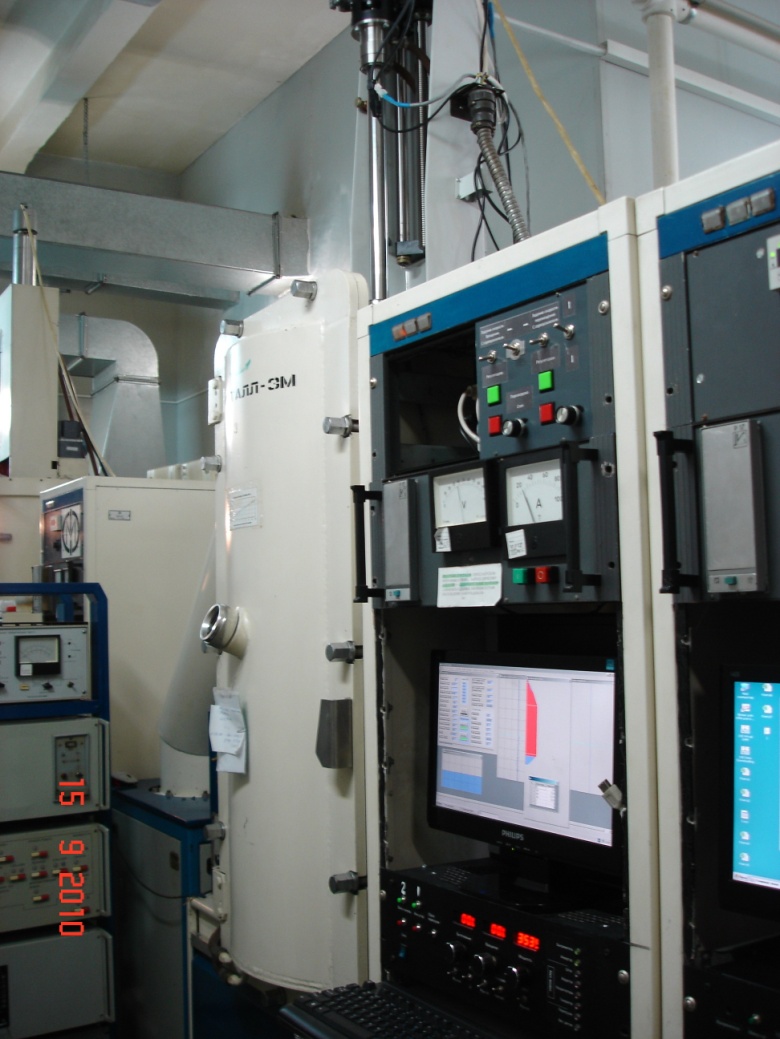}
\caption{A photograph of a CRYSTAL-3M Crystal puller (at JSC Fomos-Materials).}
  \label{fig:fig3-2}
\end{figure}

The Czochralski crystal-growing technique is a universally used method for production of high-quality
tungstate and molybdate crystals. For details see ref.~\cite{Nassau62}, the very first review of this technique.
For reviews of the history and status of the development of the Czochralski method see ref.~\cite{Brantle13}. 
A sketch of the Czochralski method is shown in Fig.~\ref{fig:fig3-1}; a photograph of an operational system
is shown in Fig.~\ref{fig:fig3-2}. Natural \camo occurs in nature as tetragonal-stolzite,
scheelite-type, and monoclinic-raspite crystals. X-ray diffraction measurements of synthetic, Czochralski-grown
\camo crystals have been determined to be scheelite-type crystals with a tetragonal symmetry
with a space group of $\mathrm{I4_{1/a}}$.

\begin{figure}\centering
\includegraphics[height=0.3\textwidth]{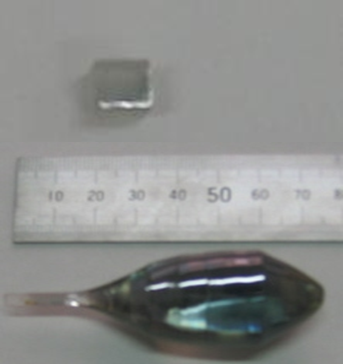}
\includegraphics[height=0.3\textwidth]{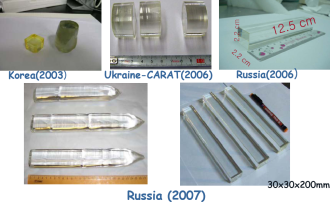}%
\caption{(left) The first \CMO crystal grown in Korea. (right) Various \CMO crystals grown in Korea, Ukraine and Russia.}
  \label{fig:fig3-3}
\end{figure}

Our initial \camo crystals were produced in Korea at Pusan National University's Crystal Bank.
Subsequently, larger and better-quality \camo crystals were produced at the Innovation Center of the Moscow
Steel and Alloy Institute (ICMSAI) in Moscow, Russia and at the Institute of Materials (IM) in Lviv, Ukraine
(see Fig.~\ref{fig:fig3-3}).

An extensive R\&D program on the optimization of growth techniques for the large-sized \camo
crystals was carried out in the framework of ISTC program \#3293~\cite{Korzhik08}. As part of this program,
crystals as large as 30$\times$30$\times$200 mm$^3$ were successfully grown, as shown in Figs.~3.4 and 3.5.
On the basis of this R\&D program, we established crystal growing techniques that are satisfactory for the
AMoRE experiment.

\begin{figure}\centering
  \includegraphics[width=0.7\textwidth]{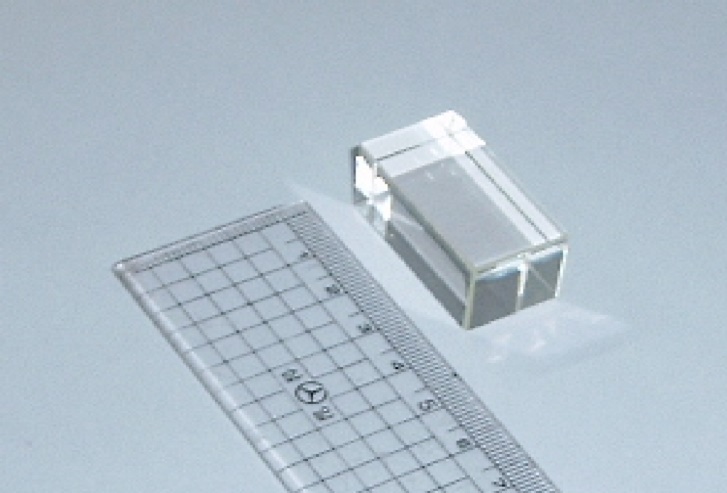}
\caption{The first \camo crystal from the ISTC project. Crystals with volumes as large as 15 cm$^3$ and
light yields of $\sim$ 400 photons/MeV were produced.}
\label{fig:fig3-4}
\end{figure}

\begin{figure}\centering
\includegraphics[height=0.3\textwidth]{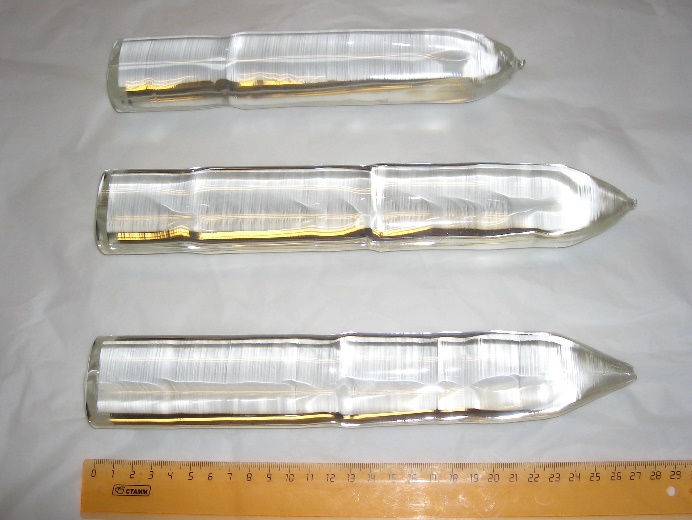}
\includegraphics[height=0.3\textwidth]{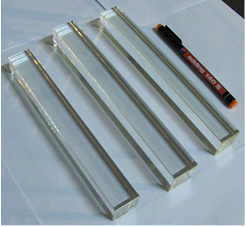}\\
\includegraphics[height=0.3\textwidth]{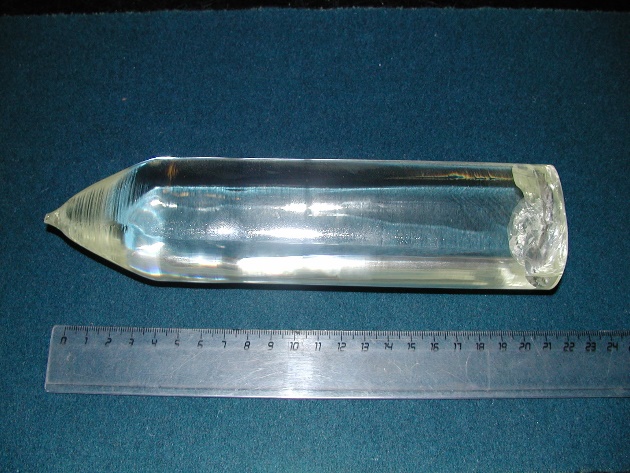}
\includegraphics[height=0.3\textwidth]{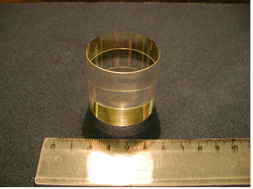}%
  \caption{Crystals grown as part of ISTC program \#3293 (Lead organization: ITEP (Moscow)).}
\label{fig:fig3-5}
\end{figure}

The technique for growing single \camo crystals can be briefly summarized as a series of steps:
\begin{enumerate}
\item Raw material preparation:
\begin{itemize}
\renewcommand{\labelitemi}{-}
\item Preparation of chemicals and chemical labware
\item Purification of Ca and Mo containing components
\item \camo raw material synthesis, drying and calcination
\end{itemize}
\item Melting (compactization) of raw material
\item Initial Czochralski growing of raw \camo single crystals
\item Final Czochralski growing of \camo single crystal and initial annealing
\item Principal annealing under high temperature in an oxygen atmosphere
\item Cutting, lapping and polishing to form the scintillation elements.
\end{enumerate}
\noindent

\subsection{Synthesis of \camo raw material}
There are two well-known techniques to synthesize
the \camo raw material (charge):
\begin{enumerate}
\item solid-phase synthesis of the oxides (CaO and $\rm MoO_3$) mixed in a stoichiometric ratio,
\item a co-precipitation reaction: 
\begin{equation}
\mathrm{(NH_4)_2MoO_{4} + Ca(NO_3)_2  \to CaMoO_4 + 2NH_4NO_3 }\,.
\nonumber
\end{equation}
\end{enumerate}
The advantages of the co-precipitation reaction are:
\begin{itemize}
\renewcommand{\labelitemi}{-}
\item a guaranteed stoichiometry of the crystal material;
\item  the possibility of including additional purification steps in the process;
\item ``the remains'' of the reaction (NH$_4$O$_3$) are easily removed by washing and heat treatment.
\end{itemize}
In the framework of the Russian government's Federal Aiming Program (FAP), we used a different Ca-compound
(calcium formate: Ca(HCOO)$_2$) instead of $\mathrm{Ca(NO_3)_2}$ because of its much higher purity~\cite{Danevich07}.

\noindent

\subsection{CaMoO$_4$ crystal growth} 

The \camo melting temperature (T=1445$^\circ$C) allows
for the crystals to be pulled in normal atmosphere from crucibles made from platinum ($T_\mathrm{melt} = 1769^{\circ}$C),
or in an oxygen-free atmosphere from an iridium crucible ($T_\text{melt} = 2454^\circ$C). Both approaches were
investigated during our preliminary tests. After analyses of the resulting samples, it was decided to use the
first method and pull crystals from a platinum crucible with diameter 90~mm and height 70~mm. In these experiments,
we also optimized the growth direction relative to the
crystallographic axis of the crystal. The crystal structure consists
of complex layers perpendicular to the $c$-axis. Each layer has a two-dimensional, CsCl-like cubic arrangement
with a Ca cation and $\mathrm{MoO_4}$ anion surrounded by eight ions of opposite sign (see Fig.~1 in
ref.~\cite{Senyshyn06}). There are at least two optimal directions for the seeding and further growth of the
\camo crystal. Crystals of similar quality can be pulled in the directions close to the $a$- or
the $c$-axis. The crystals grown along the $a$-axis exhibit an elliptically shaped cross-section with a
$\sim$1.3 ratio of maximal to minimal diameter.  We found that by tilting the $a$-axis by $\sim$25 degrees relative to the pulling
direction, we were able to produce a nearly circular cross-section,  having a diameter ratio of $\sim$1.15. 

\section{Luminescence and scintillation properties of \camo crystals }
\subsection{Luminescence properties of \camo crystals}
\begin{figure}\centering
\includegraphics[width=\textwidth]{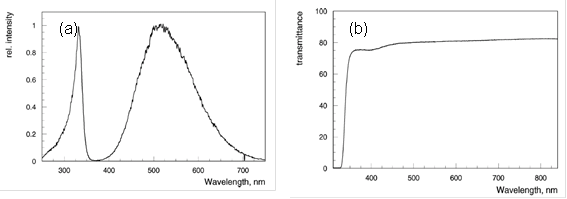}
  \caption{(a) The room temperature emission spectrum of a \CMO crystal. (b) The transmittance of the same \CMO
crystal (also at room temperature).}
\label{fig:fig3-6}
\end{figure}

The luminescence and transmission properties for some of the \camo crystal samples were investigated.
Figure~\ref{fig:fig3-6}(a) shows the excitation and emission spectra for one of the samples. The emission spectrum
has a broad peak between 400~nm and 700~nm with a maximum at around 520~nm. Figure~\ref{fig:fig3-6}(b) shows the
transmittance, which demonstrates that the crystal is highly transparent to the scintillation
light. 

\begin{figure}
  \centering
  \includegraphics[width=0.85\textwidth]{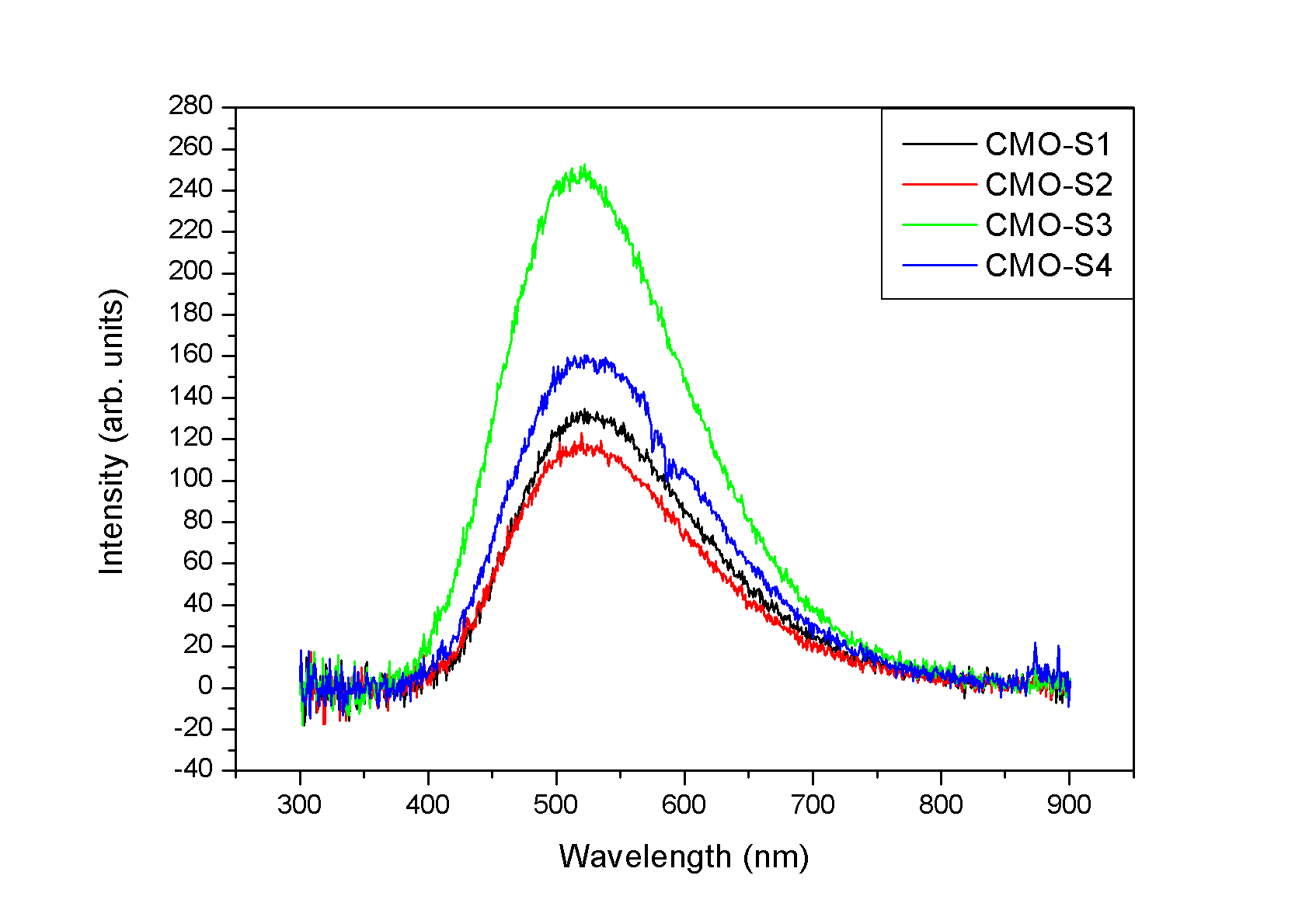}
  \caption{The room-temperature emission spectra for four different \camo crystals.}
\label{fig:fig3-7}
\end{figure}

Room-temperature, X-ray-induced emission spectra were measured with a QE65000 fiber optic spectrometer
(Ocean Optics Co.). The luminescence spectra for different crystals are shown in Fig.~\ref{fig:fig3-7}. They
all have broad emission bands in the 400 to 700 nm range with a peak emission at 520 nm, results that are in good
agreement with the measured photo-luminescence spectrum. Although the light yields are different for different
crystals, the shapes of the spectra are all quite similar.

\subsection{Light yield measurements}
Green-extended, three-inch RbC photocathode photo-multiplier tubes (PMT) (model D726Uk from Electron tube Ltd.)
were directly attached to the test crystal surfaces. The RbC photocathode has a response that is well matched
to the 520 nm peak emission of the \camo crystals; the effective quantum efficiency (QE) is 14\%.
\begin{figure}
  \centering
  \includegraphics[width=0.8\textwidth]{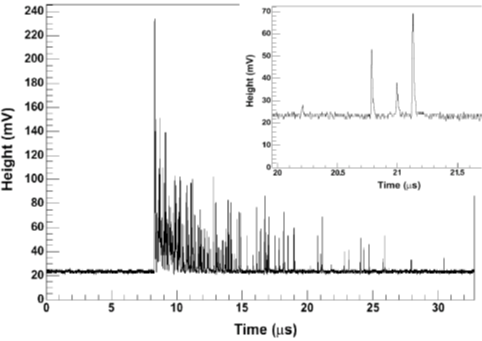}
  \caption{A typical \camo signal pulse. The inset shows a time-expanded view of the $t\approx 20~\mu s$ region. }
\label{fig:fig3-8}
\end{figure}

 Single photoelectron (SPE) signals can be identified at low energy by means of a 400 MHz FADC that is used
to digitize the PMT output pulses. In scintillators with long decay times, the identification and counting of
individual SPEs can reduce noise effects and improve the energy resolution. A typical \camo response
is shown in Fig.~\ref{fig:fig3-8}. We investigated the scintillation characteristics of the crystals by counting
the number of SPEs detected during a 24~$\mu$s time window using an offline clustering algorithm that is specially
developed to isolate SPEs~\cite{KIMS06,KIMS07a}. Scintillation light yields of various \camo crystal samples from
different manufacturers were tested. The number of detected SPEs measured for 662~keV $\gamma$-rays
from a $^{137}$Cs radioactive source ranged from 0.3~to~0.6~photelectrons/keV, depending on the crystal. The
light yield of one of the best-quality crystals was also measured at a temperature of $6^{\circ}$C. In this
case, the FADC time window was extended to 82~$\mu$s because of the longer decay time for cooled crystals. For this
cooled crystal, we obtained 1.04~photoelectrons/keV from $\mathrm{^{137}Cs}$ radioactive source measurements. If only
photoelectron statistics are considered, the energy resolution of this cooled crystal at the 3~MeV \zeronubb endpoint
of $^{100}$Mo is inferred to be 4\% FWHM. 

\subsection{Absolute light yield measurement}
We also measured the number of electron-hole (e-h) pairs and the absolute light yield of some of the \camo
crystals with a UV-sensitive, windowless large-area avalanche photodiode (LAAPD) produced by Advanced
Photonics Co.~\cite{Hass}.  This has close to 80\% quantum efficiency for visible and near-infrared photons. 
The number of e-h pairs and the absolute light yield of one of the best crystals were measured at room temperature.
The sample crystal was optically coupled to the LAAPD that was read out by a spectroscopic amplifier with a 10~$\mu$s
shaping time constant. The crystal was irradiated with 5.5~MeV $\alpha$ particles from a $\mathrm{^{241}Am}$ source
that produced a clearly identifiable peak; an $\alpha/\beta$ quenching factor (QF) of 0.20 was used for the light
yield estimation. After calibration with an $^{55}$Fe source, the number of e-h pairs of the crystal was determined to
be $3,500 \pm 350$~e-h/MeV. This absolute light yield was adjusted to $4,900 \pm 490$~photons/MeV after correction for
the light collection efficiency and losses in the Teflon reflector (90\%). SPE signals from the PMT were also used
to infer the absolute light yield, even though this method has more systematic effects that have to be
considered~\cite{Hass}. The room temperature light yield obtained this way was $4,500 \pm 1,000$ photons/MeV. 

This large absolute light yield at room temperature, which is about 10\% of that of the most efficient scintillating
cystals such as CsI(Tl), is sufficient to enable crystal-quality tests and radiopurity measurements to be carried
out at ordinary temperatures.  This greatly simplifies detector material development and manufacturing quality
control during the production of large numbers of crystals.

\subsection{Light yield comparison of different \camo crystals}
\begin{figure}
 \centering
  \includegraphics[width=0.8\textwidth]{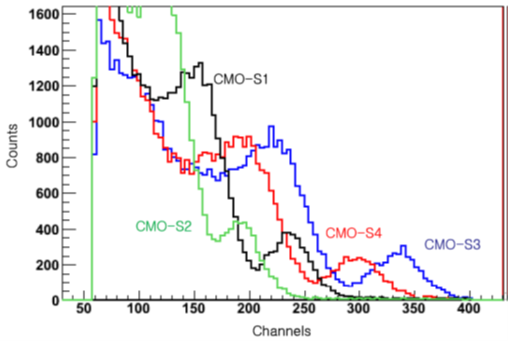}
  \caption{Responses of different crystals to 662 keV $\gamma$-rays from a $\mathrm{^{137}Cs}$ radioactive source. }
\label{fig:fig3-9}
\end{figure}
The relative light yields of small (1 cm$^3$) \camo crystals produced in different places were
compared using 662~keV $\gamma$-rays from a $\mathrm{^{137}Cs}$ radioactive source (see Fig.~\ref{fig:fig3-9}).
The crystal produced at IM shows highest light yield while the crystal produced at ICMSAI had the lowest light yields;
these results are consistent with X-ray luminescence measurements.  Crystals recently produced by ICMSAI show similar
performance as the IM-produced crystals.

\subsection{Room-temperature energy resolution of \camo crystals}
\begin{figure}\centering
 \includegraphics[width=\textwidth]{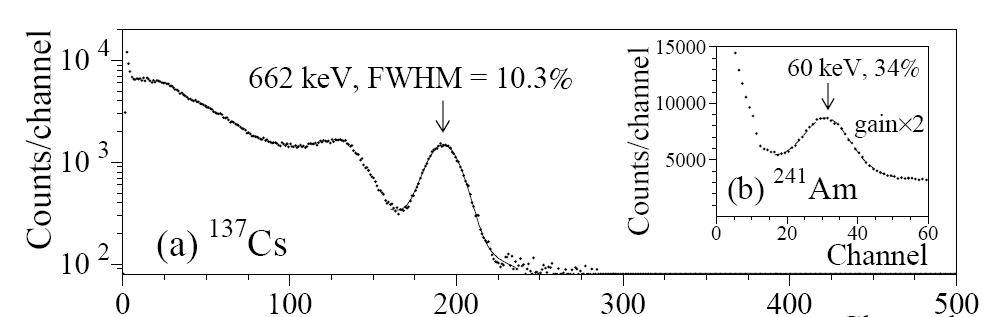}
\caption{Response of the IM-produced \CMO crystal to a $\mathrm{^{137}Cs}$ source. 
The inset shows the low-energy response of the same crystal to 60 keV $\gamma$-rays from a $\mathrm{^{241}Am}$ source.}
 \label{fig:fig3-10}
\end{figure}
The energy resolutions of \camo crystal samples from different manufacturers have been compared
using 662~keV $\mathrm{^{137}Cs}$ $\gamma$-rays.  The room temperature energy resolutions of the tested crystals
range between 10.3 to 14\% FWHM. The best energy resolution was obtained with the IM-produced \camo crystal
as shown in Fig.~3.10~\cite{Annenkov08}. The energy resolution for the crystal measured at 6$^\circ$C
was 11.9\%. 

\subsection{Pulse shape discrimination (PSD)}
\begin{figure}\centering
  \includegraphics[width=0.8\textwidth]{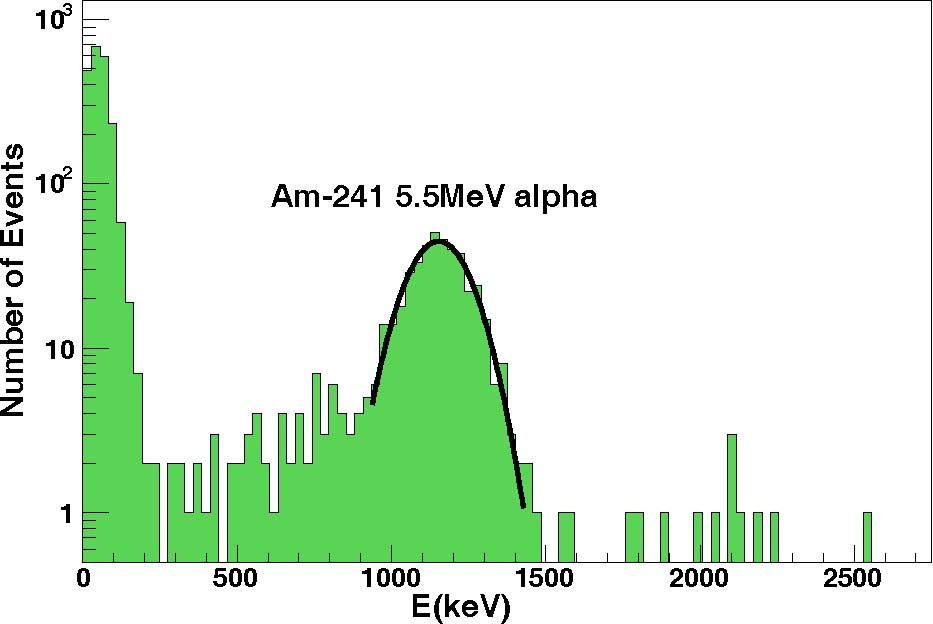}
\caption{Response of a \CMO scintillator to 5.5~MeV $\mathrm{^{241}Am}$ alpha particles. The energy scale
was calibrated with $\gamma$ rays.}
  \label{fig:fig3-11}
\end{figure}
\begin{figure}\centering
 \includegraphics[width=0.75\textwidth]{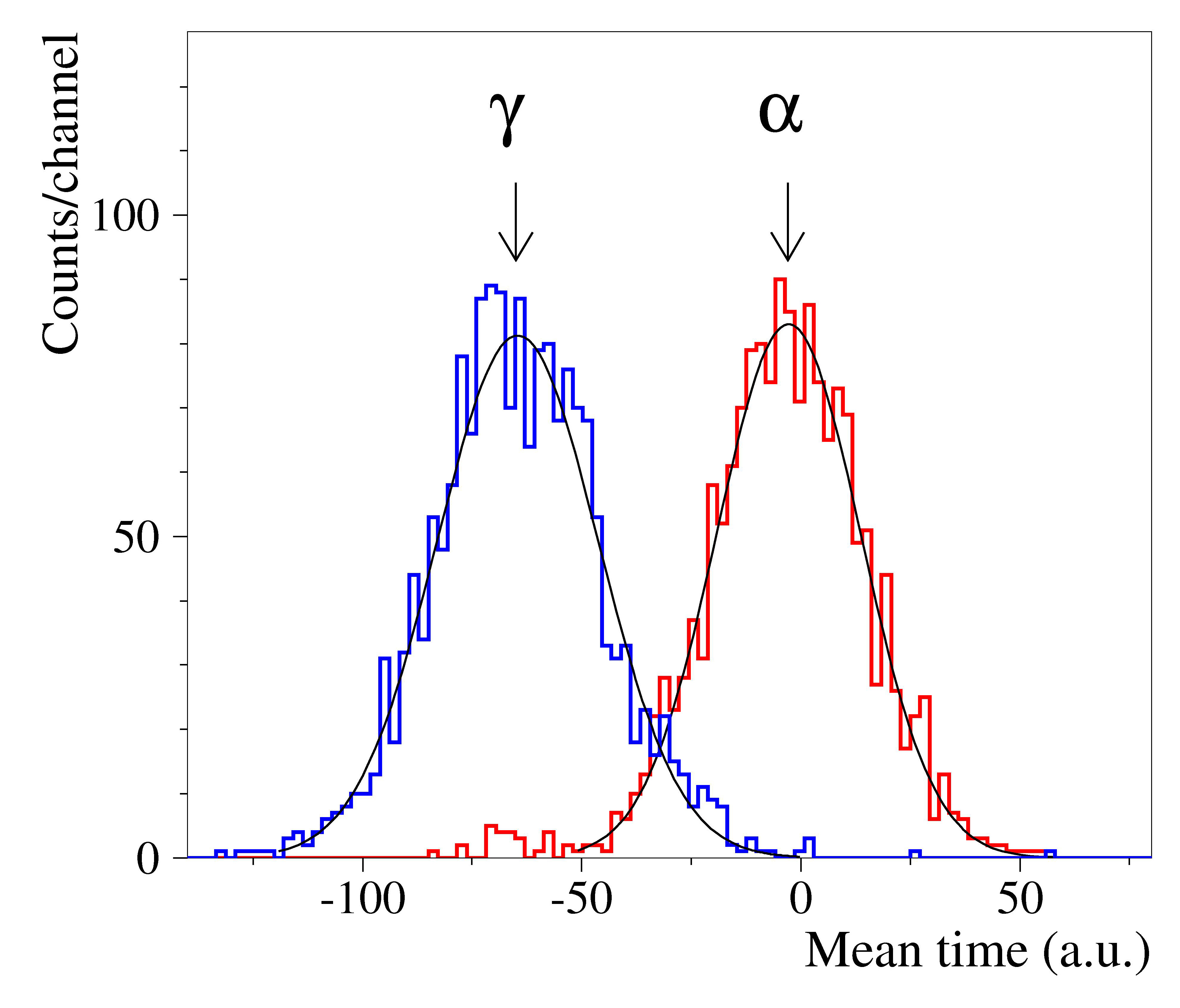}
\caption{Mean-time distributions for alpha and gamma radiation.}
\label{fig:fig3-12}
\end{figure}

Scintillation light signals from highly ionizing particles, such as alphas, usually have different light yields and
decay-time characteristics than those produced by electron- and gamma-radiation-induced signals. The resulting
pulse-shape differences can be used to separate $\alpha$-induced signals from $\beta /\gamma$-ray-induced signals.
Even though the 5.5~MeV alpha particles from a $^{241}$Am  source stop in the crystal, the visible energy that is
recorded by the PMT is only about 20\% of that for a $\gamma$-ray of the same energy, as shown in
Fig.~\ref{fig:fig3-11}.  This is referred to as the alpha-particle quenching factor.  The decay time of
alpha-induced signals is also quite distinct from that for  $\gamma$-ray-induced signals.  A simple energy-weighted
mean-time was used to characterize the decay time of scintillation signals. The mean-time-determined Pulse Shape
Discrimination (PSD) between alpha-induced and gamma-induced signals is demonstrated in
Fig.~\ref{fig:fig3-12}~\cite{Annenkov08}. PSD is a powerful tool for rejecting backgrounds that are induced
by alpha particles produced by contaminants from the U and Th decay chains.

\subsection{Temperature dependence of the light yield of \camo crystals}
The temperature dependence of the relative light output and decay time of CaMoO$_4$ crystals was studied by Mikhailik et al.~\cite{Mikhailik07}, as shown in Fig. 3.13 . At liquid nitrogen temperatures, the light output is six times larger than at room-temperature. The scintillation efficiency at 
temperatures below 1K is not known and should be measured. 
\begin{figure}\centering
  \includegraphics[width=\textwidth]{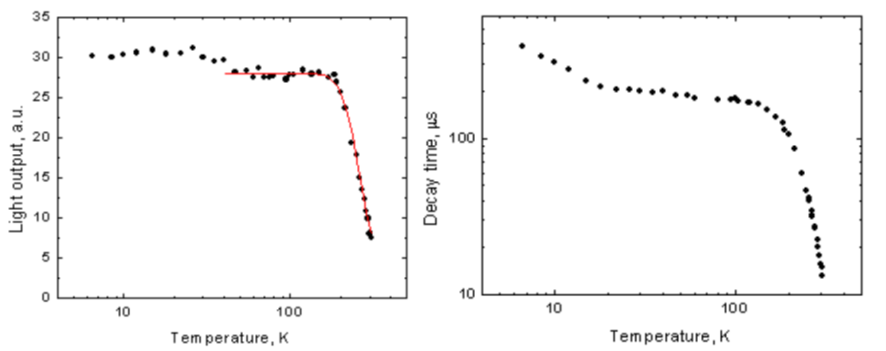}
\caption{The temperature dependence of the light output (left) and decay time (right) 
of \camo crystal scintillators~\cite{Mikhailik07}. }
  \label{fig:fig3-13}
\end{figure}

The high detection efficiency and good energy resolution of \camo crystal scintillators make them very suitable
sources and detectors for \zeronubb-decay searches. As mentioned above, successful experiments with
$\mathrm{CdWO_4}$ crystals~\cite{Danevich03,Polischuk15} demonstrated the applicability of scintillating crystal
techniques for $0\nu\beta\beta$-decay searches. 

\section{$\mathrm{^{48}}$Ca-depleted, $\mathrm{^{100}}$Mo-enriched $\mathrm{^{48depl}Ca^{100}MoO_4}$ crystals }
\camo crystals produced from enriched \cafortyeight could be also used to search for \zeronu decays of
\cafortyeight.  However, for AMoRE the \twonubb decay of $\mathrm{^{48}Ca}$ poses an irreducible
background.  The natural abundance of $\mathrm{^{48}Ca}$ is 0.187\%
and the half-life for \twonubb $\mathrm{^{48}Ca}$ decay is $4.2\times 10^{19}$~yrs. Since the $Q$-value
for $\mathrm{^{48}Ca}$ is 4272~keV, this decay would produce a serious background counting rate in the \zeronubb
signal region for $^{100}$Mo. For a \CMO crystal made with natural Ca, the background level from $\mathrm{^{48}Ca}$
decay at the  \mohundred $Q$-value (3034~keV) would be 0.01 counts/keV/kg/yr. Therefore, the concentration of
$\mathrm{^{48}Ca}$ in the \CMO crystals has to be reduced by at least a factor of 100 to get its associated background
below our ultimate, AMoRE-II, goal of $10^{-4}$ counts/keV/kg/yr.

A 4.5~kg quantity of $\mathrm{^{48}Ca}$-depleted ($\leq 0.001\%$) calcium carbonate powder with
ICP-MS-measured $\mathrm{^{238}U}$ and $\mathrm{^{232}Th}$ contaminations below 0.2 \ppb and 0.8 \ppb,
respectively, was produced by the Russian ElectroKhimPribor Integrated Plant (EKP).  However, a $\gamma$-spectroscopy analysis of the
$\mathrm{^{48depl}CaCO_3}$ powder showed the specific activities due to $\mathrm{^{226}Ra}$ and its
progenies to be at the few hundred mBq/kg level. Therefore, this $\mathrm{^{48depl}CaCO_3}$ material
was subjected to additional purification.

A 8.25~kg quantity of $\mathrm{^{100}}$Mo-enriched (96\%) molybdenum in the form of  MoO$_3$ powder was produced by
the JSC Production Association Electrochemical plant (Zelenogorsk, Russia) by a gas centrifugation technique.
The enriched material is very pure with respect to radioactive elements: the results of ICP-MS measurements
show that the concentrations of $\mathrm{^{238}U}$ and $\mathrm{^{232}Th}$ in the oxide are below 0.07 \ppb and
0.1 \ppb, respectively.

\section{Growth of large radio-pure $\mathrm{^{48depl}Ca^{100}MoO_4}$  crystals}
\begin{figure}\centering
  \includegraphics[width=\textwidth]{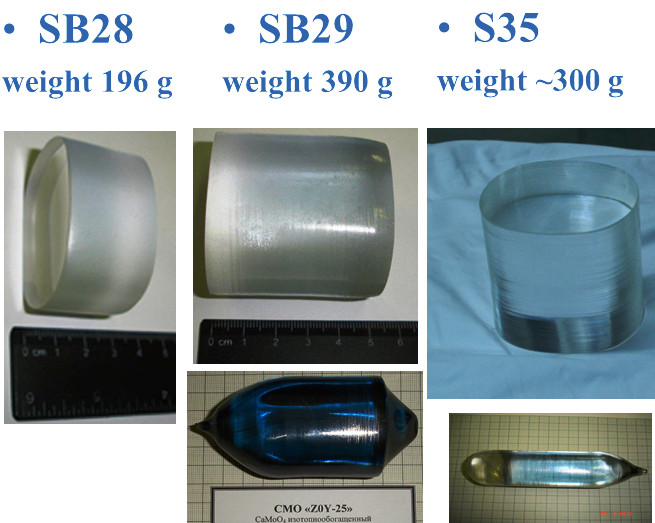}
\caption{$\mathrm{^{48depl}Ca^{100}MoO_4}$ crystals grown at the Innovation Center of the Moscow Steel and Alloy
Institute (ICMSAI) in Russia.}
\label{fig:fig3-14}
\end{figure}
The $\mathrm{^{48depl}Ca^{100}MoO_4}$ crystal growing process includes the following stages:
\begin{enumerate}
\item	initial powder ICP-MS analyses;
\item	preparation of pellets -- 550~g each;
\item	initial charge for crystal growing preparation including a small mass excess of MoO$_3$;
\item	growth of the initial crystallized charge - crystals up to 550~g each;
\item	initial crystallized charge for final-crystal growing preparation;
\item	crystallizer assembley and final-crystal growing;
\item	two crystal annealing procedures;
\item	production of \CMO scintillation elements according to the mechanical specifications by cutting, lapping and polishing.

\end{enumerate}
\begin{figure}\centering
 \includegraphics[width=0.65\textwidth]{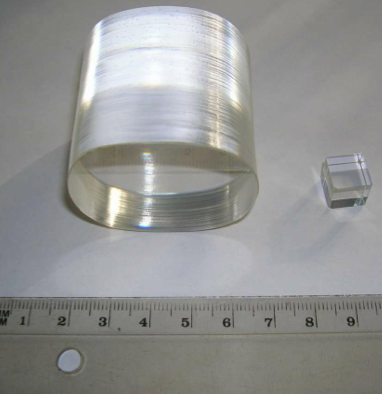}
\caption{A $\mathrm{^{48depl}Ca^{100}MoO_4}$ single crystal after annealing in oxygen and a cubic sample cut from the same crystal boule.}
\label{fig:fig3-15}
\end{figure}
A commonly used technique for the synthesis of the \camo raw material (charge) is the
co-precipitation reaction discussed above. This reaction method offers some essential advantages, including
the possibility of additional purification of the starting compounds and better control of the final-product
stoichiometry. $\mathrm{^{48depl}Ca^{100}MoO_4}$ crystals have been pulled by JSC Fomos-Materials (Russia) in air
from a platinum crucible by means of the Czochralski technique. The technology of $\mathrm{^{48depl}Ca^{100}MoO_4}$
single crystal production (the so-called double crystallization procedure, or re-crystallization) can be
summarized as the following sequence of steps: the charge of $\mathrm{^{48depl}Ca^{100}MoO_4}$ in powder form is
heated to the melting temperature to convert it into pellets with density similar to the density of crystals;
the pellets are loaded into a crucible, melted, and the raw crystal is pulled; raw crystals produced in this
way are loaded into the crucible, melted again, and the final crystal is pulled. 

The technology that was developed for the production of $\mathrm{^{48depl}Ca^{100}MoO_4}$ single crystals can be
summarized by the following sequence of consecutive steps:
\begin{enumerate}
\item The charge of $\mathrm{^{48depl}Ca^{100}MoO_4}$ in powder form is loaded in crucible
and heated to the melting temperature. 
\item The seeding starts at a rotation speed of 5 -- 12 min$^{-1}$ and the
raw crystal is produced at a high pulling speed (3 -- 5~mm/hr). 
\item The resulting (raw) crystals are loaded into the crucible, melted and the final crystal
is pulled at a slower speed (2 -- 3~mm/hr).
\item The produced crystal is subjected to initial annealing heat
  treatment while still in the setup for 12 hours. 
\end{enumerate}
The $\mathrm{^{48depl}Ca^{100}MoO_4}$ crystals produced in this way have an elliptic cylindrical shape with major and minor diameters of $45\sim 55$~mm and 40~--~50~mm respectively, 
lengths of 40~--~60~mm (from an original overall length of $\sim$100 mm), and a
total mass of $\approx$ 0.55~kg.

\begin{figure}\centering
 \includegraphics[width=0.8\textwidth]{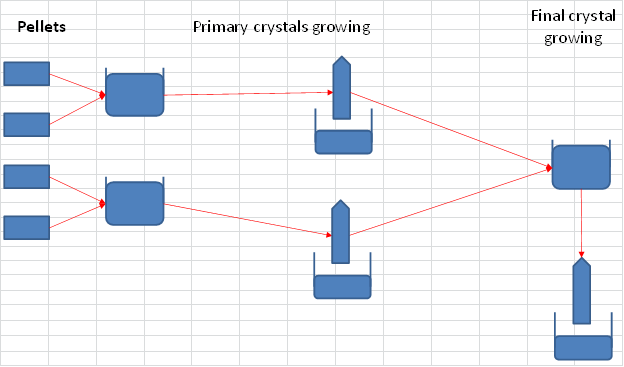}
  \caption{Crystal growth process scheme at Fomos Materials}
  \label{fig:fig3-16}
\end{figure}

The possible evaporation of MoO$_3$ during the crystal growth can result in the formation of
defects. The best possible optical transparency and light output are ensured by growing the crystals from recrystallized
raw materials with an approximate 1.0\% mass-excess of $\mathrm{MoO_3}$ added to the charge. The as-grown crystal
has a notoriously strong blue coloration due to oxygen depletion and a deficiency of Mo$^{6+}$ ions (see Fig.~\ref{fig:fig3-14}). 
It has been demonstrated that this coloration can be removed by prolonged annealing in an oxygen
atmosphere~\cite{Belogurov05, Mikhailik07}. Almost transparent $\mathrm{^{48depl}Ca^{100}MoO_4}$ crystals were obtained
after the annealing (see Figs.~\ref{fig:fig3-14} and \ref{fig:fig3-15}).   The crystal growth
process developed by Fomos Materials is shown in Fig.~\ref{fig:fig3-16}.

\section{Radioactive contaminants in \camo crystal scintillators}
\label{CMO_radio}
\subsection{Low-background setup at Y2L}
Radioactive contamination levels in the \CMO crystals were studied in
a specially designed low-radiation setup at the Yangyang Underground
Laboratory (Y2L).

\begin{figure}\centering
 \includegraphics[width=\textwidth]{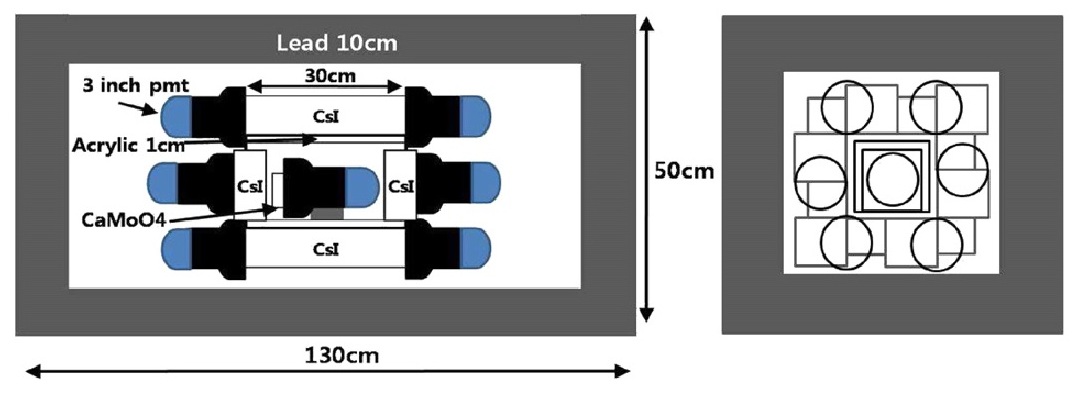}
  \caption{A schematic drawing of the CsI(Tl)-crystal shield $4\pi$-veto cavity.}
\label{fig:fig3-17}
\end{figure}

\begin{figure}\centering
 \includegraphics[width=0.9\textwidth]{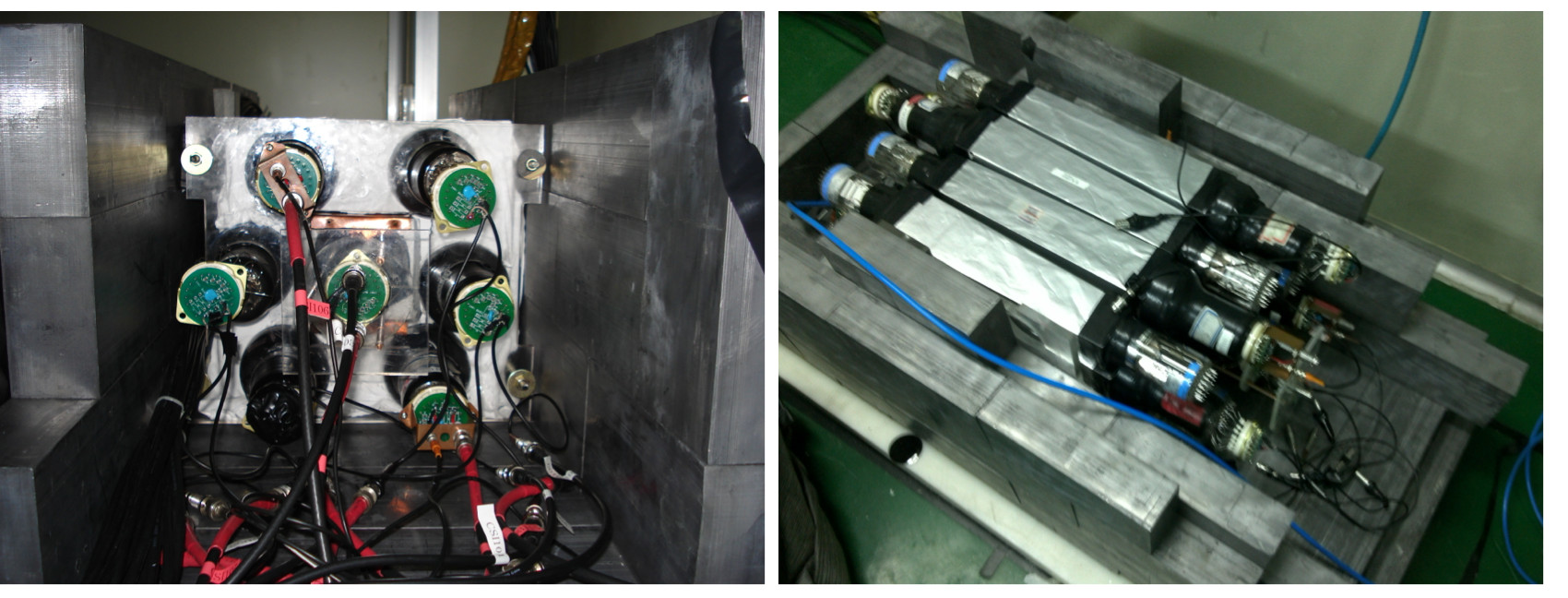}
\caption{The $4\pi$ CsI(Tl) active shielding structure at the Y2L laboratory.}
\label{fig:fig3-18}
\end{figure}

The  \CMO crystals being tested were fixed by an acrylic support inside a cavity in an array of
CsI(Tl) crystal scintillators that are used to veto external gamma and neutron radiation and residual cosmic muons.
The CsI(Tl) crystals that are used to form the ``barrel'' region of the cavity are trapezoidal with a length of 30~cm,
a larger area of 6.5~cm~$\times$~6.5~cm, and smaller area of 5.5~cm~$\times$~5.5~cm. Twelve crystals that are
read out by twelve, 3-inch PMTs form the barrel. For the data reported here, each PMTs was attached to a pair
of crystals, forming an interleaved zigzag pattern at the end of the array, in which each PMT monitored two crystals,
and each crystal was monitored by two PMTs, one at each end. The two end-cap regions were covered by crystals
that each have only one readout PMT, as shown in Fig.~\ref{fig:fig3-17}. The CsI(Tl) crystal veto array was
surrounded by a 10~cm~thick passive lead shield. The acrylic cavity was flushed with a Nitrogen gas flow of 4 L/min in order to eliminate ambient radon~\cite{KRISS_CMO}. Photographs of the 4$\pi$-veto system are shown
in Fig.~\ref{fig:fig3-18}.

Recently, this $4\pi$-veto system was upgraded to improve the veto efficiency. Now each of the twelve crystals in
the barrel section have their own pair of PMTs that are distinct for each crystal. In addition, a 20~cm~thick
polyethylene shield was placed outside of the lead shielding to attenuate external neutron backgrounds. 

Radioactive contamination of the \mohundred-enriched S35, SB28, NSB29 and SS68 crystals by thorium and radium was studied.
As indicated in Figure~\ref{fig:fig3-14}, the SB28, SB29 and S35 crystals were grown at ICMSAI. Since the transmittance of
SB29 was poor, it was re-annealed at Fomos Materials and the crystal was re-named as NSB29.  To avoid the need for a re-annealing
procedure to cure a possible oxygen deficiency, the SS68 crystal was grown with a low-level doping of Nb.  
The data were analyzed with the time-amplitude method described in the following section. The method selects fast sequences of
$\beta-\alpha$ decays in the $\mathrm{^{238}U}$ chain and $\alpha-\alpha$ decays in $\mathrm{^{232}Th}$ chain.  Background data were
accumulated for more than 70 days for the S35 crystal and 40 days for the SB28 crystal.  The data were taken in the Y2L $4\pi$-gamma
veto system, which severely attenuated external backgrounds.

\subsection{Contamination of  \CMO crystals by  thorium and radium.}
The time-amplitude analysis method enables the identification of subchains containing short lived isotopes in the \Th, \U[235] and \U chains. Because of the very low counting rates, the
probability of signals from two uncorrelated decays to occur in a
small time interval is quite low.  Thus, time correlations between consecutive events can provide unambiguous signatures for specific isotopes (see Table~\ref{tab:3.1}).
In these analyses pulse-mean-time PSD results  are also used to distinguish between $\beta$-induced and
$\alpha$-induced events.

\begin{figure}\centering
 \includegraphics[width=0.8\textwidth]{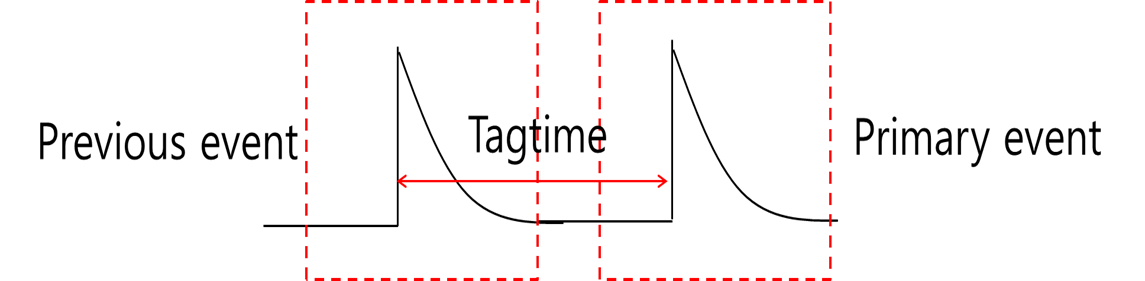}
\caption{Time difference between two signals.}
\label{fig:fig3-19}
\end{figure}

\begin{table}
  \caption{Summary of decays of short-lived isotopes and their coincident progenitor decays in the $^{238}$U, $^{235}$U and $^{232}$Th decay chains.}
  \input{tables/table3-1.tex}
  \label{tab:3.1}
\end{table}
As can be seen in Table~\ref{tab:3.1}, Polonium isotopes in the decay chains have relatively short half-lives.
After a Po nuclei is produced via Bi, Rn or Bi decays, it quickly decays to Pb with a characteristic lifetime.
Figure~\ref{fig:fig3-19} illustrates the terms used below and their definitions.

For example, to identify \Po[214] in \U[238] decay chain, we examine a
100~$\mathrm{\mu s}$ to 600~$\mathrm{\mu s}$ time window after a detected $\beta$-decay.
If this $\beta$ originated from $\Bi[214]\!\to$~\Po[214] beta decay,  this time interval
includes 57.6 \% of subsequent $\Po[214]\!\to$~\Pb[210] alpha
decays. Thus, most of the alpha decay events that occur in this time
window should have a kinetic energy that is equal to the 7.83 MeV
$Q$-value for $\Po[214]\!\to$~\Pb[210] alpha decay and a tag-time
distribution should display a 164~$\mathrm{ \mu s}$ \Po[214]-lifetime dependence.

Using selection efficiencies that were computed by GEANT4 simulations, we converted the measured
\Po[215] (\U[235] chain),  \Po[214] (\U chain), and \Po[216] (\Th chain) 
decay rates to contamination levels in the \Ca[48]-depleted, \Mo[100]-enriched \camo crystals
that are currently on hand, namely  SS68, NSB29, S35 and SB28.  The results are listed in Table~\ref{tab:3.2}.
The measured contamination levels of crystal S35 are relatively high, so we now use it as a control
sample to estimate the alpha/beta discrimination efficiency.

\begin{table}\centering
  \caption{Radioactive contamination of \Ca[48]-depleted,
    \Mo[100]-enriched \camo crystals by \Ac[227], \Ra[226] and \Th[228].  All units are in $\rm \mu Bq/kg$. For comparison data are also shown for the \camo crystal produced by  IM (Lviv, Ukraine)  with natural isotopic composition. It should be stressed that the IM crystal was produced wit no special attempts to purify the initial materials from radioactive contamination.}
  \input{tables/table3-2.tex}\\
  \label{tab:3.2}

\end{table}

%% file: tables/table3-1.tex
\begin{tabular}{|>{\centering}p{3cm}|>{\centering}p{3cm}>{\centering}>{\centering}p{3cm}>{\centering}p{3cm}|}																	
\hline																	
		&	$\rm	^{232}Th~ family	$ &	$\rm	^{235}U~family	$	 &	$\rm	^{238}U~ family $			\tabularnewline		
		& (\Th[228] sub-chain)       & (\Ac[227] sub-chain)    & (\Ra[226] sub-chain) \tabularnewline \hline
	Progenitor 	&	$\rm	^{220}Rn	$& 	$\rm	^{219}Rn	$	& $\rm	^{214}Bi	$				\tabularnewline		
		Isotope &	$\rm Q=	6.40	~MeV$ 	&	$\rm Q=	6.95	~MeV$	&	$\rm Q=	3.27	~MeV$		\tabularnewline		
		(prompt decay) &	$T_{1/2}$$\rm=	55.6~s$ 	&	$T_{1/2}$$\rm=	3.96~s$ &	$T_{1/2}$$\rm=	19.9	~min.$			\tabularnewline		\hline
	Short-lived daughter  	&	$\rm	^{216}Po	$ 	&	$\rm	^{215}Po	$ &	$\rm	^{214}Po	$			\tabularnewline		
		isotope &	$\rm Q=	6.91	~MeV$	&	$\rm Q=	7.53	~MeV$ &	$\rm Q=	7.83	~MeV$			\tabularnewline		
		(delayed decay) &	$T_{1/2}$$\rm=0.145 ~s$ 	&
                $T_{1/2}$$\rm=	1.78	~ms$ &	$T_{1/2}$$\rm=	164	~\mu s$			\tabularnewline		\hline	
\hline																	
\end{tabular}																	

%% file: tables/table3-2.tex
\newcolumntype{C}{>{$\rm}c<{$}}
\begin{tabular}{|c|C|C|C|}\hline												
       & \Th[228]       & \Ac[227]       & \Ra[226]                     \tabularnewline	
       &  \Th[232]~ Chain & \U[235]~ Chain &  \U[238]~ Chain     \tabularnewline	\hline
SS68	 &  30 \pm 5      &   200 \pm 14	  &	60 \pm 8     		\tabularnewline	\hline
NSB29  &	80 \pm 9      &	700 \pm 26	  &	200 \pm 14   		 \tabularnewline	\hline
S35	   &	500 \pm 22    &	1200 \pm 35	  &	4400 \pm 66	 	 \tabularnewline	\hline
IM (Lviv, Ukraine)	 &	230 \pm 15    &	90 \pm 10	    &	1500 \pm 39	 	 \tabularnewline	\hline
SE1	   &	50 \pm 15     &	60 \pm 8	    &	40 \pm 12	   	   \tabularnewline	\hline
SB28	 &	70 \pm 8      &      -         &	80 \pm 9 	        	   \tabularnewline	\hline			\end{tabular}												

%% file: tex/cryogenic_particle_det.tex
\chapter{Cryogenic particle detection}
Cryogenic particle detectors have been developed in response to the need for detectors
with superior energy resolution and ultra-low thresholds in nuclear and particle physics.
Recent developments of these detectors have demonstrated energy resolution and threshold
performance levels that exceed the extreme limits of conventional semiconductor-based detectors.
Cryogenic detection techniques have been adopted for rare-event searches and have become 
important measurement tools for many other applications~\cite{Enss05}.

In general, the energy deposited in matter by the interaction of radiation with the atoms in
matter can be converted into measurable effects such as ionized charged particles, scintillation
light, and phonons. The ionization can be measured by collecting the charges with an applied
electric field. Scintillation light is, at least in principle, easily measured with photon
sensors such as photomultiplier tubes or photo-diodes. However, the major portion of the energy
deposition is converted into the phonons. This suggests that phonon measurements can be more effective
than ionization or scintillation measurements.

However, the accurate measurement of phonons created by particle or radiation absorption is not trivial,
particularity at room temperatures, since huge numbers of phonons exist naturally in any condensed
material, with a statistical frequency/energy distribution that is determined by the temperature of the
material. When an absorber is thermally attached to a heat reservoir the thermal energy of the absorber
fluctuates, resulting in a phonon distribution that changes with time. These thermal fluctuations can
easily overwhelm the phonons created by radiation absorption. At low temperatures, however, the available
thermal energy (i.e., the heat capacity) is greatly reduced, as are the fluctuations. Moreover, because
the heat capacity of the absorber is typically lower at low temperatures, the resulting temperature
increase caused by radiation-induced energy is increased.

The intrinsic energy resolution of the detector is determined by the fluctuations of the produced quanta in
the energy-loss process. The energy needed to produce an electron-hole pair or a scintillation photon is
typically a few to 10 eV. However, in the phonon measurement case, the average energy of a phonon at a
temperature {\it T} is about $k_BT$ where $k_B$ is the Boltzmann constant. At 10 mK, $k_BT$ is close to $10^{-6}$ eV.

In the case of a thermal detector attached to an ideal temperature measurement device, statistical fluctuations
in the thermal energy limit the intrinsic resolution. If the total thermal energy of a detector with a heat
capacity $C$ is approximated as $CT$, the average number of energy quanta can be expressed by $N \approx CT/k_BT$.
Thus, the statistical thermal noise or the ultimate limit on the energy resolution due to thermal energy
fluctuations is $(\delta E)_\mathrm{rms} \approx CT(N)^{1/2} \approx (k_BT^2C)^{1/2}$. For a 1~kg \camo detector
at T=10 mK, the fluctuation limit on the resolution is about 20 eV (FWHM). 

\section{Principle of thermal detection in low temperature calorimeters}
When a particle interacts with a solid-state absorber, its energy is transferred to the electrons and nuclei
in the material of the absorber. Most of this transferred energy is eventually converted into the thermal
energy of the solid. If a suitable thermometer is attached to the absorber, the temperature change caused
by the initial energy transfer can be measured. Typical low temperature calorimeters consist of two parts: 
one is an absorber to make initial interactions with the particles; the other is a temperature sensor that
measures the temperature change of the absorber. Usually, the absorber and the temperature sensor are in
good thermal contact, while one of them is connected to a thermal reservoir, or heat bath, by a weak thermal
link.  The bath should have a sufficiently large heat capacity so that its temperature, typically well below
1~K, does not change with time.

One of the commonly used temperature sensors for low temperature calorimeters is the thermistor, which is a
critically doped semiconductor operating at a temperature that is near, but below, its metal-insulator transition.
Neutron transmutation doped (NTD) Ge thermistors are commonly used for low-temperature, rare-event searches.
The thermistors are fairly easy to use because they can be operated with conventional electronics, such as Field-Effect
Transistors (FETs), and do not require sophisticated superconducting electronics. They are typically current-biased
and radiation-induced voltage changes across the thermistor are amplified with a FET located at a higher temperature
level. NTD Ge sensors are widely used as thermal detectors for various crystals because of their reproducibility
and their uniformity in doping density.  Mass production of NTD sensors is also possible. The CUORE, LUCIFER and
LUMINEU double-beta experiments all use NTD Ge thermistors as their temperature sensors.

Transition Edge Sensors (TESs) are one of the most highly developed type of cryogenic thermometers. A TES is a
superconducting strip operating at its superconducting-normal transition temperature. The superconducting strip
is often made from a thin pure superconducting film (W) or from a bilayer of superconductor and a noble metal,
such as Mo/Au, Mo/Cu or Ti/Au. The transition temperature width of the films is typically a few mK or less. The
resistance in the normal state is usually a few tens of~m$\Omega$.  Well developed devices have a very sharp
transition with a transition width that is narrower than 1~mK.  Thus, at the transition, a small change in
temperature produces a large change in resistance.  This makes it a very sensitive thermometer, but one that only
works in a very narrow range of operating temperatures.

One advantage of using a TES for particle detection is that the superconducting sensor can be directly evaporated
on to the surface of an absorber. This direct contact provides an efficient heat transfer from the absorber to the
TES, and this results in a much faster response time for TESs than that for NTD sensors. It is also suitable for
detecting athermal phonons that directly deposit their energy in the sensor. TESs have been used as the temperature
sensor for the CDMS and CRESST dark-matter search experiments. 

\section{Magnetic Metallic Calorimeter (MMC)}
\begin{figure}
 \includegraphics[width=0.7\textwidth]{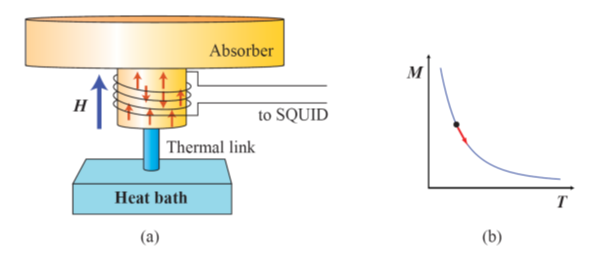}
\caption{(a) A simplified MMC setup with an absorber. (b) Typical M-T characteristics of an Au:Er sensor in a
magnetic field. Energy input into an absorber results in a change in the magnetization of the MMC sensor attached
to the absorber.}
\label{fig:fig4-1}
\end{figure}

Metallic Magnetic Calorimeters (MMCs) utilize a magnetic material whose magnetization is a function of temperature.
The sensor material is a gold alloy with small concentration (100-2000 \ppm) of erbium, denoted as Au:Er. The diluted
magnetic ions in the metallic host have paramagnetic properties that can be approximated as a spin 1/2 system with
a Lande g~value of 6.8. The magnetization is inversely proportional to the temperature, a simple paramagnetic relation
known as Curie's law. This means that a measurement of magnetization can be used to measure the temperature of a
paramagnetic material.  This makes a magnetic thermometer attached to an absorber effectively a 
``magnetic calorimeter'' for particle detection. Au:Er maintains its paramagnetic properties at tens of mK temperatures.
A simplified MMC setup is illustrated in Fig.~\ref{fig:fig4-1}.

In early applications of MMCs for particle detection, the Au:Er sensor was placed inside the loop of a SQUID. In this
early design, the SQUID loop itself was used as a pickup coil~\cite{Fleischmann05}. The state-of-the-art development
of SQUID technologies guaranteed an accurate and fast measurement of the magnetization change with low noise for any
temperature change caused by energy absorption. The SQUID converts the change of magnetic flux into a measurable
voltage signal on the basis of a quantum interference measurement operating at low temperatures.

The detection principle of an MMC can be characterized as $E \to \Delta T \to \Delta M \to \Delta\Phi \to \Delta V$,
where $E, T, M, \Phi$ and $V$ indicate the physical quantities of energy, temperature, magnetization, magnetic flux and
voltage, respectively. An MMC setup with a two-stage SQUID system achieved an energy resolution of 1.6 eV FWHM for X-rays
from a $\mathrm{^{55}Fe}$ source in Heidelberg~\cite{Heidelberg-1.6eV}. This detection method has recently been adopted
by KRISS for MeV-region alpha spectrometry and $Q$-spectrometry measurements for radionuclide analysis. Moreover, MMCs permit
the investigation of the thermal properties of crystal absorbers and the detectors' responses to a wide range of energy
inputs because of their superior energy resolution, fast response time, and flexible operating temperatures and magnetic
fields. By choosing the dimensions of the temperature sensor and the concentration of the magnetic material in it
appropriately, the detector can be scaled up to a large mass because an optimal detector design that  minimizes the 
energy-sensitivity degradation caused by the larger heat capacity can be easily achieved~\cite{Fleischmann05}.

\begin{figure}
 \includegraphics[width=\textwidth]{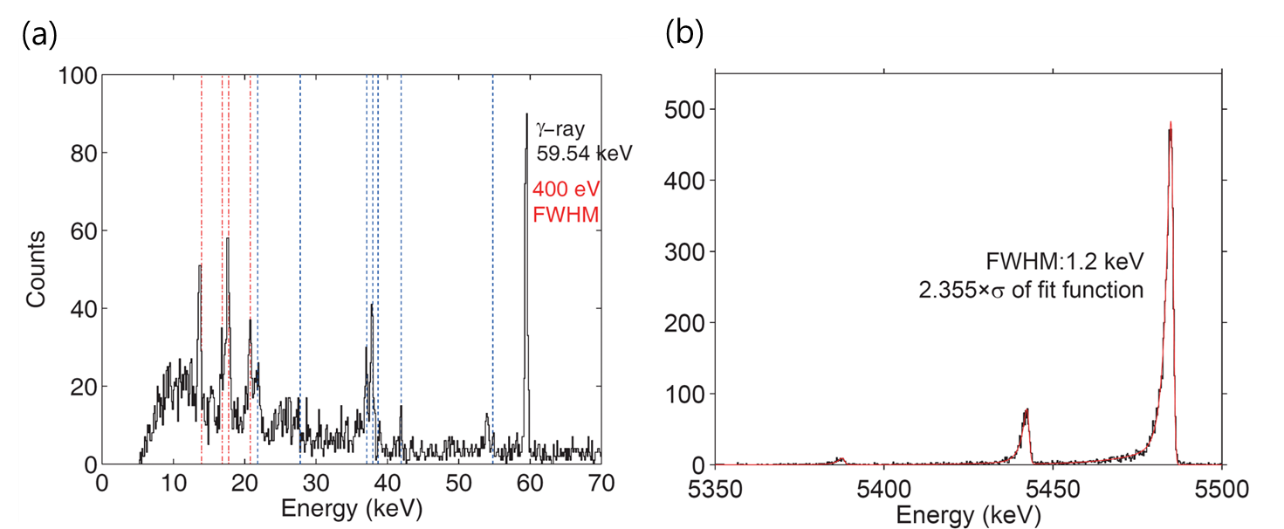}
\caption{The $\mathrm{^{241}Am}$ spectrum for an MMC setup with a gold foil absorber. (a) The below 70~keV low-energy spectrum.  
(b) Alpha peaks in the 5.5~MeV region. The resolution at 60~keV is of 0.4 keV~FWHM. The vertical lines in (a)
indicate the expected positions of low energy X-ray and conversion electron lines~\cite{WSYoon_alpha}.}
\label{fig:fig4-2}
\end{figure}

A meander-type MMC sensor has been developed to be used with an absorber with a large heat
capacity~\cite{KRISS_alpha, WSYoon_alpha}. This ``meander-type'' MMC was first tested with a
$2 \times 2 \times 0.07\ \mathrm{mm}^3$ gold foil. The volume of the metal absorber is smaller than typical dimensions of
crystal absorbers. However, the heat capacity of this absorber is 0.2~nJ/K and 0.4~nJ/K at 10~mK and 20~mK, respectively,
while a 60~cm$^3$ \camo crystal has heat capacities of 0.17~nJ/K and 1.4~nJ/K at these same temperatures. The performance
of the detector is shown in Fig.~\ref{fig:fig4-2}. In measurements with an external $\mathrm{^{241}Am}$ alpha source, this detector
had a measured energy resolution of 1.2 keV FWHM for 5.5~MeV alphas. This resolution, which was inferred from a fit with a
Gaussian-width function that takes source straggling effect into account by means of exponential convolutions,
was the best measured resolution ever reported for an $\mathrm{^{241}Am}$ alpha spectrum. Moreover, simultaneously,
the lower-energy 60~keV gamma line was clearly seen, along with other low energy X-ray and conversion-electron lines. The
resolution of the 60~keV gamma line was 400~eV FWHM.  These measurements indicate promising possibilities for high performance
applications of the MMC technique with large absorbers, with very good energy resolution over a very wide dynamic range.

\section{Phonon measurement with large \camo crystals}
\begin{figure}\centering
 \includegraphics[width=0.7\textwidth]{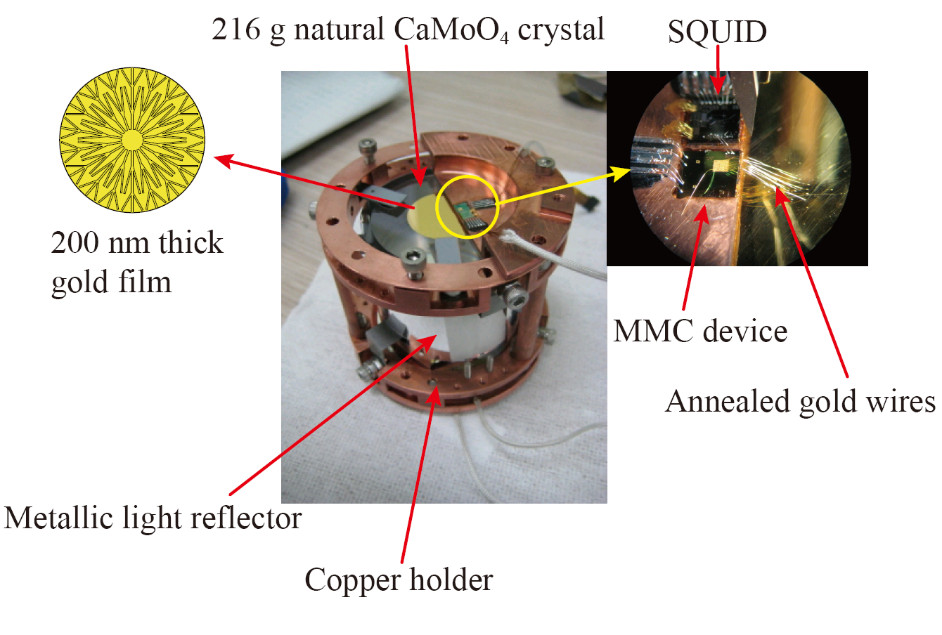}
\caption{A \camo detector setup with an MMC phonon sensor.}
 \label{fig:fig4-3}
\end{figure}

An early stage of MMC development for the AMoRE project used an 0.6~~cm$^3$ \camo crystal to test the applicability of the medium for use in a 
\zeronubb search~\cite{KRISS_CMO}. This successful experiment showed simultaneous good detector performance in both the keV and MeV
energy regions. The next experiment used a larger crystal absorber corresponding to a factor of 80 volume increase.  A cylindrical
\camo crystal, 4~cm in diameter and 4~cm in height, and a mass of 220~g, was instrumented with a meander-type MMC sensor similar to
the one described in the previous subsection. Figure~\ref{fig:fig4-3} shows pictures of the detector setup. The crystal was
mechanically supported by 12 Teflon-coated phosphor-bronze springs (one was eventually not used in order to accommodate the
meander and SQUID chip). In order to make a good thermal connection between the crystal and the temperature sensor, a
gold film was evaporated on one of the flat crystal surfaces. Annealed gold wires were attached between the film and the
Au:Er section of the meander chip. A meander-type pickup loop for the MMC sensor was connected to the input pads of a
current-sensing dc-SQUID. The measurement circuit formed a superconducting loop that produced a current change in response
to a temperature-rise-induced magnetization change in the MMC sensor. Details of the experimental setup are described
in ~\cite{GBKim}.

The low temperature measurement with the MMC and the crystal was carried out with a dilution refrigerator in a surface-level
laboratory. The cryostat of the refrigerator was surrounded by a 10-cm-thick lead shield (with an open top) to reduce
environmental gamma-ray background.

The rise time of the MMC signals was about 1~ms at 40~mK for both electron- and alpha-induced events, with a slightly
faster rise time for alphas. When irradiated with an external \Th[232] source, the 2615~keV peak was 11~keV wide (FWHM);
the 583~keV peak width was 6.5~keV (FWHM)~\cite{GBKim}.

\begin{figure}\centering
 \includegraphics[width=0.9\textwidth]{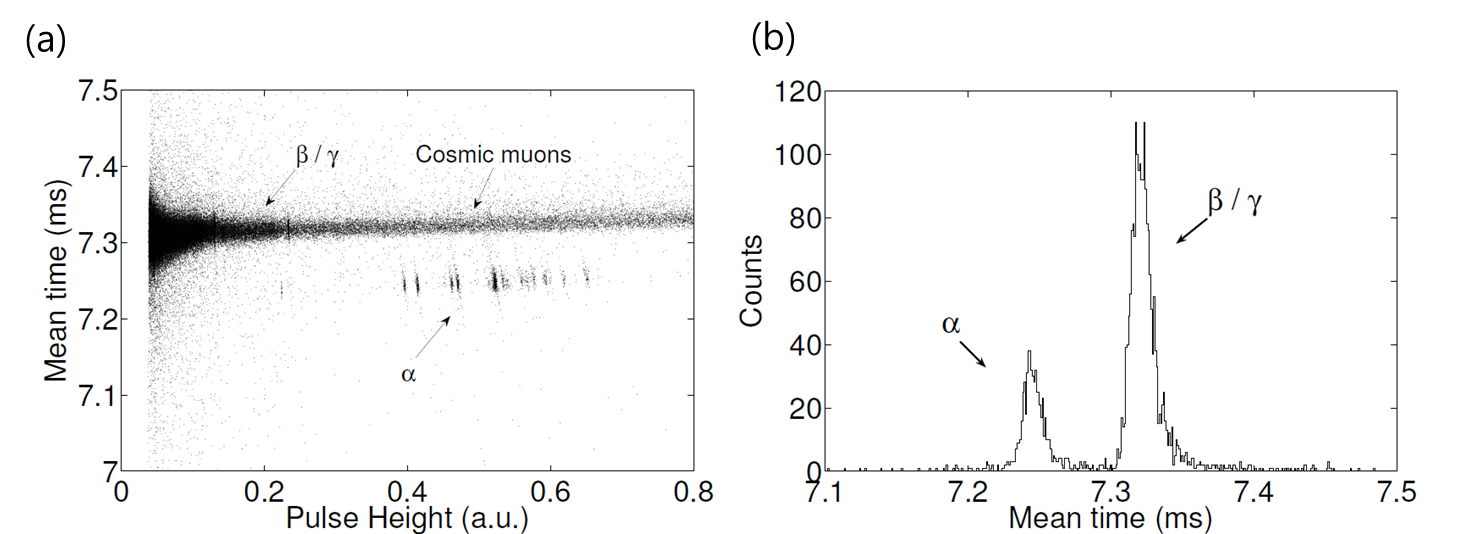}
\caption{(a) A scatter plot of the mean-time {\it vs} pulse height obtained from a 95~hr background measurement in a
surface laboratory.   $\alpha$ and $\beta /\gamma$ (including cosmic muons) events are clearly distinguishable by their
mean-time values. (b) The mean time distribution in  the 4 -- 5~MeV region of alpha-equivalent energy~\cite{GBKim}.}
\label{fig:fig4-4}
\end{figure}

The two-dimensional scatter plot in Fig.~\ref{fig:fig4-4}(a) shows the distribution of signal height versus the mean-time
pulse-shaping parameter MT, where MT is equivalent to the one-dimensional center-of-mass of the pulse when its shape is
treated as the density function of a 1-D object.  The pulses in the distribution can be grouped into two distinct event
types according to their MT values. The horizontal band of signals above MT=7.3 ms is produced by electrons and gammas
incident on the \camo crystal. These $\beta /\gamma$ events include signals generated by cosmic-ray muons passing through
the crystal and environmental gamma ray backgrounds. The group of signals below the cosmic-ray muons are
produced by alpha particles.

The distribution of MT parameters in the 4 -- 5~MeV region of alpha-equivalent energies has two distinct peaks as
shown in Fig.~\ref{fig:fig4-4}(b).  The $\alpha-\beta /\gamma$ discrimination power was determined to be $7.6\sigma$ by
fitting each peak with a normal Gaussian function, although small high-mean-time tails are noticeable on the
right-hand sides of each peak.  These results indicate the pulse shape discrimination  with high separation power
can be realized with phonon signals only.

\section{Development of low-temparature photon sensors}

Two of the most important parameters that characterize a scintillating bolometer are the light yield and scintillation
quenching factor. An MMC-based light sensor is being developed by our group in order to explore/exploit indications
that the low-temperature QF-based light-phonon separation power of \camo detectors will be superior due to their large
light output.

\begin{figure}\centering
 \includegraphics[width=\textwidth]{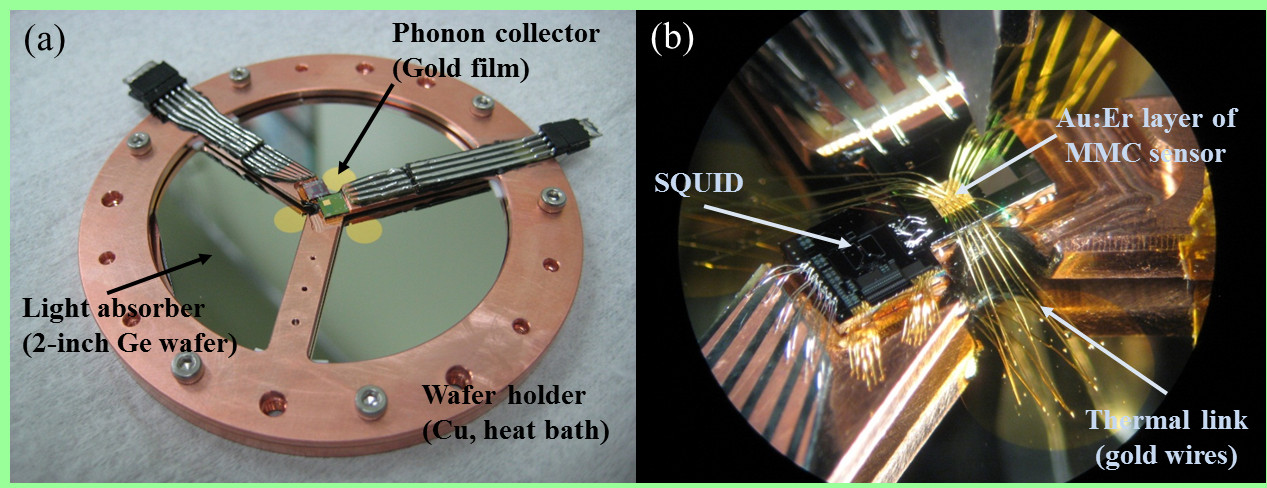}
\caption{(a) The prototype cryogenic photon detector. The bottom side of the wafer faces a \camo crystal to measure scintillation
light produced therein.  (b) A magnified image near the center of the detector~\cite{HJLee}. }
\label{fig:fig4-5}
\end{figure}

The images in Fig.~\ref{fig:fig4-5} show a completed light sensor, in which a two-inch diameter, 0.5-mm-thick polished
Ge wafer is used as a photon absorber.  The wafer was fixed with three small Teflon clips at its edge that were clamped
with two flat copper rings. The top ring had three spokes in the middle, whereas the bottom ring (not visible in the
photograph) had an open hole for light collection.

The performance of the light detector was investigated in a surface-level cryogen-free adiabatic demagnetization
refrigerator (ADR). The ADR cryostat was surrounded by a 5-cm-thick lead shield. Two sets of experiments were performed
using $\mathrm{^{55}Fe}$ and $\mathrm{^{241}Am}$ radioactive sources. With the $\mathrm{^{55}Fe}$ source, an energy resolution
with FWHM = 545~eV was obtained for 6~keV X-rays.  The temperature-dependence of the $\mathrm{^{241}Am}$ source's 60~keV
gamma-induced signal sizes was studied.

\begin{figure}
 \includegraphics[width=\textwidth]{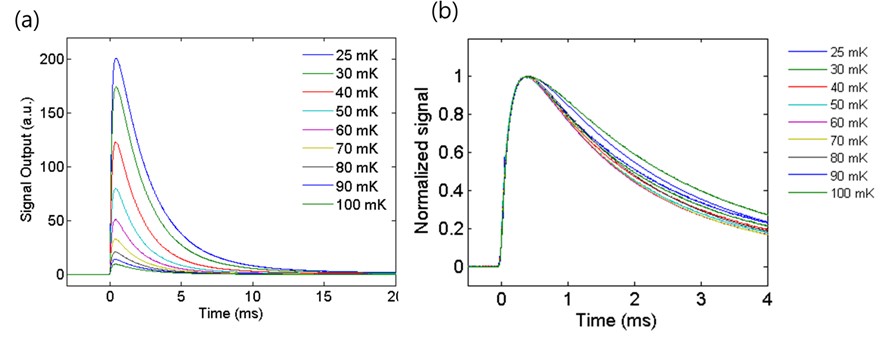}
\caption{(a) Typical 60~keV gamma ray signals at various ADR temperatures. (b) The same signals normalized by
maximum pulse height. 
The temperature-independent rise-time is $\simeq 0.2$~ms~\cite{HJLee}. }
\label{fig:fig4-6-1}
\end{figure}

As the operating temperature of the light detector was decreased, the 60~keV signal sizes increased, as shown
in Fig.~\ref{fig:fig4-6-1}a. This is expected because the wafer heat capacity decreases and the MMC sensitivity
increases with decreasing temperatures. However, the signal rise times are found to be almost constant at about 0.2~ms
for all of the measured temperatures as shown by the pulses normalized by their maximum pulse height in
Fig.~\ref{fig:fig4-6-1}b. Heat-flow
via athermal phonons is likely responsible for the temperature-independent time constant. Photon
signals from MMC have a faster rise-time than that for phonons, which should increase the efficiency for
distinguishing real events from random overlapping events, which will be serious backgrounds for AMoRE, especially
random coincidences of \twonubb \mohundred decays.

\section{Simultaneous measurement of phonon and photon signals}
\label{Sec:Cryo:Simul}

Simultaneous measurements of phonon and photon signals were made using combinations of the phonon and photon
detectors described in the previous sections.  These were done in a dilution refrigerator located in a
surface-level laboratory.   A doubly enriched \doublecamo crystal (SB28) with a mass 196~g was used as the absorber.
The optical properties of this crystal and its internal background levels from room-temperature measurements are described in Chapter~3.

\begin{figure}\centering
 \includegraphics[width=\textwidth]{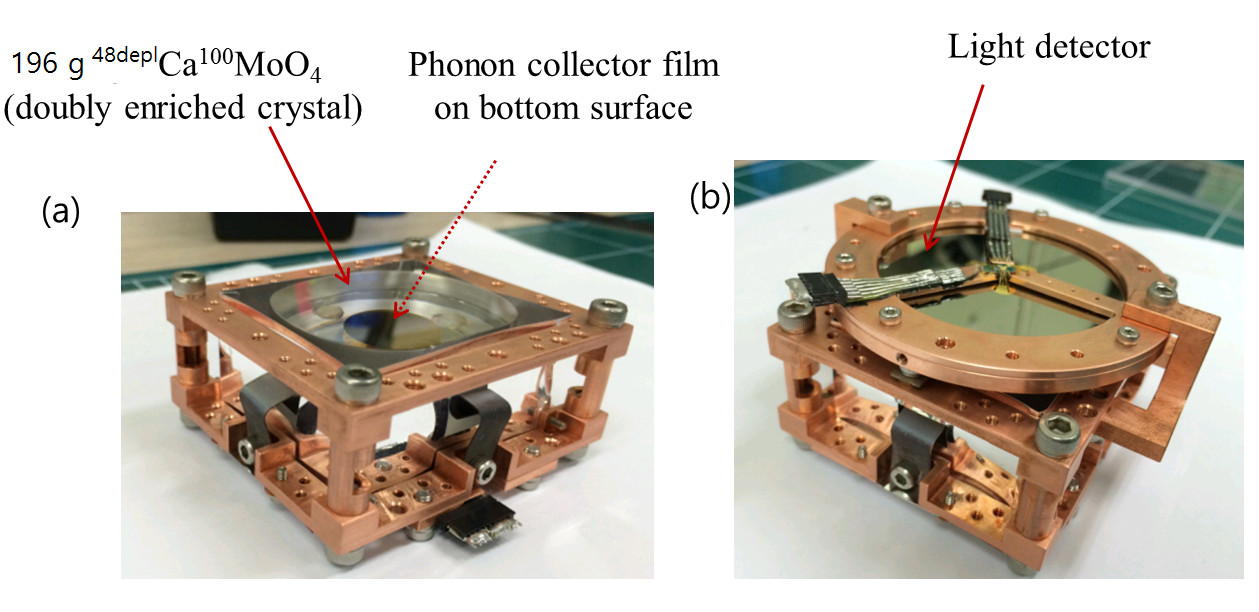}
\caption{The setup for simultaneous measurements with phonon and photon sensors. (a) The phonon collector and MMC sensor are
located on the bottom of the \camo crystal; (b) the Ge light-photon detector is placed on the top. }
\label{fig:fig4-6-2}
\end{figure}

A copper sample-holder for the elliptical cylindrically shaped crystal has a rectangular-prism shaped structure,
as shown in Fig.~\ref{fig:fig4-6-2}a.  The phonon collector is located on the bottom surface of the crystal shown
in the image.  The photon detector described in the previous section was placed on the top of the crystal 
to measure the scintillation light (see Fig.~\ref{fig:fig4-6-2}b).  A light reflecting
foil covers all sides of the sample holder other than the top surface.

\begin{figure}\centering
 \includegraphics[width=\textwidth]{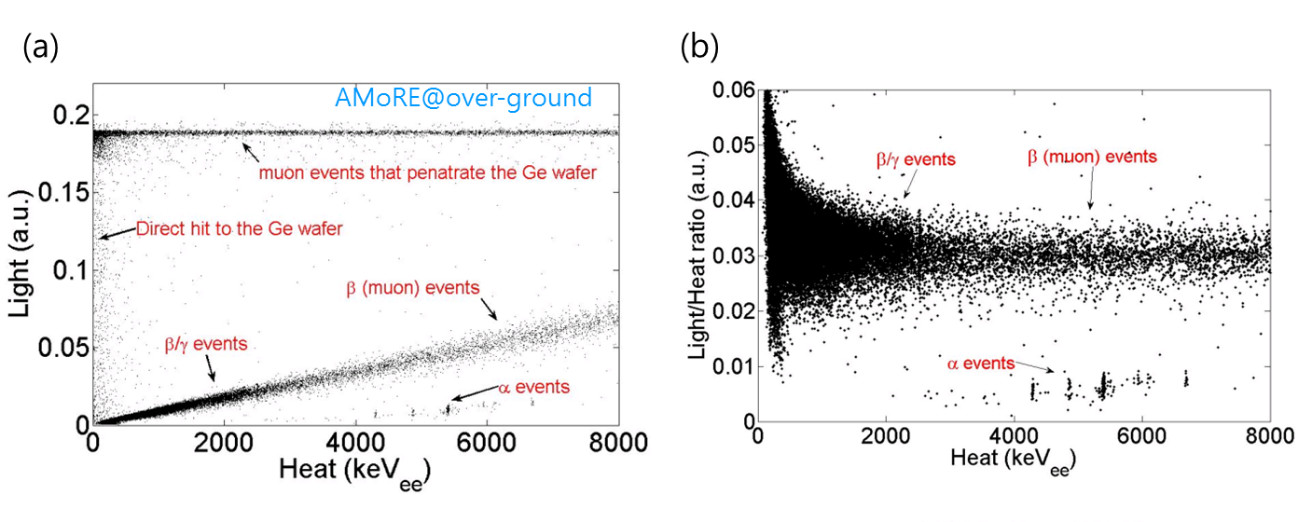}
\caption{Surface-level phonon-photon measurements with the 196~g \doublecamo bolometer in coincidences with a Ge-based light detector. (a) A two-dimensional scatter plot of signal sizes from the two sensors. (b)
The signal size ratios; $\alpha$- and $\beta /\gamma$-(including cosmic muons) induced events are clearly separated. }
\label{fig:fig4-7}
\end{figure}

Figure~\ref{fig:fig4-7}a shows a two-dimensional scatter plot of the signal sizes from the two sensors. Roughly four groups
of signals appear in the simultaneous measurement.  Cosmic-ray muons passing through the Ge wafer and the \camo crystal
produce the upper horizontal band in the scatter plot; the cosmic-ray-induced photon-sensor signals were saturated by the
muon energy deposits. Environmental backgrounds that are absorbed only in the photon detector produce the vertical band
near zero phonon signal sizes. The $\beta /\gamma$ events absorbed in the \camo crystal show a linear relation between
the two signal sizes. Muons passing through the \camo crystal but not the Ge wafer, extend this $\beta /\gamma$ event
behavior to higher energies. Alpha-induced signals also show a linear relation between the two sensors
but with smaller light signals than those for $\beta /\gamma$ signals.  These relative differences of the phonon
and photon signals for alpha and $\beta /\gamma$ events can be clearly seen in Fig.~\ref{fig:fig4-7}b, where the ratio
of the two sensors are plotted. A separation power of  $8.6\sigma$ for signals over a wide region of alpha-equivalent
energies (4~MeV$< E_{\alpha} <$7~MeV) is found. 

There is a clear light-yield difference between $\alpha$- and $\beta /\gamma$-induced events in the \camo crystal.
This effect, together with PSD from the phonon sensor alone, will provide a powerful tool for attaining
our goal of a ``zero background''  \zeronubb decay search experiment.

%% file: tex/experimental_design.tex
\chapter{Experimental design}

\section{Overview}
 We have recently completed the  assembly of the ``AMoRE-Pilot'' experiment that consists of an array of the five 
$\mathrm{^{48depl}Ca^{100}MoO_4}$ R\&D crystals that are currently on hand, with a total  mass of
approximately 1.5~kg.   The goals of AMoRE-Pilot are: a better understanding of the background conditions;
gain experience with the assemby and operation of the experiment; and reach the current state-of-the-art level of
$\zeronubb$ half-life sensitivity for $^{100}$Mo.
After about a year of data taking, we will move to the first phase of the experiment, dubbed AMoRE-I, that will
consist of approximately 5~kg of $\mathrm{^{48depl}Ca^{100}MoO_4}$ crystals in the same cryostat and shielding
as AMoRE-Pilot with modifications based on what we learn from our experience with the Pilot run.
This will start at the beginning of 2017 and will run for about three years.  During this time, we will continue
to pursue our vigorous program of R\&D on the deep purification of \camo and studies of other candidate crystals.
Based on our R\&D studies  and  our experience with the early AMoRE-I data, we will make decisions on the
crystal production and procurement for the second phase, AMoRE-II.  This phase of the experiment will start in early
2020 with about 70~kg of crystals instrumented in a new, larger cryostat situated in a new, deeper laboratory. 
This will be subsequently upgraded in mass to a maximum of 200~kg if the background levels are low enough to
warrant this.
 
\section{AMoRE-Pilot}

\subsection{Experimental arrangement}
\begin{figure} \centering
\includegraphics[width=0.8\textwidth]{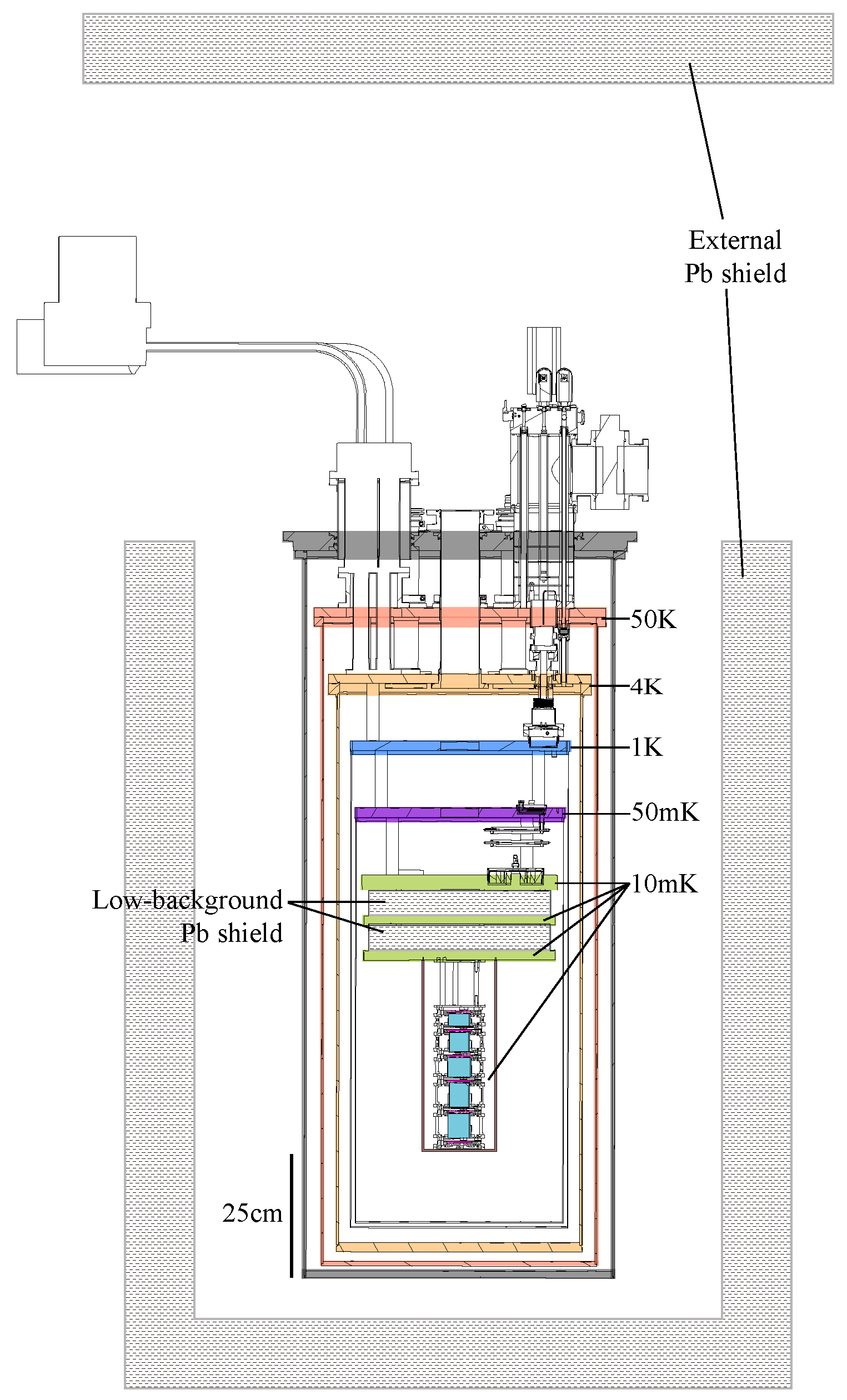}
\caption{A schematic view of the AMoRE-Pilot experimental arrangement.}
\label{fig:fig5-1}
\end{figure}

Figure~\ref{fig:fig5-1} provides a schematic view of the cryostat interior for the AMoRE-Pilot
experiment that has been assembled and is currently being commisssioned.  The cryostat has five
different temperature stages and two separate vacuum chambers.  The 4K stage defines the boundary
between the inner and outer vacuum volumes.   The cryostat is surrounded by an external, 15-cm-thick
lead shield in the low-temperature room in the A5 tunnel of Y2L. 

\begin{figure}\centering
\includegraphics[width=0.5\textwidth]{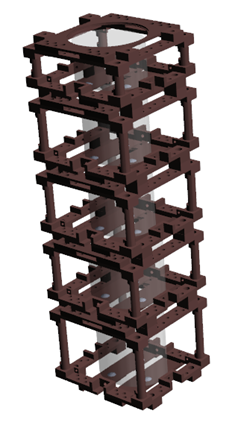}
\caption{A schematic view of the AMoRE-Pilot array of crystals with copper holders.}
\label{fig:fig5-2}
\end{figure}

The AMoRE-Pilot and the AMoRE-I experiments use the same dilution refrigerator with a cooling power of 1.4~mW at 120~mK 
and a base temperature of 10~mK.  The refrigerator can cool and maintain a \camo crystal array with a total mass
as large as 10~kg at temperatures as low as 10~mK.  A pulse-tube-based cryo-cooler is used to bring the system down
to 4~K, cooling the leak-tight inner vacuum chamber walls, the \CMO crystals, and the inner shielding to 4~K by helium
heat-exchange gas.  The  AMoRE-Pilot array of 300 -- 400~g crystals with their copper frames (see Fig.~\ref{fig:fig5-2})
is located in the dilution refrigerator's mixing chamber where the base temperature can be brought down to as low as 10~mK.
A sketch of the Pilot crystal array is shown in Fig.~\ref{fig:fig5-2} and the individual crystal dimensions and masses are
listed in Table~\ref{tab:cmo5}. 

\begin{table}
\caption{The dimensions and masses of the AMoRE-Pilot $\mathrm{^{48depl}Ca^{100}MoO_4}$ crystals.}
\input{tables/table5-1.tex}
\label{tab:cmo5}
\end{table}

The dilution refrigerator's mixing chamber is shielded from background radiation from the outer
stainless-steel walls of the cryostat by a 1-cm-thick assembly of four concentric copper cylinders 
that surround the bottom and sides of the array as shown in Fig.~\ref{fig:fig5-1}.
Radiation from radioactive contaminants in the upper layers of the cryostat are attenuated by
a double layer of 5-cm-thick, 40-cm-diameter disks of low-background lead (total mass $\simeq 150$~kg)
mounted on a 1-cm-thick low-background copper plate that is supported from the lowest temperature
stage of the cryostat.  The \camo crystal detector array is  encased in a  2-mm-thick cylindrical structure made
from the same low-background lead.  This cylinder, which is superconducting at the operating temperature of
the experiment, shields the detector from external magnetic fields.  An additional magnetic shield is
provided by two layers of mu-metal placed between the cryostat's outer vacuum vessel and the external
lead radiation shield (not shown in Fig.~\ref{fig:fig5-1}).  This magnetic shielding is necessary for
reducing possible electronic noise that is produced in the detector's readout wires from ground-vibration-induced
wire motions.   

\subsection{Inner shielding}
The lead shielding that is inside the cryostat is designed to reduce backgrounds due to scattered
gammas from the gap of the external shield and from radioactive contaminants of the cryostat components, such as
electrical wires, cooling pipes, and other sources from the upper layers of the cryostat.
Two layers of standard-thickness lead bricks from ancient lead taken from sunken Roman ships that was purchased from
and machined by LEMER PAX in France are used; the total lead thickness is 10~cm. The supporting disk is made from 1-cm-thick OFHC
copper plate with U/Th levels below 50~\ppt. The radioactivity levels in the ancient lead were measured underground in the Modane
laboratory in France and are listed in Table~\ref{tab:shieldinglead}.

\begin{table}
\caption{The measured radioactivity for the internal shielding lead.}
\input{tables/table5-2.tex}
\label{tab:shieldinglead}
\end{table}

\subsection{Outer shielding}
\label{Sec:Design:OuterShield}

\begin{figure}
\includegraphics[width=0.8\textwidth]{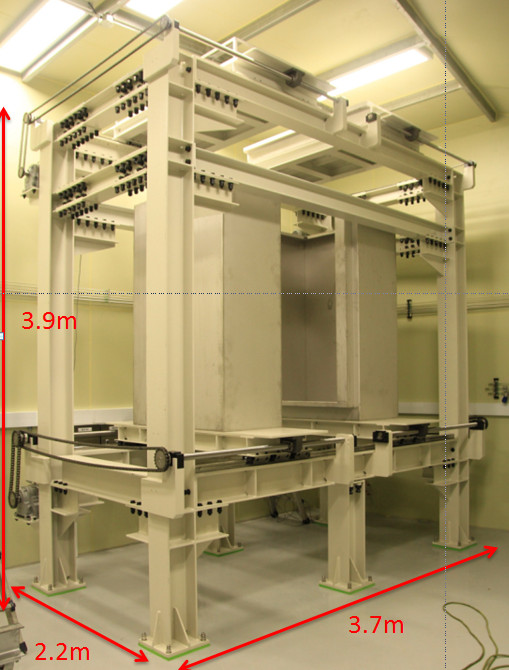}
\caption{The support structure for scintillating counters, external lead shielding and cryostat.}
\label{fig:shielding1}
\end{figure}

\begin{figure}
\includegraphics[width=0.8\textwidth]{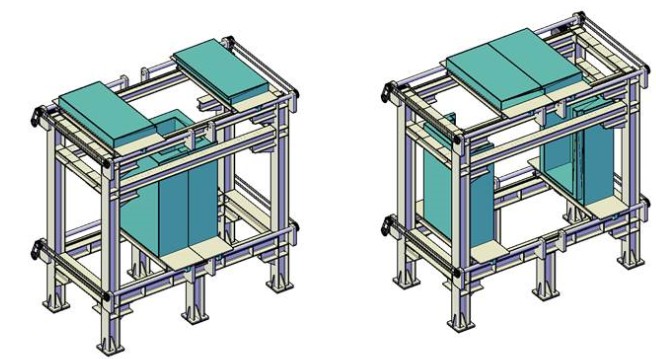}
\caption{The lead-shielding structure. The top (left) and main (right) structure can slide in and out independently.}
\label{fig:shielding2}
\end{figure}

The cryostat is situated inside a supporting structure made from steel H~beams, as shown Fig.~\ref{fig:shielding1}. This
structure supports an array of cosmic-ray veto scintillation counters and an external shield formed from a total of 15~tons
of lead bricks that attenuate gammas from the surrounding environment such as the rock walls of the A5 cavity.   The main
lead-shielding structure  sits on a 3-mm-thick stainless-steel bottom plate and surrounds the bottom and sides of the cryostat
with transverse dimensions of 1.5$\times$1.1~$\rm m^2$ and a height of 1.75~m. The structure is built in two halves that can
be independently moved apart to provide access to the cryostat. The top is covered by a separate lead ceiling at the top of
the support structure, with a gap between it and the sidewall shield for electrical wires and cooling pipes that run from
the cryostat to the readout and cooling control systems elsewhere in the experimental hall (see Fig.~\ref{fig:shielding2}). 
The transverse dimensions of the top lead shield are determined to be sufficient to prevent any external gamma
to have an unshielded, direct path to the crystal array.

The shield is assembled from standard-sized lead bricks, dimension of 50~mm $\times$ 100~mm $\times$ 200~mm, that
were purchased from JL Goslar GmbH in Germany. They were cleaned and machined by a local company in Korea.  The total
thickness of the external lead shield is 15~cm. The $\rm^{232}$Th and $^{\rm 238}$U contents
in the brick were measured to be 3.8 \ppt and 6.9 \ppt, respectively,
with an Inductively-Coupled-Plasma Mass Spectrometer (ICP-MS). The $^{\rm210}$Pb activity
was measured to be $\rm(59\pm6)~Bq/kg$ at PTB (Physikalisch-Technische Bundesanstalt) in Germany.

\subsubsection{Cosmic-ray Muon Veto Counters}
\begin{figure} \centering
\includegraphics[width=0.8\textwidth]{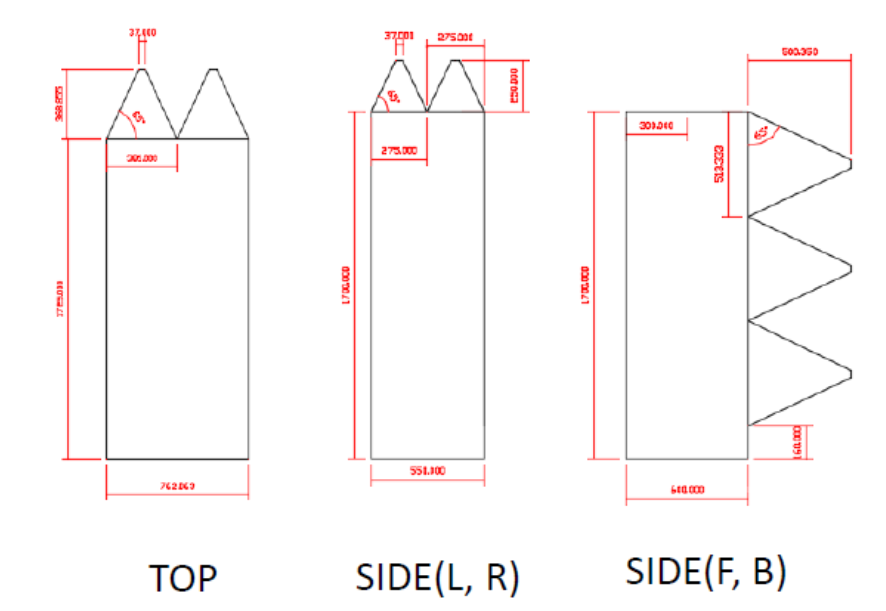}
\caption{Different sizes of plastic scintillators that cover the shielding structure. }
\label{fig:fig5-5}
\end{figure}

\begin{figure} \centering
\includegraphics[width=0.8\textwidth]{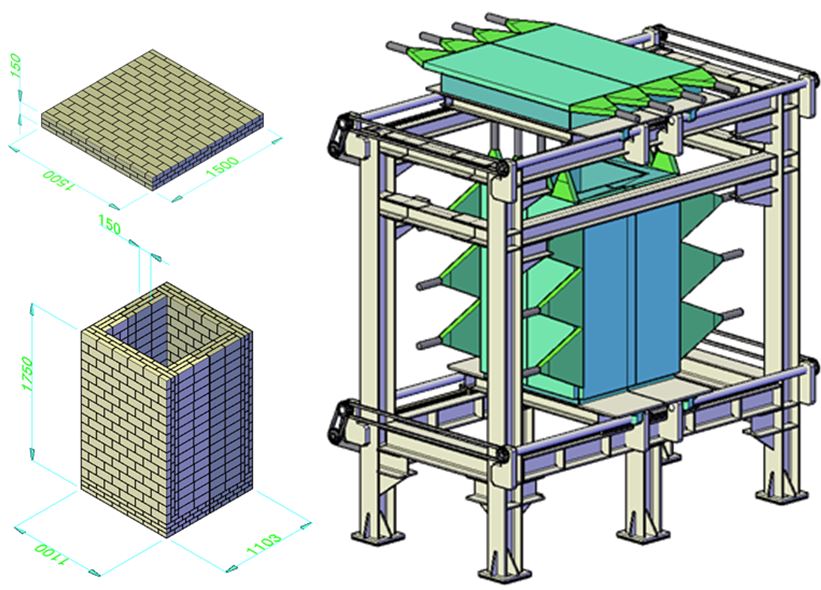}
\caption{A schematic view of the muon veto system.}
\label{fig:fig5-6}
\end{figure}

The top and sides of the external lead shield are surrounded by a cosmic-ray muon veto system based on 5-cm-thick 
plastic scintillators. Each scintillator panel (EJ-200 by Eljen Company) is coupled to 2~inch PMTs via light guides as shown in
Fig.~\ref{fig:fig5-5}.  The muon-veto scintillation counters are installed on the outside of the shielding structure
as shown in Fig.~\ref{fig:fig5-6}.

\section{AMoRE-I}

\begin{figure}
\begin{center}
\includegraphics[width=0.3\textwidth]{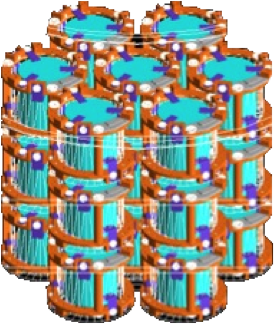}
\caption{A schematic view of a possible AMoRE-I crystal array.}
\label{fig:fig5-7}
\end{center}
\end{figure}

The AMoRE-I experiment will use about 5 kg of double enriched
\doublecamo detectors (about 15 crystals). The crystals are under preparation
by the Fomos-Materials company in Russia. A schematic view of the planned crystal array is shown in Figure
~\ref{fig:fig5-7} for a 10~kg configuration. With this system, 5~kg of natural \camo crystals
may be used together with the 5~kg of enriched crystals. 

The cryostat and the shielding structure will be the same as
AMoRE-Pilot experiment; our current intent is to phase from AMoRE-Pilot to AMoRE-I
experiment by adding more crystals from Fomos Materials. We plan to begin the AMoRE-I
experiment in 2017. The main purpose of the AMoRE-I experiment is to verify that the background levels
are consistent with a zero-background environment.  At the same time, we will finalize the crystal choice
for the AMoRE-II experiment, that is, whether we will use \camo crystals
for the next phase or switch to another molybdate crystal, such as ${\rm ZnMoO_{4}}$ or ${\rm Li_{2}MoO_{4}}$ crystal
scintillators.  The decision will be made while we run the AMoRE-I experiment.

\section{AMoRE-II}

\begin{figure}
\begin{center}
\includegraphics[width=0.2\textwidth]{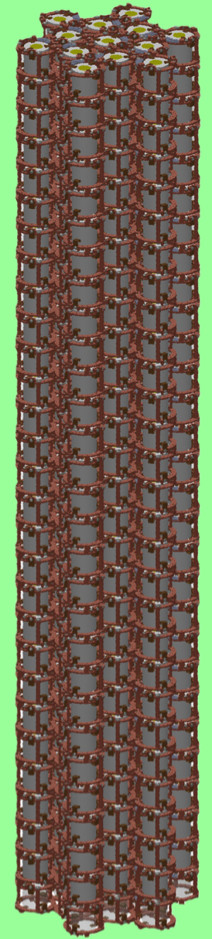}
\caption{A schematic view of a possible configuration of the AMoRE-II crystal array.}
\label{fig:fig5-8}
\end{center}
\end{figure}

For the AMoRE-II experiment, we aim to ultimately operate with a 200~kg array of
molybdate crystals; a schematic diagram of a possible crystal arrangement in which the AMoRE-I configuration is
stretched into long towers is shown in Fig. \ref{fig:fig5-8}. The configuration can
be changed, and the arrangement shown in Figure \ref{fig:fig5-8} is only one possibility. Our goal is to settle
on the AMoRE-II design in the middle of year 2019.

For a 200~kg array of the crystals, we will need a much larger cryostat than that being
used for AMoRE-Pilot and AMoRE-I experiments. The design of the larger cryostat
will begin in 2017 and continue for about a year. 

\subsection{Crystals for AMoRE-II}

To maintain a zero-background environment that fully exploits the increase in detector mass, the crystals
for AMoRE-II must have internal background levels that are an order-of-magnitude below
those of the current state-of-the-art crystals that will be used in
AMoRE-I. Moreover, to reach our total mass goal at reasonable cost, we
have to be able to recover the precious $\rm ^{48depl}Ca$ and \mohundred isotopes from
the scrap material produced in the crytal-growing and cutting
processes with high efficiency.  To accomplish these tasks we are
engaged in an aggressive R\&D program aimed at improving chemical
purification and recovery processes. As part of this we have set up
our own chemical purification laboratory and recruited a PhD-level
chemist specialized in low-level impurity extraction to manage it and
supervise a small staff of purification laboratory (mostly PhD students). In
addition, we have constructed our own crystal growing facility to research
crystal-growing techniques. This is managed by a PhD physicist with a specialty in crystal-growing.

%% file: tables/table5-1.tex
\newcolumntype{Y}[1]{>{\centering}p{#1}}															
\begin{tabular}{|p{5.1cm}|Y{2cm}Y{1.2cm}Y{1.2cm}Y{1.2cm}Y{1.2cm}|}															
\hline															
				&	SB28	&	NSB29	&	S35	&	SS68	&  SS81	\tabularnewline	\hline
		Height (mm)	&	25.5	&	51	&	40	&	40	&	52 \tabularnewline\hline
		Major axis (mm)	&	49.5	&	50	&	45	&	53	&	45	\tabularnewline	\hline
		Minor axis (mm)	&	40.5	&	42.5	&	40	&	47.2	&	41 \tabularnewline\hline
	        Mass (g)	&	196	&	390	&	256	&	352	&	354	\tabularnewline	
\hline															
\end{tabular}															

%% file: tables/table5-2.tex
\newcolumntype{Z}[1]{>{\centering$\rm}p{#1}<{$}}								
\newcolumntype{Y}[1]{>{\centering}p{#1}}								
\begin{tabular}{|Y{3.5cm}Y{3.5cm}Y{3.5cm}|}								
\hline								
$\rm	^{210}Po~  \newline[Bq/kg]	$&$\rm	Total~\alpha~rate\newline [cm^{-2}h^{-1} ]	$&	Total $\gamma$ from U, Th, K and Cs [$\rm mBq/kg$]		\tabularnewline	\hline
$\rm	(0.30 \pm 0.08)	$&$\rm	0.01	$&$	 <7.4 	$	\tabularnewline	\hline
\end{tabular}								

%% file: tex/simulation_tools.tex
\chapter{Simulation tools and background estimations}
\label{chap:Sim}
To estimate background conditions in the AMoRE experiments, we are performing simulations  with the GEANT4 toolkit~\cite{GEANT4,GEANT4b}.  Since radiation originating within the \camo will probably be the dominant source of backgrounds, internal \camo crystal backgrounds from the full \U, \Th, and $^{\rm 235}$U decay chains, as well as from \K[40], were simulated and their effects on the \mohundred \zeronubb decay signal region were investigated. For the AMoRE-I  experimental  configuration, backgrounds from materials in the nearby detector systems, including the internal lead shielding plate, G10 fiberglass, and the outer lead shielding box, were simulated, as well as backgrounds from more remote external sources, such as the surrounding rock material. Simulation results were normalized to measured activity levels. Activities of \U, \Th, \K[40], and \U[235] were measured by HPGe detectors or inductively coupled plasma mass-spectroscopy.  The ICP-MS measurements were all performed by the KAIST Analysis Center for Research Advancement (KARA). For all simulations, except as otherwise stated below (with particular exception for the \Pb[210] subchain) we presently assume that the decay chains of \U, \U[235], and \Th are each in equilibrium, thus all related activities within the chains are simply equal to the \U, \U[235], and \Th activities multiplied by the branching ratios for decay of the daughter isotopes.

For backgrounds that originate from decays outside of the cryostat, such as from the lead shield or rock shell,
only gamma de-excitations were found to produce signals in the crystals. Random coincidences of radiation from different background sources were found to have the most significant effect on the \mohundred \zeronubb decay signal region.  These are reported here explicitly for sources where they were found to be significant.  When not explicitly stated, random coincidences are not included. Cosmic-ray induced backgrounds were also simulated. Other sources of background (cosmogenic $^{88}$Y, residual \Ca[48] in \doublecamo, and \Bi[214] in copper) are not expected to contribute significantly to the background near the signal region, but they will be considered in the future.

Results from background simulations described in this section are distinguished as $\alpha$- or $\beta /\gamma$-like. A separation power of 8.6$\sigma$ is demonstrated in Sec.~\ref{Sec:Cryo:Simul} for separation of $\alpha$- and $\beta/\gamma$-like events, implying that significant $\alpha$ background rates in the region of interest, ${\rm ROI}\equiv 3034\pm 10$~keV, can be safely rejected.  Results are reported in Table~\ref{tab:major_bkgds} and can be compared to the AMoRE-I background goal of 0.002~counts/keV/kg/yr.

\section{Geometry of simulation configuration}

\begin{cfigure1c} [h]
\begin{tabular}{ccc}
\includegraphics[width=0.3\textwidth, trim=10 10 10 10,clip ]{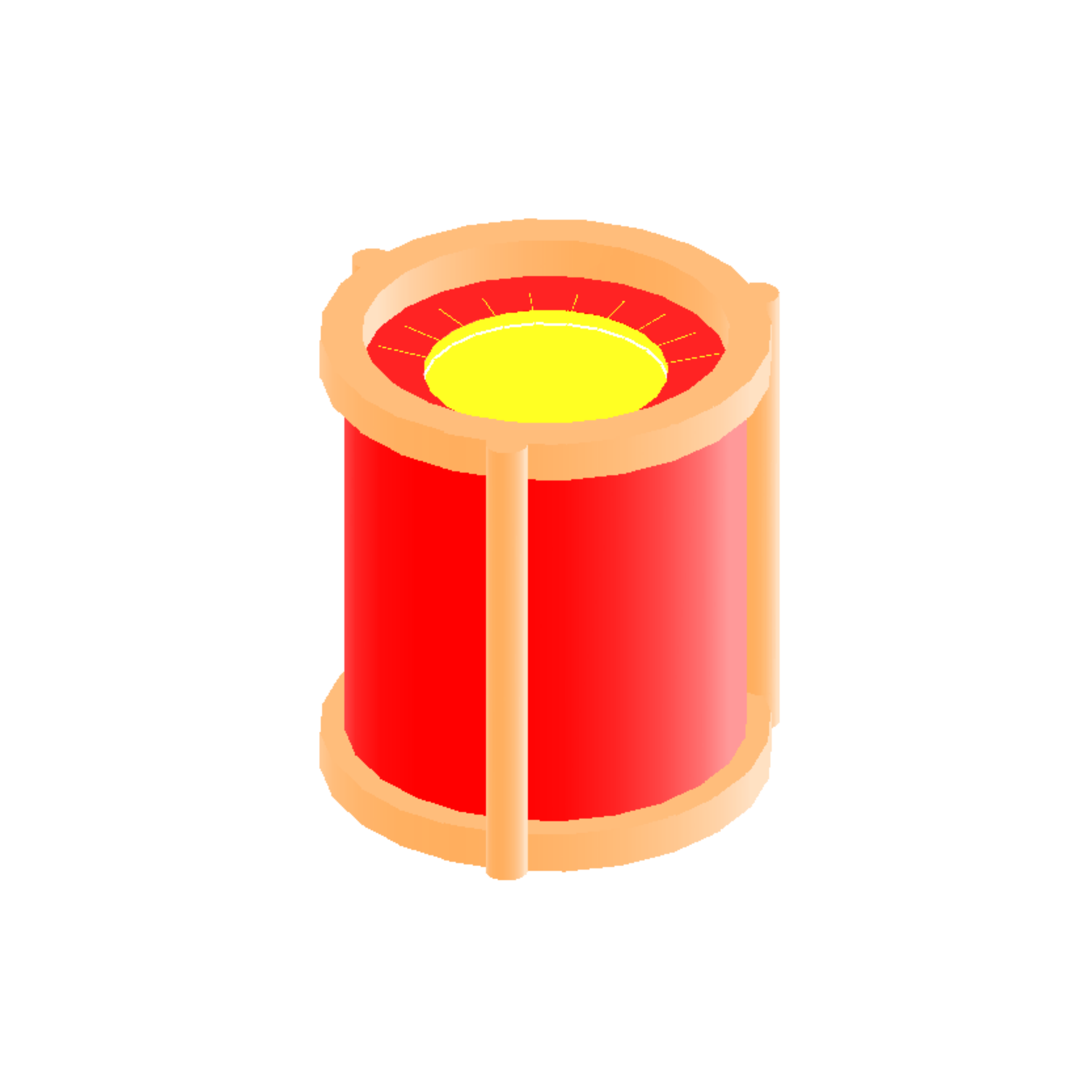} &
\includegraphics[width=0.3\textwidth, trim=10 10 10 10,clip ]{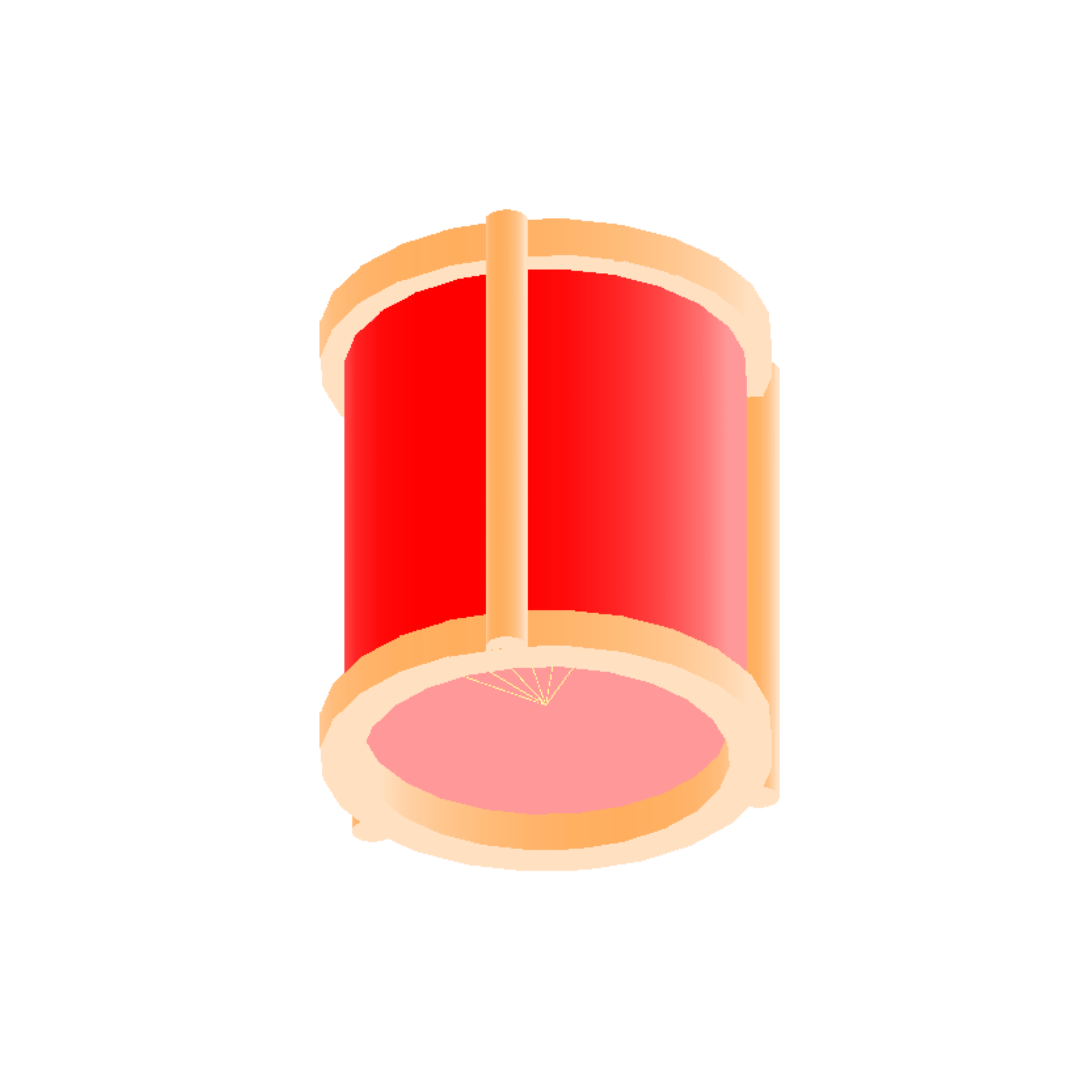} &
\includegraphics[width=0.3\textwidth, trim=10 10 10 10,clip]{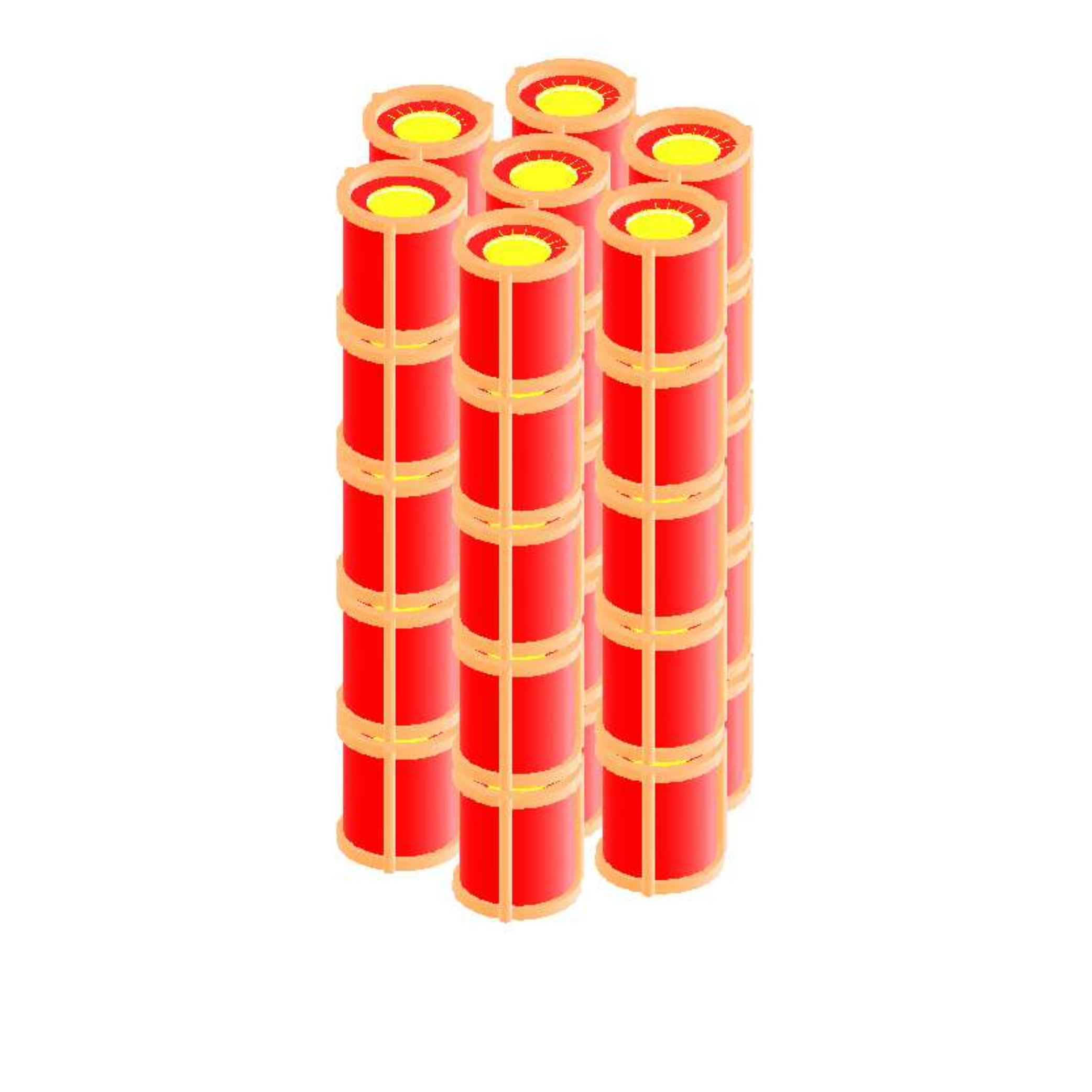} \\
(a) & (b) & (c) \\
\end{tabular}
\caption[Detector Geometry]{
(a) \camo crystal with Vikuiti reflector and Cu supporting frame;
(b) bottom view of \camo crystal;
(c) the 35 \camo crystals were stacked up in 5 layers and 7 columns.
}
\label{fig:sim-2}
\end{cfigure1c}

\begin{cfigure1c}[h]
\begin{tabular}{cc}
\includegraphics[width=0.45\textwidth, trim=110 10 110 130,clip ]{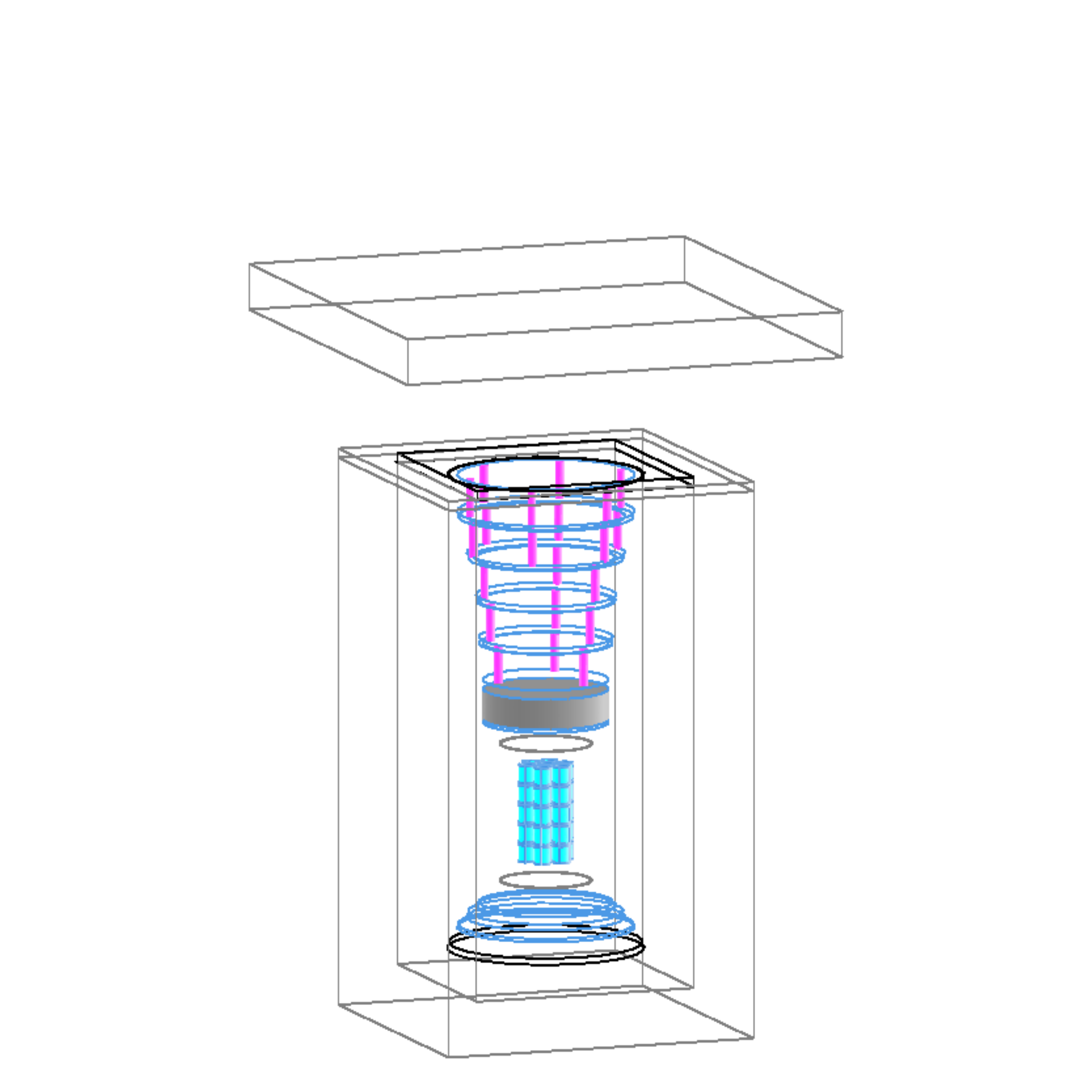} &
\includegraphics[width=0.55\textwidth, trim=60 20 110 130,clip]{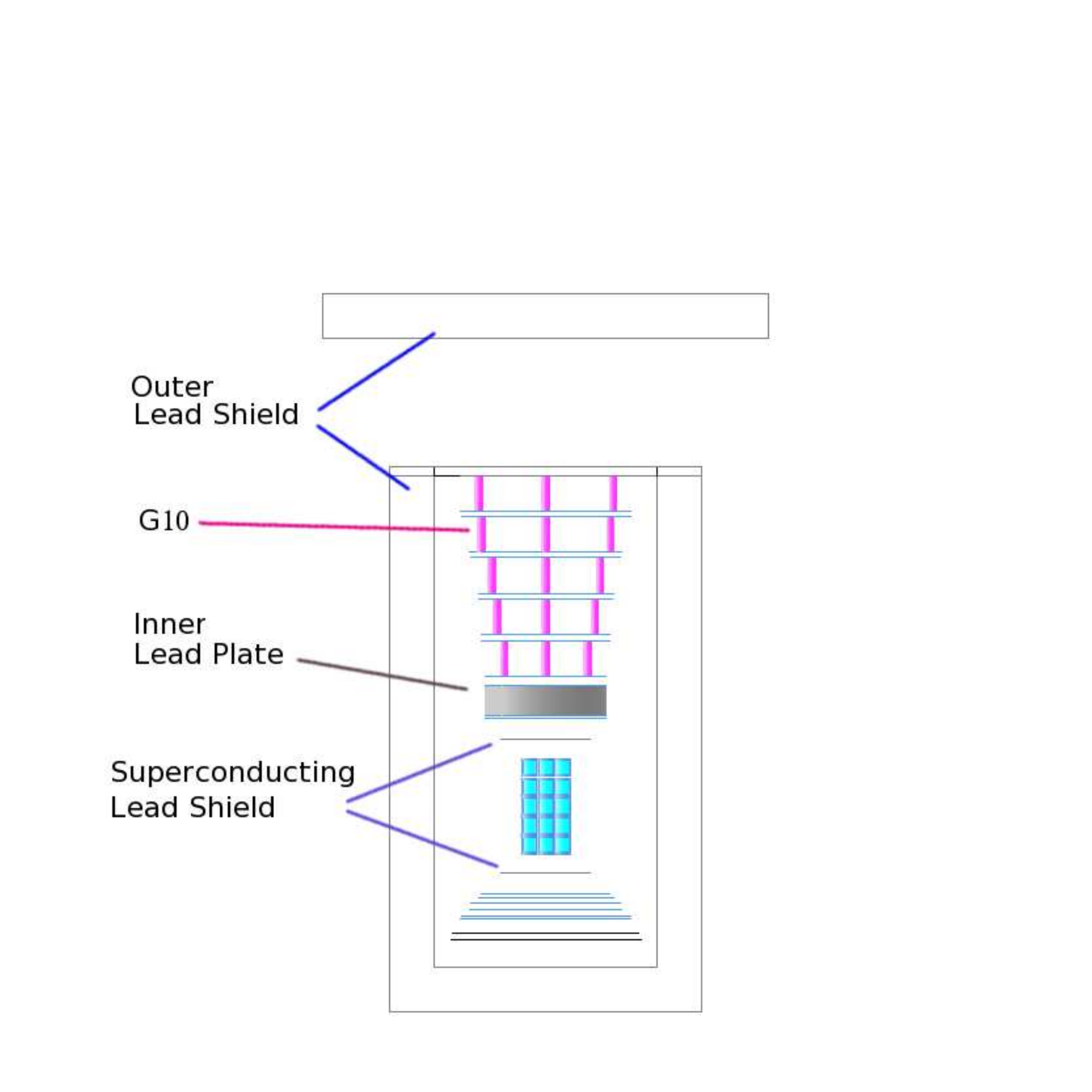} \\(a) & (b) \\
\end{tabular}
\caption[fig:sim-1]{
Geometry of the AMoRE-I Monte Carlo Simulation. The outer vacuum cylinder (OVC) and  four Cu shields surround the \camo  crystals.  Inside the  cryostat, the  Inner  Lead plate  (gray)  on top of 
a Cu plate  (blue)  is located above the crystals.  This lead plate attenuates radiation from the G10 support
rods (magenta) and other materials in the upper cryostat.  
}
\label{fig:sim-1}
\end{cfigure1c}

The detector geometry of the AMoRE-I simulation includes the \CMO crystals, the shielding layers internal to the
cryostat, the external lead  shielding box, and the outer rock shell. An array of 35 \CMO crystals is located
inside the cryostat.  Each crystal has a cylindrical shape with a 4.5~cm diameter, 4.5~cm height, and a mass of
310~g.  The  total simulated crystal mass is 10.9~kg. The 35 crystals are arranged in seven vertical columns,
each with five crystals stacked coaxialy, with one center column surrounded by six external ones. The side
surface of each crystal is covered by a Vikuiti ESR reflecting film.

As discussed above, each crystal is contained in a copper mounting frame as shown in Fig.~\ref{fig:sim-2}.
A 10-cm-thick lead plate (diameter 40.8~cm and mass 148.3~kg) placed on a 1-cm-thick copper plate is located
just above the \CMO crystal array to attenuate backgrounds from cryogenic piping, electrical connectors, and
other internal cryostat structures. The array is contained inside of four concentric copper  cylinders with
a total Cu thickness of 10~mm, all within an outer stainless-steel vacuum cylinder that is 5~mm thick.
Sequential top plates are connected with G10 tubes. In the current version of the simulation, a simple top
plate without any features or structures is used.  Realistic structures and features have been positioned
on each of the top plates for use in future simulations.  

The cryostat is located inside a 15-cm-thick external lead shield. The top plate of the lead shield was placed
above the lead shield and covering an area of $\rm 150\times 150~cm^{2}$. A 50-cm-thick rock shell surrounds the
lead shield and cryostat, representing the rock cavern at the YangYang Underground Laboratory (Y2L) where the
AMoRE-I experiment will operate.

\section{Internal background in \camo}
We  simulated  the full \U, \Th and \U[235] decay chains, as well as \K[40], with contamination
taken to be uniformly distributed inside the thirty-five crystals. The assumed radioactivity levels of the contaminants, listed in Table~\ref{tab:5-1}, were based on the measurements described above in Section~\ref{CMO_radio}.  About $10^6$ events each for \U, \Th, \K[40] and \U[235], and  about $10^{7}$ \Pb[210] events were simulated.

\begin{table}[!htb]
\begin{center}
\caption{Levels of radioactive contaminants based on SB28 measurement.}
\label{tab:5-1}
\begin{tabular}{c|c|c|c|c|c}\hline
&	\Pb[210] & \U & \Th & \K[40] & \U[235] \\ \hline
\multicolumn{1}{ c|  }{Activity}	& \multirow{2}{*}{10} & \multirow{2}{*}{0.1} & \multirow{2}{*}{0.05} & \multirow{2}{*}{1} & \multirow{2}{*}{1} \\  
\multicolumn{1}{ c|  }{[mBq/kg]} & & & & & \\ \hline
\end{tabular}
\end{center}
\end{table}

Background rate estimates were determined from the  numbers of events  in the \Mo[100] \zeronubb ROI, as shown in Table~\ref{tab:5-2}.  Most of the events in the signal region are found to originate from $\alpha$ decays.  The $\beta$-decay-induced events are mostly from the \Th chain and originate from \Tl[208]. Because the \Tl[208] half-life is only 3.1 minutes, decays of \Tl[208] can be rejected using time correlations with the $\alpha$-signal from the preceding \Bi[212]$\!\to$~\Tl[208]~$\alpha$ decays.  
Rejection of $\beta$ events that occur within 15 minutes after a 6.207~MeV $\alpha$ event in the same crystal, results in a 94\% veto efficiency for \Tl[208]-induced $\beta$ events in the signal region, while introducing a negligible dead-time.

\begin{table}[!htb]
\begin{center}
\caption{Simulated background rates (counts/keV/kg/yr) in the ROI for sources within the crystals, before analysis cuts.}
\label{tab:5-2}
\begin{tabular}{c|c|c|c|c|c}\hline
&	\Pb[210] & \U & \Th & \K[40] & \U[235] \\ \hline
Total anti-coincidence rate	& 0.023	& 0.0020	& 0.0278	& 0	& 0.012 \\ \hline
$\alpha$ event rate	& 0.023	& 0.0017	& 0.0005	& 0	& 0.012 \\ \hline
$\beta$ event rate	& 0	& 0.0003	& 0.0273	& 0	& 0 \\ \hline
\end{tabular}
\end{center}
\end{table}

\begin{figure}[h]\centering
\includegraphics[height=0.42\textwidth]{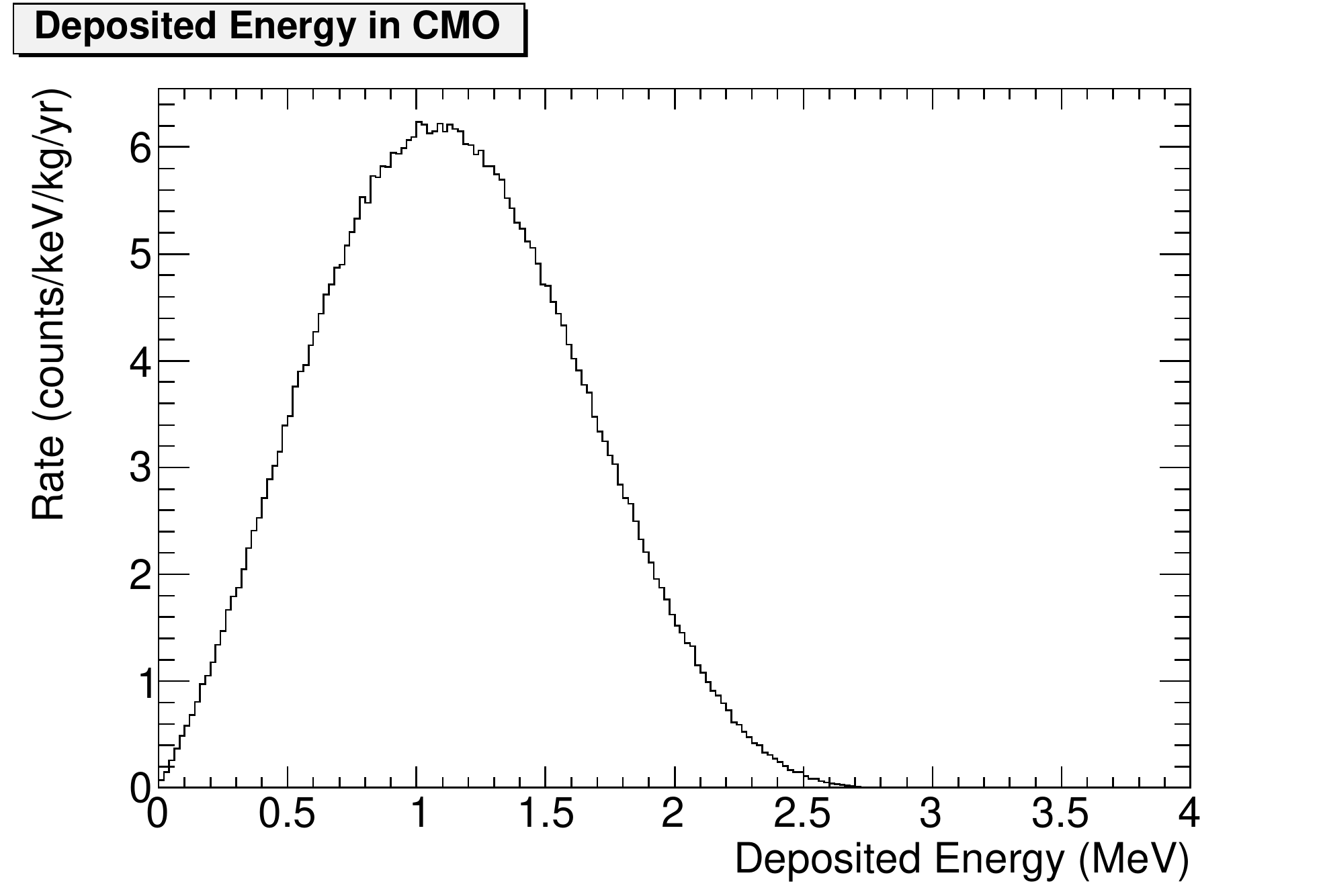}
\includegraphics[height=0.42\textwidth]{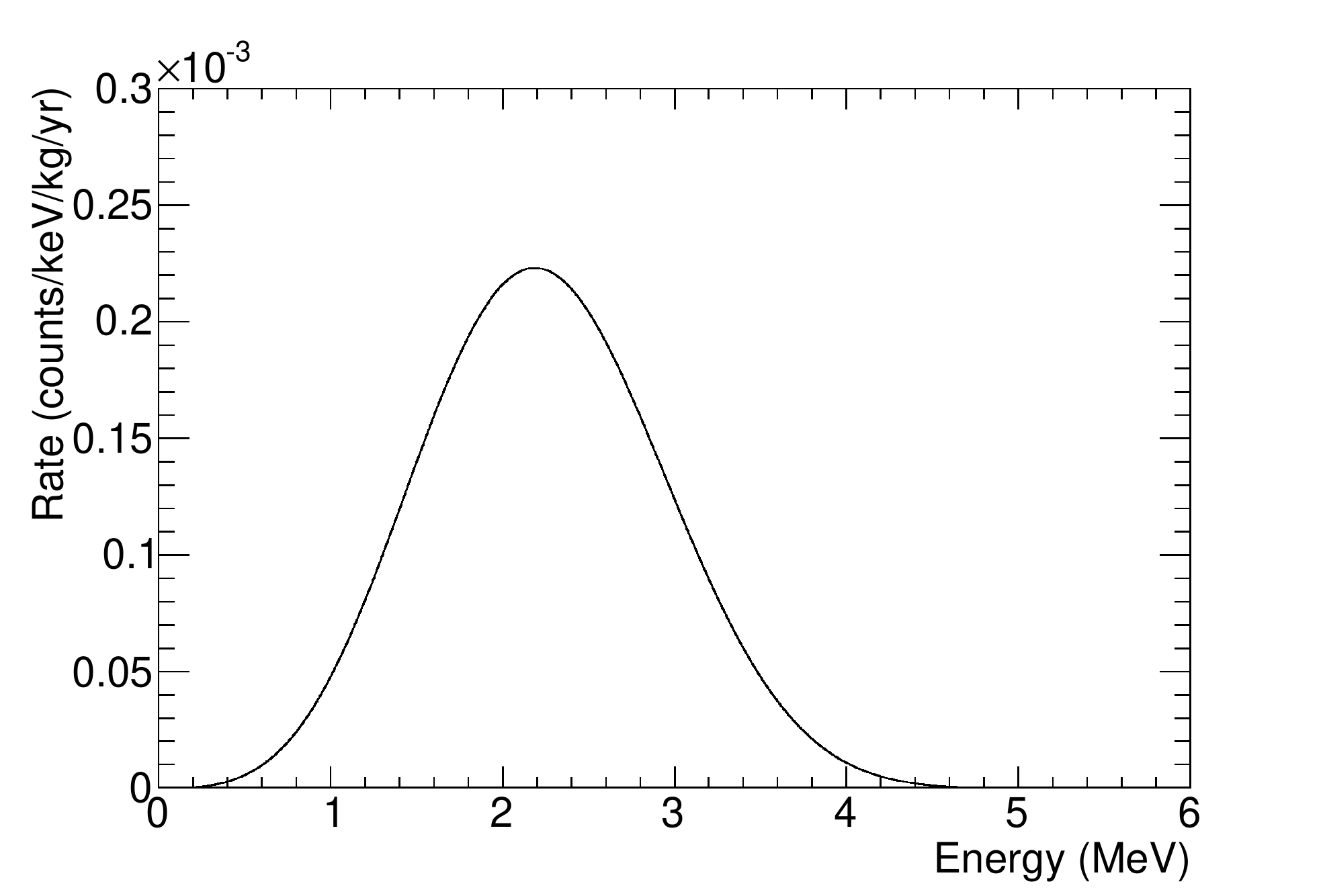}%
\caption[ConvolutionBB]{The energy distribution (bottom) of random coincidences of two \twonubb decays of \Mo[100].
The random coincidence spectrum is derived by convolving two  \twonubb spectra (top), and was normalized according
to the \twonubb rate assuming an 0.5~ms coincidence window.}
\label{fig:sim-3}
\end{figure}

The \twonubb decay in a \camo approaches zero rate at the end-point energy, but random coincidence of these
events can sum together (pileup) creating backgrounds for the \zeronu decay signal.  
The expected rate of \twonubb decay in a single \camo crystal is 0.00284 counts/s, 
as listed in Table~\ref{tab:5-4} with an energy distribution
as shown in Fig.~\ref{fig:sim-3} (left).

\begin{table}[!htb]
\begin{center}
\caption[ExpectedRate]{
Expected rate of \twonubb decays.
}
\label{tab:5-4}
\begin{tabular}{lcc} \\
\hline
  & one crystal & 35 crystals  \\
\hline
 Mass (kg) &  0.31 & 10.87    \\
\hline
\twonubb rate (counts/s) & 2.84 \tm $10^{-3}$ &  9.93 \tm $10^{-2}$ \\
\hline
\end{tabular}
\end{center}
\end{table}

When $5\times10^{8}$ \twonubb events were simulated in the \camo crystals (310~g each), 
corresponding to about 335~years of operation,  \sm 98.6$\%$ of events deposited energy in one crystal (single-hit events) and the remaining events produced hits in multiple crystals. The random coincidence rate of two \twonubb decays was calculated by convolving two single-hit \twonubb decay energy distributions  (see Fig.~\ref{fig:sim-3}),  and the accidental rate, assuming an 0.5 ms coincidence window, is 1.18 \tm $10^{-4}$ counts/keV/kg/yr in the ROI. This is substantially lower than the AMoRE-I background goal of 0.002~counts/keV/kg/yr.

\section{Backgrounds from materials in detector system}
\label{subsec:sim_pbs}
We simulated the full decay chains of backgrounds in the outer lead shielding box and in other items such as
the inner lead plate,  the copper support frame for the crystals, the G10 fiberglass structural elements of
the cryostat, and the superconducting-lead magnetic shield. 
The assumed concentraions of radioactive background nuclides were based mostly on
ICP-MS measurements~\cite{So13} and were used to normalize the simulation results.

\begin{table}[!htb]
\begin{center}
\caption{Levels of radioactive contaminants in materials inside the cryostat. Values marked with ''$^*$'' were arbitrary assumptions. }
\label{tab:5-5}
\begin{tabular}{p{3cm}|c|c|c|c|c}\hline
&	\Pb[210] & \U & \Th & \K[nat] & \U[235] \\ \hline
\multicolumn{1}{ c|  }{Crystal-supporting}	&  \multirow{2}{*}{-}	&  \multirow{2}{*}{0.027 \ppb}	&  \multirow{2}{*}{0.051 \ppb}	&  \multirow{2}{*}{-}   &  \multirow{2}{*}{-} \\
\multicolumn{1}{ c|  }{copper frame}	&	&	& 	&   & \\ \hline
Inner plate lead from Lemer PAX T2FA & 0.3 Bq/kg & 1$^*$ \ppt & 1$^*$ \ppt & 9$^*$ \ppb  & 1$^*$ \ppt  \\ \hline
G10 fiberglass	& - & 1932 \ppb & 12380 \ppb & 360 \ppm  & - \\ \hline
Superconducting shield, Lemer PAX T2FA lead, & 0.3 Bq/kg & 1$^*$ \ppt & 1$^*$ \ppt & - & - \\ \hline
Outer shield lead from JL Goslar Gmbh & 59$\pm$ 6 Bq/kg & 6.9 \ppt & 3.8 \ppt & - & - \\ \hline

\end{tabular}
\end{center}
\end{table}

The ICP-MS technique measures concentrations of \K[39] in the material. The concentrations of \K[40] were
calculated by assuming the ratio of concentrations of \K[39] to \K[40] is equal to the natural abundance ratio.  
The concentrations shown in Table~\ref{tab:5-5} were based on our measurements, some of which were provided by the vendors.

In order to reduce backgrounds from the lead shield inside the cryostat, we purchased T2FA lead bricks from
Pax Lemer with a certified activity of \Pb[210] of 0.3 Bq/kg of \U and \Th of $< 1$~\ppt.  The concentration
of \K is not yet known but decays of \K[40] are expected to produce a negligible background contribution. 
The low $Q$-value of these decays implies that they can only contribute to the \camo \zeronubb ROI by way of
accidental coincidence. 

We itemize below the items inside the detector shielding for which radioactive backgrounds have been measured:
\begin{itemize}
  \item \CMO copper support frame (the yellow elements in Fig.~\ref{fig:sim-2});
  \item G10 fiberglass support tubes for cryostat (the magenta elements of Fig.~\ref{fig:sim-1});
  \item lead plate inside cryostat (the gray item in Fig.~\ref{fig:sim-1});
  \item superconducting lead magnetic shield (this surrounds the inner crystal array). \\
\end{itemize}

In order to attenuate gammas that originate from the surrounding rock a 15-cm-thick
lead shielding box surrounds the cryostat. The top lead plate is located 46.4~cm above
the top of the lead side-walls. The total mass is 15.6~tons.  The lead was purchased from JL Goslar
GmbH in Germany.  The lead came with a \Pb[210] certification from PTB, the National Standards Laboratory of Germany
and was measured by ICP-MS at KARA for \U and \Th contamination.  Activities are listed in Table~\ref{tab:5-5}. 
The expected number of \U, \Th decays originating from the lead shield were calculated based on the inferred
activities and natural isotope abundance ratios.  
For \U and \Th no simulated events were found in the experimental ROI.

The expected rate of \Pb[210] decays in the lead shield is 1.08 \tm $10^{11}$~counts/day ($\rm 3.943 \tm 10^{13}$~counts/yr) when the concentration of \Pb[210] in the lead shield is on the level of 60 Bq/kg, (The lead vendor provided a certification of $59\pm 6$~Bq/kg). We simulated $\rm 4.4 \tm 10^{9}$~\Pb[210] events uniformly distributed within the lead shielding. This corresponds to the expected number of events for 1.25~hr for a concentration of 60~Bq/kg. The expected event rate of \Pb[210] from the lead shield in \camo is $\rm 8.9 \tm 10^{5}$~counts/keV/kg/yr. No events were produced in the signal ROI.

\begin{figure}[h]
\begin{center}
\includegraphics[width=0.7\columnwidth]{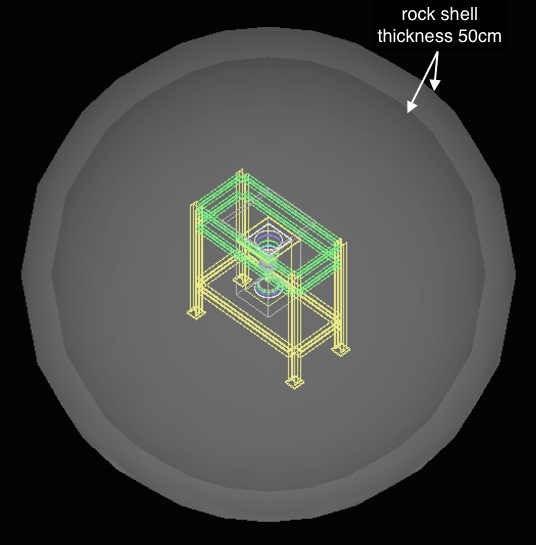}
\caption{Geometry of the rock shell surrounding the setup.}
\label{fig:sim-6}
\end{center}
\end{figure}

To check the effectiveness of the outer shield at attenuating gammas from the surrounding rock, and to optimize the
size of the top lead plate, mono-energetic $\gamma$'s of 2615~keV  (emitted from \Tl[208]) and 1461~keV (from \K[40]) 
were generated randomly in a 50-cm-thick layer of the surrounding rock shell (Figure~\ref{fig:sim-6}). The total mass
of the 50~cm rock shell is $\rm 5.7\times 10^{8}$~g. The measured concentrations of \U, \Th and K in the rock
material are given in Table~\ref{tab:Concentrations}.

\begin{table}[!htb]
\begin{center}
\caption{Concentrations of \U, \Th, and K in the rock sample from A5 tunnels in Y2L, measured by ICP-MS.}
\label{tab:Concentrations}
\begin{tabular}{c|c|c|c}\hline
&	\U & \Th & \K[nat] \\ \hline
Concentration (\ppm)	& 2.7	& 9.6	 & 2100 \\ \hline
\end{tabular}
\end{center}
\end {table}

From these simulations we conclude that a top lead plate with transverse dimensions 150~cm $\times$ 150~cm would
be sufficient to block $\gamma$'s from entering through the gap in the shielding.
The rate of events generated in the detector from the rock is found to be smaller than the rate of events arising
from \Pb[210] in the shield,  but the mean energies of gammas in the rock-produced gamma spectra are higher.
No events were produced in the signal ROI.  The rate of \Mo[100] \twonubb decays in random coincidence with
$\gamma$'s from \Th in the rock was estimated to be 1.01 \tm $10^{-4}$ counts/keV/kg/yr in the signal ROI.

\section{Cosmic ray induced background}
We have simulated cosmic muons and neutrons induced 
by muons passing through the detector system. We did not simulate the environmental neutrons produced by 
radioactive decays in underground rock since these neutrons have energies below 10 MeV and can be shielded
with thick polyethylene blocks at outside of the detector.

\subsection{Muon energy spectrum}

\begin{figure}[h]
\begin{center}
\includegraphics[width=0.6\columnwidth]{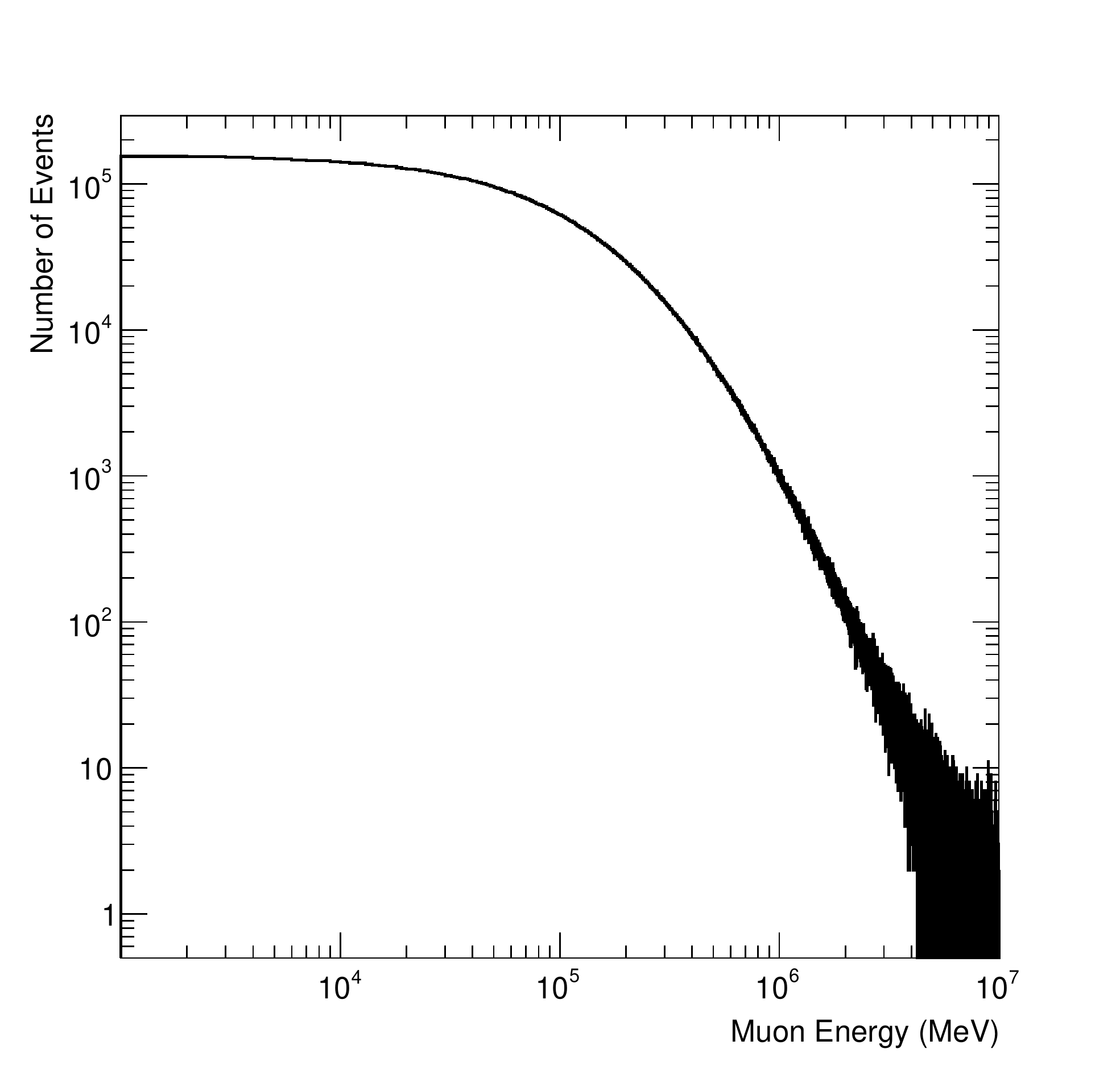}
\caption{Muon energy spectrum at Yangyang.}
\label{fig:sim-4}
\end{center}
\end{figure}

High energy neutrons are produced primarily by muons passing through the materials in the detection system. 
These neutrons generate secondary neutrons via hadronic interactions in the surrounding materials.
We simulated all the primary and secondary processes using GEANT4 simulation program starting with muons passing
through the detector system.  We used the muon energy spectrum provided by Mei and Hime~\cite{Mei06}:

\begin{equation}
  \frac{dN}{dE_{\mu}} = Ae^{-bh(\gamma_{\mu}-1)} \times (E_{\mu} + \epsilon_{\mu}(1-e^{-bh}))^{-\gamma_{\mu}}\, ,
\end{equation}
where $h$ is the rock slant depth in km.w.e.  We used the  parameters provided by
Groom \emph{et~al.}~\cite{Groom01} for $\epsilon_{\mu}$, $b$, and $\gamma_{\mu}$.  For Y2L, $h$=1.8~km.w.e. Figure~\ref{fig:sim-4}
shows the muon energy spectrum generated at this depth.
The average muon energy is 201~GeV, which is consistent with the measured values of Mei and Hime~\cite{Mei06}.

\subsection{Schematic layout of simulation geometry}

\begin{cfigure1c} [h]
\begin{tabular}{cc}
\includegraphics[width=0.5\textwidth, trim=1 1 1 1, clip]{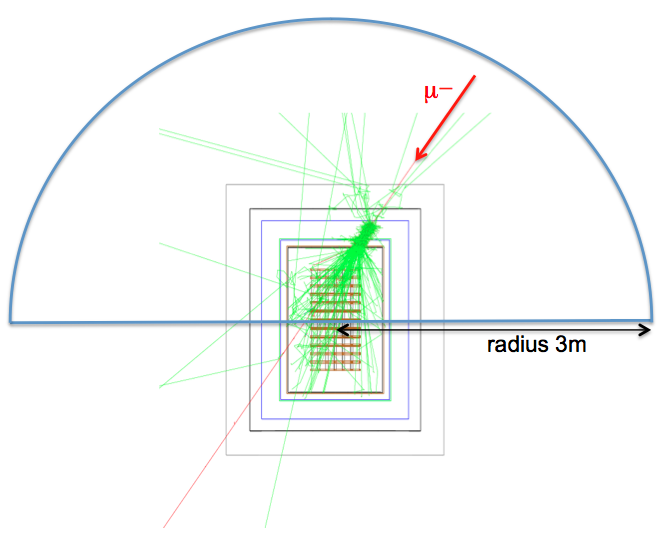} &
\includegraphics[width=0.4\textwidth, trim=1 1 1 1, clip]{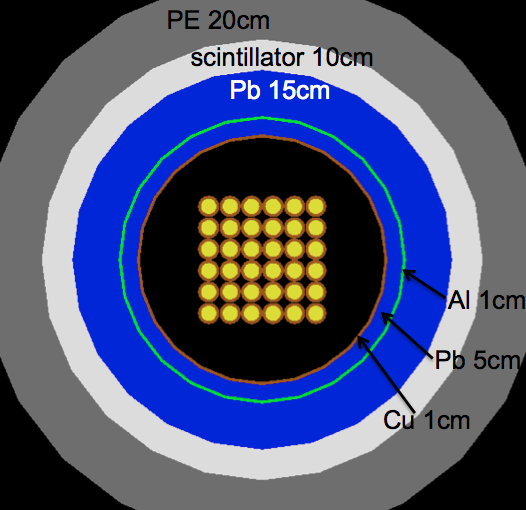} \\(a) & (b) \\
\end{tabular}
\caption[fig:sim-5]{
(a) Schematic layout of simulation geometry.
(b) Shielding configuration
}
\label{fig:sim-5}
\end{cfigure1c}

Muons were generated in a manner which produced all muons that pass within 1.5~m of the detector center, selected
randomly for position and angle from the distribution of $\frac{dN}{d\Omega\,dA}$ proportional to $\cos^{2}\theta$,
as shown in Figure~\ref{fig:sim-5}(a). The procedure insured that all muons were generated with vertices at least
3~m from the detector center.   Figure~\ref{fig:sim-4} shows the energy spectrum of generated muons at the hemisphere.
The muon flux at Yangyang is taken to be $\rm 2.7\times 10^{-7}~muons/cm^{2}/s$ 
or $\rm 3.7\times 10^{7}~muons/(3~m\times3~m)/(48.3~years)$.

For the conceptual shielding configuration for AMoRE-II, we modeled 432~\camo crystals, each with a diameter of
5~cm, a height of 5~cm, and a mass of 426~g. The total mass of crystals is 184~kg. The target crystals
are located inside a 1~cm copper shield which is surrounded by lead and polyethylene~(PE) shields as shown in
Fig.~\ref{fig:sim-5}(b). Plastic scintillators with 10 cm thickness are located outside of the lead shield.

\subsection{Muon and muon-induced neutron backgrounds}
In the AMoRE-II conceptual shielding configuration, muon and muon-induced neutron single-crystal background rates
are found to be $\rm \sim$1.1$\times 10^{-4}~{\rm counts/keV/kg/yr}$, after applying a veto cut for events
that are coincident with a hit in a muon scintillation counter.  The veto efficiency for these events was taken to be 90\%.

\section{Other backgrounds}
\begin{itemize}
   \item  Other sources of background (cosmogenic \Y[88], residual \Ca[48] in the \CMO crystals, and \Bi[214] 
in the copper) are not expected to contribute significantly to the background near the signal ROI. Nevertheless,
they will also be considered in the future.
   \item Vikuiti reflector \\
 We also simulated the effects of \U and \Th contaminants and their daughters, in reflecting foils that surround
the \CMO crystals. An intitial candidate for this part is 3M Vikuiti.  Our HPGe counting of this material presently sets limits of about
 $< 1.6 $~mBq/kg in the \U chain, $< 0.9 $~mBg/kg for the \Th chain, and $< 15 $~mBq/kg for \K[40].  Preliminary simulation results are reported in Table~\ref{tab:5-6}.  These results are very recent and potassium has not yet been included in the analysis.

\begin{table}[!htb]
\begin{center}
\caption{Simulated background rates from Vikuiti reflector~(counts/keV/kg/yr) using measured activity limits. These results are very recent and potassium has not yet been included in the simulation analysis. \Th includes $^{228}$Ra and $^{228}$Ac with assumption of sub-chain equilibrium. We assume complete chain equillibrium for the \U chain.
}
\label{tab:5-6}
\begin{tabular}{c|c|c|c|c} \hline
\multirow{5}{*}{Background source} &   & \multicolumn{3}{ c  }{Backgrounds in ROI} \\ 
&  & \multicolumn{3}{ c  }{[$10^{-3}$ cnt/keV/kg/yr]} \\ \cline{3-5}
&  Activities &   $\alpha$ events & $\beta$ events & after applying \\
&  &   &  & cuts for events \\ \hline \hline
Vikuiti reflector	
& \U[238]:  $ 1.6 $~mBq/kg & 7.4 & - & 0.007  \\
& \Ra[226]:  $ 1.6 $~mBq/kg & 10 & 1.3 & 0.28  \\
&  \Th: $ 0.74 $~mBg/kg &  2.2 & - & 0.002  \\ 
&  \Th[228]: $ 0.90 $~mBq/kg &  4.7 & 0.77 & 0.30  \\  
& \K[40]:  $ 17$ ~mBq/kg & - & - & - \\ \hline
Total & & 24 & 2.1 & 0.58 \\ \hline
\end{tabular}
\end{center}
\end{table}
\end{itemize}

\section{Summary of Background Estimation}
Table~\ref{tab:major_bkgds} summarizes the major backgrounds expected in the AMoRE-I experiment.
\U and \Th results are listed for major parts.  Estimates of some \K concentrations are also listed. 
\K can contribute to backgrounds only through accidental pile-up, which is still being studied.

\begin{table}[!htb]
.
\begin{center}
\caption{Summary of backgrounds in major components of AMoRE-I, estimated with measurements and simulation.
Limits are 95\% C.L. Perfect alpha rejection is assumed (see Sec.~\ref{Sec:Cryo:Simul}).  Here,
items that have simulated times that are still less than a year and have not produced any counts in the
ROI have not been translated into limits.}
\label{tab:major_bkgds}
\newcolumntype{M}[1]{>{\centering}m{#1}}
\begin{tabular}{M{2.5cm}|M{1.5cm}|M{2.1cm}|M{1.5cm}|M{1.5cm}|M{1.4cm}|M{1.5cm}} \hline
& & & & \multicolumn{3}{M{4.5cm}}{Backgrounds in ROI ($3.03 \pm 0.01$~MeV) [$10^{-3}$ cnt/keV/kg/yr]} \tabularnewline \cline{5-7} 
&  Isotope & Activity [mBq/kg] & Simulated Time [years] & $\alpha$ events & $\beta$-like events & after applying cuts for events \tabularnewline \hline \hline
\multirow{4}{2.5cm}{\centering Internal \CMO} & \Pb[210] & 10  & 31 & 23.0 & - & - \tabularnewline 
& \U & 0.1  & 335 & 1.7 & 0.3 & 0.3	\tabularnewline
&  \Th & 0.05  & 698 &0.5 & 27.3 & 1.5	\tabularnewline 
& \K & 1  & 232 & $<$0.056 & $<$0.056 & $<$0.056	\tabularnewline 
& \U[235] & 1  & 32 &12.0 & $<$0.41 & $<$0.41	\tabularnewline \hline
Random coincidence of two \twonubb	& & & 335 & $<$0.039 & 0.12 & 0.12 \tabularnewline \hline
\multirow{2}{2.5cm}{\centering \CMO copper frame}	& \U &  0.330 & 96 &0.63 & 0.45 & $\sim$ 0. \tabularnewline
&  \Th & 0.207 & 207 & 0.46 & 1.43 & $\sim$ 0.7  \tabularnewline  \hline
\multirow{3}{2.5cm}{\centering G10 fiberglass support tubes} &\U & 240\,000 & 0.1 & - & - & - \tabularnewline&  
\Th & 50\,000  & 0.03 & - & - & -  \tabularnewline  
&  \K & 1.2~  & 0.4 & - & - &  -  \tabularnewline  \hline
\multirow{3}{2.5cm}{\centering SC lead shield} & \Pb[210] & 300 & 1.0 & 92.8 & - & -  \tabularnewline 
& \U & .012 & 4425 &0.007 & 0.004 & $\sim$ 0.001  \tabularnewline
&  \Th & .0041 & 17884 &0.003 & 0.002 & $\sim$ 0.001  \tabularnewline  \hline
\multirow{3}{2.5cm}{\centering Inner lead plate} & \Pb[210] & 300 & 0.09 & - & - & -  \tabularnewline 
& \U & .012 &  10 &  $<$1.3 & $<$1.3 & $<$1.3  \tabularnewline
&  \Th & .0041 & 10 &  $<$1.3 & $<$1.3 & $<$1.3  \tabularnewline  \hline
\multirow{3}{2.5cm}{\centering Outer lead shield} & \Pb[210] & 59\,000 & 1.25 hr & - & - & -  \tabularnewline 

& \U & .085 & 0.28 &  - &  - & -  \tabularnewline
&  \Th & .015 & 0.53 & - &  - &  -  \tabularnewline  \hline
$\gamma$ from rock in random coincidence with \twonubb  & \Th & 39\,000  & 1.2 hr & - &  0.10 & 0.10 \tabularnewline \hline
Muons and muon-induced neutrons & & & 48 & - & 1.1 & 0.11  \tabularnewline \hline
Total & & & & 133.8 & 33.9 & 5.9 \tabularnewline \hline
\end{tabular}
\end{center}
\end{table}

%% file: tex/enriched_materials.tex
\chapter{Supply, purification and recovery of enriched materials}

\section{$\rm {}^{100}Mo$}
Molybdenum isotopes are separated by centrifuge techniques, which are characterized by high
productivity and reasonable cost. This technique can work for chemical elements that form volatile
compounds at room temperature. Fortuitously, Molybdenum forms the volatile MoF$_6$ compound at room
temperature. There are several facilities at the world that have the capability for the separation
of large quantities of Molybdenum: JSC Production Association Electrochemical plant (Zelenogorsk),
JSC Ural Electochemical plant (Novouralsk) and JSC Siberian Chemical Plant (Seversk) in Russia,
and URENCO (Almelo) in the Netherlands. These plants have dealt with the enrichment of Uranium and
also with enrichment of stable isotopes of some other chemical elements.

An 8.25~kg quantity of \mohundred (enriched to 96\%) was produced by the JSC Production Association
Electrochemical Plant (ECP) in Zelenogorsk, Russia, by the gas centrifugation technique. This enriched
material, supplied in the form of $\mathrm{^{100}MoO_3}$, is very pure with respect to radioactive
contaminants: the results of ICP-MS measurements show that the concentrations of \U and \Th in the
$\mathrm{^{100}MoO_3}$ powder do not exceed 0.07~\ppb and 0.1 \ppb,~respectively.

The general scheme for \mohundred production at the ECP is as follows:
\begin{centering}
  \begin{itemize}
  \item \textbf{Fluoridation:}\\
    $\mathrm{^{nat}Mo + 3*F_2 \to ^{nat}MoF_6}$\\
(gaseous impurities: $\mathrm{^{235,238}UF_6 + ^{nat}WF_6 + ^{nat}GeF_4}$ + ...)\\
    Purification of: $\mathrm{ThF_4, RaF_2, PaF_4, AcF_3, KF}$ - no volatile high-pressure fluorides at RT).
    \item \textbf{Centrifugation:}\\
1st stage: $\mathrm{^{nat}MoF_6 \to  ^{98-94}MoF_6 + (^{235,238}UF_6 + ^{nat}WF_6 +\mathrm{^{100}MoF_6)}}$\\
2d stage: $\mathrm{^{100}MoF_6} + (\mathrm{^{100}MoF_6} + \mathrm{^{235,238}UF_6 + ^{nat}WF_6})$.
\item \textbf{Conversion (``wet'' chemistry):}\\
$\mathrm{^{100}MoF_6 + H_2O + HNO_3 \to (evaporation~\&~calcination)~\to~^{100}MoO_3}$\\
$\mathrm{^{100}MoO_3 + H_2O+ NH_3 \to (evaporation~\&~calcination)~\to~^{100}MoO_3}$.
  \end{itemize}
\end{centering}

The \mohundred production capacity at ECP is, at present, $\sim$28~kg/yr, in form of
$\mathrm{^{100}MoO_3}$ with an isotopic enrichment IE~$\geq$~95\%. At this rate, it would take about
3.5~years to produce 100 kg of enriched \mohundred. It is not clear at this time if there is another
plant capable of enriching \mohundred at this rate.  According to our underground background measurements at
Y2L, the quality of the \moothree powder (from the point of view of chemical purity and \U and \Th contamination)
is sufficient for $\mathrm{CaMoO_4}$ crystal growing.

The cost to the AMoRE Collaboration is  90$\sim$100~\$/g for 100~kg quantities (mass of Mo) with an enrichment
to 95\%. There is strong ``cost competition'' between the application of Mo isotopes for Nuclear Medicine (\Mo[98] and
\mohundred for production of \Mo[99] /\Tc[99m])

and Underground Physics. This situation presents a risk to AMoRE (the cost of \Mo[98] for Nuclear Medicine is at
the level of 300~\$/g).  The $\mathrm{MoF_6}$ working gas for centrifugation is extremely corrosive. If a 
centrifuge cascade is devoted to Mo isotope enrichment, this production should continue ``forever'', i.e., until
the time of the cascade's decomissioning.

\section{$\rm {}^{48depl}Ca$}

Calcium isotopes can be separated by electromagnetic methods that are characterized by very low productivity and
high cost. There are currently two sites in the world that possess large-scale facilities for this purpose:
the Y-12 National Security Complex (Calutrons in Oak Ridge, DOE, USA) and FSUE Electrochimpribor (Lesnoy, Russia).
At present, the Calutrons in Oak Ridge are closed. The SU-20 facility at FSUE Electrochimpribor (EKP) is in operation
and is still producing $\mathrm{^{48}Ca}$ (and $\rm ^{48depl}Ca$). In a search for existing $\mathrm{^{48}Ca}$-depleted material,
we found that EKP has a sizable amount of Ca material from which the $\mathrm{^{48}Ca}$ component was depleted by as
much as a factor of 187 compared to that  for natural calcium. The production of $\mathrm{^{48}Ca}$-depleted (or,
equivalently, \Ca[40]-enriched) material by means of industrial electromagnetic separators, has been going on
continuously in the USSR/Russia to provide $\mathrm{^{48}Ca}$-, \Ca[44]- and \Ca[46]-enriched isotopes for medical
and  physics applications since the 1950s. The \Ca[48]-depleted ``left-over'' material from these projects,
which is currently stored in the form of $\mathrm{^{48depl}CaCO_3}$, would be suitable for our needs. We are currently
making an inventory of all the world's quantities of Ca solutions that are depleted on $\mathrm{^{48}Ca}$ at the level
of less than 0.005\%.

The first batches (of 4.5~kg in total) of the calcium carbonate depleted in the $\mathrm{^{48}Ca}$ isotope ($\leq 0.001\%$)
that were produced by EKP have been characterized with mass-spectrometry (ICP-MS) and HPGe detectors. The concentrations of
\U and \Th in the powder measured by ICP-MS are below 0.2~\ppb and 0.8~\ppb, respectively. However, a $\gamma$-spectroscopic
analysis of the $\mathrm{^{48depl}CaCO_3}$ powder revealed some specific activity due to $\mathrm{^{226}Ra}$ and its progenies at
the level of a few hundred mBq/kg. Also, the content of Sr and Ba elements, which belong to the same sub-group of chemical
elements as Ra, were quite high (Sr = 25 \ppm and Ba = 26 \ppm), and this also indicates the presence of some Ra impurity.
Because of this, the $\mathrm{^{48depl}CaCO_3}$ material was subjected to additional purification.

At present, there is 32~kg of $\rm ^{48depl}Ca$ available at EKP, in Lesnoy, Russia. It has less than 0.001\% of
\Ca[48]. AMoRE-II needs about 40~kg of $\rm ^{48depl}Ca$. The small shortage could be covered by the
anticipated production of \Ca[48]-depleted Ca in the next few years.  

\section{Purification of calcium and molybdenum oxides  and recovery of enriched materials after crystal production }
The ultimate sensitivity of AMoRE is limited by internal \CMO backgrounds that are due to impurities in the form of natural
radioactive isotopes, such as \Tl[208] from the \Th chain and \Bi[214] among the $\mathrm{^{226}Ra}$ decay products in the
\U chain ~\cite{Khanbekov}. Therefore, the removal of uranium, thorium and radium insures the removal of other isotopes in
these chains and reduces internal backgrounds in the \CMO detectors. One of our approaches has been to utilize chemical
purification methods to remove impurities from the raw materials used to grow the scintillation crystals. 

In the process of making or cutting crystals, scraps (or crystal wastes) of $\mathrm{^{48depl}Ca^{100}MoO_4}$ crystals
are unavoidable. To reduce the project cost one should extract  separately materials containing calcium and those
containing molybdenum from those scraps. Subsequent mixtures of these two materials can be used to make
$\mathrm{^{48depl}Ca^{100}MoO_4}$ crystals. We have recently developed technologies to recover and purify calcium and
molybdenum from crystal wastes.

In following section, we present the deep purification methods we use for $\mathrm{^{48depl}CaCO_3}$ and $\mathrm{^{100}MoO_3}$
and the methods for their recovery from the crystal wastes.

\section{Deep purification of \caco powder}
The \caco purification process uses an inorganic sorbent MDM that is based on manganese dioxide ($\mathrm{MnO_2}$). A
chromatographic column used for this purpose is made of stainless steel. The major steps in this purification process
are discussed in the following.

\begin{figure}\centering
\includegraphics[width=\textwidth]{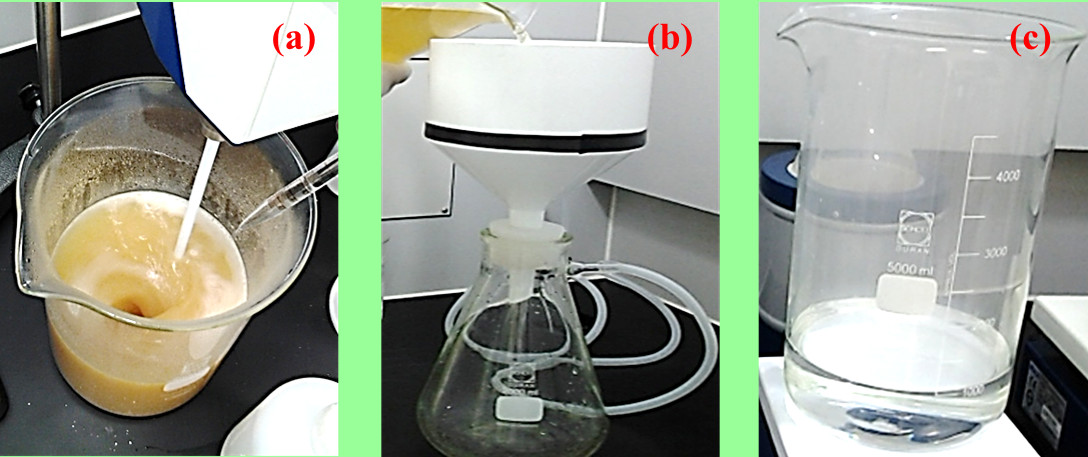}
\caption{(a) Dissolution of \caco (b) Vacuum filtration (c) Final filtrate.}
\label{fig:fig6-1}
\end{figure}

About 500 grams of calcium carbonate powder are placed in a glass beaker and mixed with a calculated amount of dilute
nitric acid (65\%) until all of the \caco is completely dissolved. The chemical reaction for the dissolution of calcium
carbonate in nitric acid is
\begin{equation}\rm
CaCO_3 + HNO_3   \xrightarrow{pH 1.5}  Ca(NO_3)_2 + H_2O + CO_2
\end{equation}
\noindent
After mixing, the solution has a light brown color due to presence of insoluble impurities as shown in
Fig.~\ref{fig:fig6-1}(a). Subsequent filtering removes these impurities as shown in Fig.~\ref{fig:fig6-1}(b). The resulting
solution after filtration is transparent and has a pH=1.5. Figure~\ref{fig:fig6-1}(c) shows the final filtrate obtained
after vacuum filtration.

\begin{figure}\centering
\includegraphics[width=0.85\textwidth]{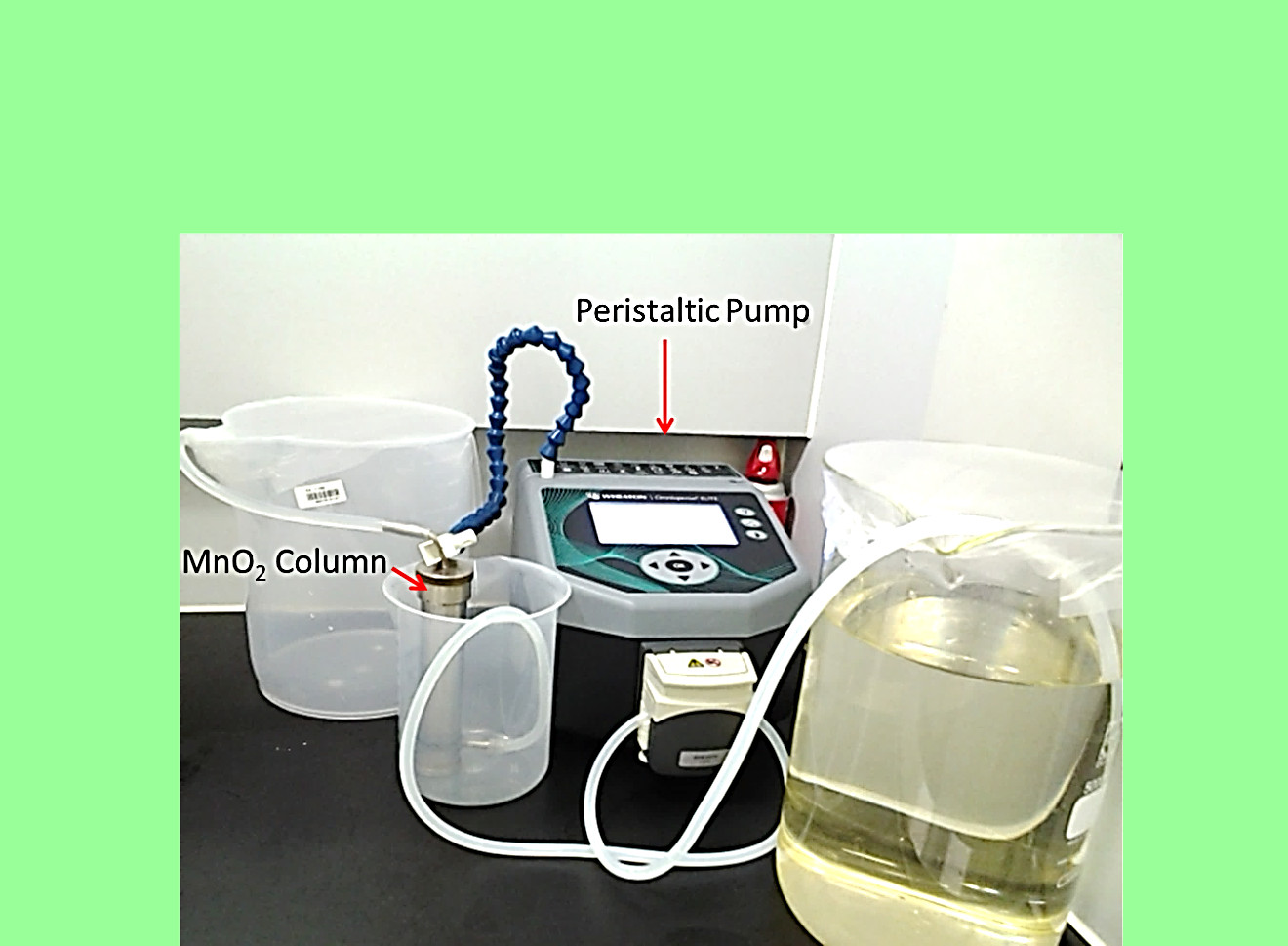}
\caption{Photographic view of purification of \caco using MnO$_2$ resin.}
\label{fig:fig6-2}
\end{figure}

\begin{figure}
\includegraphics[width=0.85\textwidth]{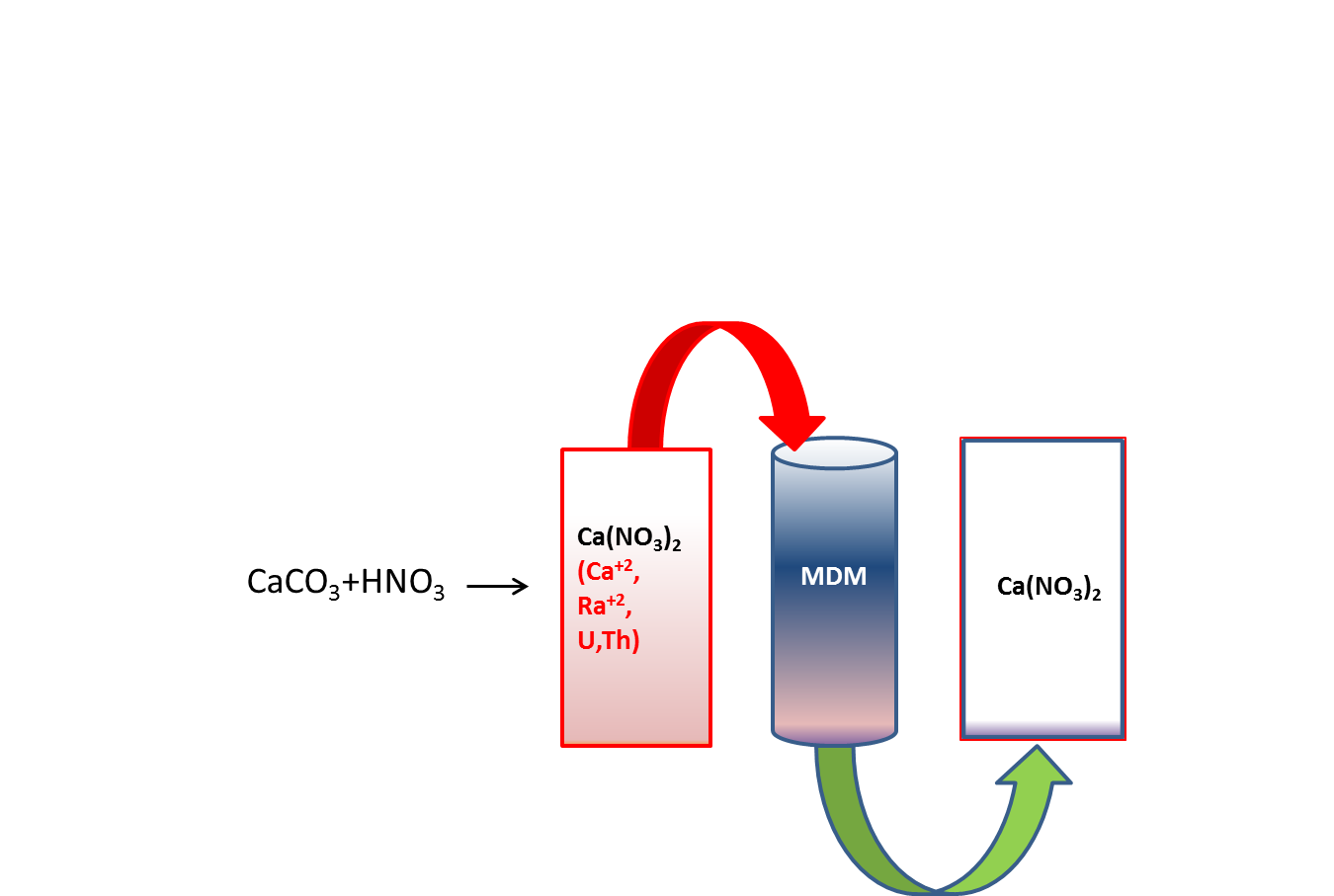}
\caption{Schematic of purification of \caco by $\mathrm{MnO_2}$.}
\label{fig:fig6-3}
\end{figure}

The filtrate is further purified by passing it through a sorption column filled with a granular MDM sorbent. The volume
of the sorbent inside the column is 60~$\rm cm^{3}$. The feed solution is forced through the column by a peristaltic pump
with a speed of 120~ml/h, as shown in Fig.~\ref{fig:fig6-2}. Filtrates after the column purification are collected and
analyzed for concentrations of iron, uranium, thorium, strontium and barium by an ICP-MS system. Figure~\ref{fig:fig6-3}
shows a schematic drawing of the ion-exchange process using an $\mathrm{MnO_2}$ resin.

After purification with the MDM sorbent, the solution is neutralized with aqueous ammonia to pH=9.0 and 50~g of
ammonium carbonate are added to precipitate the remaining quantities of impurities in the calcium carbonate. The
reaction of calcium nitrate solution with ammonium carbonate and aqueous ammonia precipitates 5\% of \caco:
\[ \rm
Ca(NO_3)_2+NH_4CO_3(5\%)+HN_4OH  \xrightarrow{pH 9} CaCO_3(Th,U,Ba..)+NH_4NO_3+CO_2
\]
The solution is filtered to collect the impure precipitate of \caco on filter paper.  
  
\begin{figure}
\includegraphics[width=\textwidth]{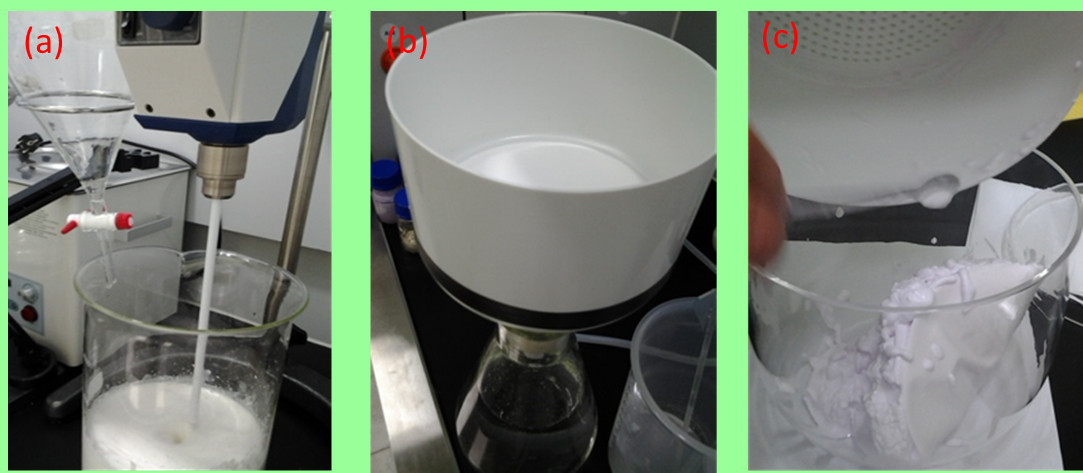}
\caption{(a) Precipitation. (b) Vacuum filtration. (c) \caco pure precipitate.}
\label{fig:fig6-4}
\end{figure}

The remaining part of ammonium carbonate (680~g) is added to the filtrate solution until all of the calcium is
precipitated. The reaction of the calcium nitrate solution with ammonium carbonate and aqueous ammonia to
precipitate 95\% of \caco is:
\[ \rm
Ca(NO_3)_2+NH_4CO_3(95\%)+HN_4OH \xrightarrow{pH 9}  CaCO_3(Pure)+NH_4NO_3+CO_2
\]
The resulting precipitate of calcium carbonate is filtered off, washed on filter paper with distilled water and
collected in a quartz crucible. Figures~\ref{fig:fig6-4}(a),~(b) and (c) show the precipitation of \caco with
subsequent filtration and collection of precipitates, respectively. 

\begin{figure}\centering
\includegraphics[width=0.9\textwidth]{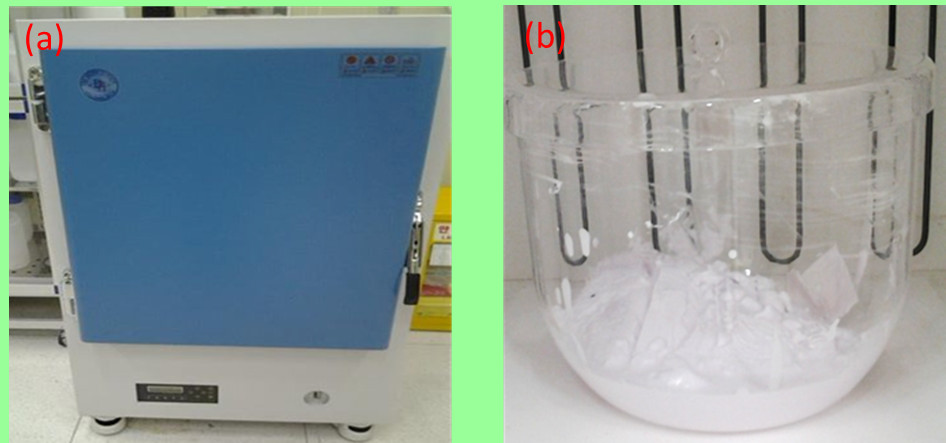}
\caption{(a) Muffle Furnace. (b) \caco calcination in quartz crucible.}
\label{fig:fig6-5}
\end{figure}

The crucible containing the \caco precipitate is placed in an electric muffle furnace and heated initially to
100$^{o}$~C for 2~hours to remove water, and then at 450$^{o}$~C for 2~hours for the final calcination of the material.
Figure~\ref{fig:fig6-5} shows the calcination of \caco in the muffle furnace. The final weight of purified \caco is
typically $\sim$440~g, which is $\sim$88\% of the initial weight. This calcium carbonate sample is analyzed for
impurities.  Table~\ref{tab:6-1} shows the ICP-MS analysis results for an initial and final calcium nitrate solution
before and after being passed through the $\mathrm{MnO_2}$-based ion exchange resin.

To quantify the level of purification, we use the decontamination factor, DF, defined as the ratio of initial-to-final
specific radioactivity resulting from the chemical separation process. It numerically quantifies the efficiency with
which impurities are removed from a chemical substance. In our case, this is the factor by which the radioactive
contamination level is reduced. Taking $\mathrm{C_{in}}$ and $\mathrm{C_f}$ as the concentration of impurities in the
initial and final materials, the decontamination factor is $\mathrm{DF=C_{in}/C_f}$. A purification process that
removes impurities will have a decontamination factor greater than one ~\cite{Kumar}.

Table~\ref{tab:6-2} lists measured decontamination factors: iron impurities are decreased by a factor of 100, strontium
by factors of 10 to 15, barium by 1,000, thorium by between 100 to 10,000, and uranium by 100 to 1,000. From the results
for barium, which is the closest family element to radium on the periodic table, it can be inferred that the decontamination
factor for radium is also quite high (at least 2--3 orders of magnitude).

\begin{table}
\caption{ICP-MS results of the analysis of the filtrates after sorbent MDM.}
\input{tables/table6-1.tex}
\label{tab:6-1}
\end{table}
\begin{table}
\caption{Purification results for the MDM sorbent (DF is the decontamination factor).}
\input{tables/table6-2.tex}
\label{tab:6-2}
\end{table}

Table~\ref{tab:6-3} shows results from a ICP-MS analysis of the purified \caco powder. The iron content in the purified
product was reduced by approximately an order of magnitude, and the uranium and thorium content by 2 orders of magnitude.
No strontium and barium purification occurred.  The purification results for thorium, strontium, and barium are entirely
inconsistent with the results shown in Table~\ref{tab:6-2}. This may be due to the insufficient purity of the reagents
used, especially ammonium carbonate.

\begin{table}
\caption{The concentration of impurities in the initial and purified calcium carbonate and decontamination factors (DF).}
\label{tab:6-3}
\input{tables/table6-3.tex}
\end{table}
Finally, the radioactive contamination of the initial (unpurified) \caco powder and of the \caco powder that was processed
through the $\mathrm{MnO_2}$ ion exchange resin were studied with a HPGe detector for background measurements at the Yangyang
underground laboratory. From the HPGe results (Table~\ref{tab:6-4}), it can be seen that decontamination factors as high as
300 can be achieved for $\mathrm{^{226}Ra}$; other radioactive elements are reduced by factors of 10 to 100.

\begin{table}
\caption{HPGe results. ``Before'' (``After'') stands for before (after) purification of \caco powder.}
\label{tab:6-4}
\input{tables/table6-4.tex}
\end{table}

\section{Deep purification of  \moothree  powder}
Molybdenum oxide (\moothree) powder is purified in the form of an ammonium molybdate solution, which
is prepared by dissolving \moothree powder in aqueous ammonia. In this section, we will describe our study
of purification methods for \moothree powder, and R\&D of MoO$_3$ purification methods that is in progress.

\begin{figure}
\includegraphics[width=\textwidth]{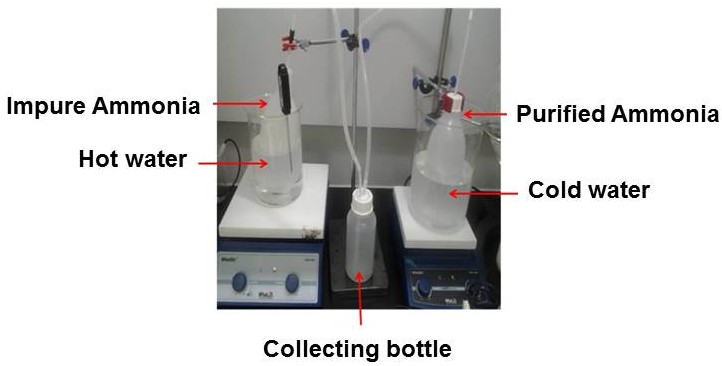}
\caption{The procedure for purification of aqueous ammonia.}
\label{fig:fig6-5-6}
\end{figure}

An ammonium molybdate solution can be prepared by mixing the MoO$_3$ powder with purified ammonia in an aqueous media
with pH=7. The first step is the purification of the ammonia.  About 500~ml of aqueous ammonia is placed in a plastic bottle
that is connected via a Teflon tube to a collection bottle. The collection bottle is, in turn, connected to a bottle that is
filled with 500~ml of deionized (DI) water. Impure ammonia in the first bottle is heated in a water bath and the water
droplets emerging from the bottle accumulate in the collection bottle. Then the third bottle, which is placed in a cold water
bath and filled with DI water, is used to collect the pure ammonia fumes. The reaction stops when temperature in the water
reaches 80$^{o}$C. This procedure is shown in Figure \ref{fig:fig6-5-6}.

In order to prepare ammonium molybdate solution, about 200 g of \moothree powder is placed in a plastic beaker and
dissolved in 50~ml of DI water to avoid dust formation with the ammonia. Then, with the pH maintained at 7, the purified
ammonia is added until all of the \moothree powder dissolves completely. The principal reactions for this procedure are as follows.

\begin{raggedright}
  At pH $\sim$ 7:\\
\indent $\rm MoO_3+NH_3+H_2O  \to   (NH_4)_6Mo_7O_{24} (solid)$ \\
At pH above 7 (8 or 9):\\
\indent $\rm (NH_4)_6Mo_7O_{24} +NH_3+H_2O   \to   (NH_4)_2MoO_4$\\
\indent $\rm Th^{+4}, Fe^{+3}, Pb^{+2}  \to   Th(OH)_4, Fe(OH)_3, Pb(OH)_2$.
\end{raggedright}
\vspace{0.1in}
The purification of the \moothree powder starts with the solution of $\rm (NH_4)_2MoO_4$. We have studied the following
purification methods:
\begin{itemize}
\item Recrystallization
\item Co-precipitation
\item Sedimentation
\item Column chromatography.
\end{itemize}

\begin{figure}\centering
\includegraphics[width=\textwidth]{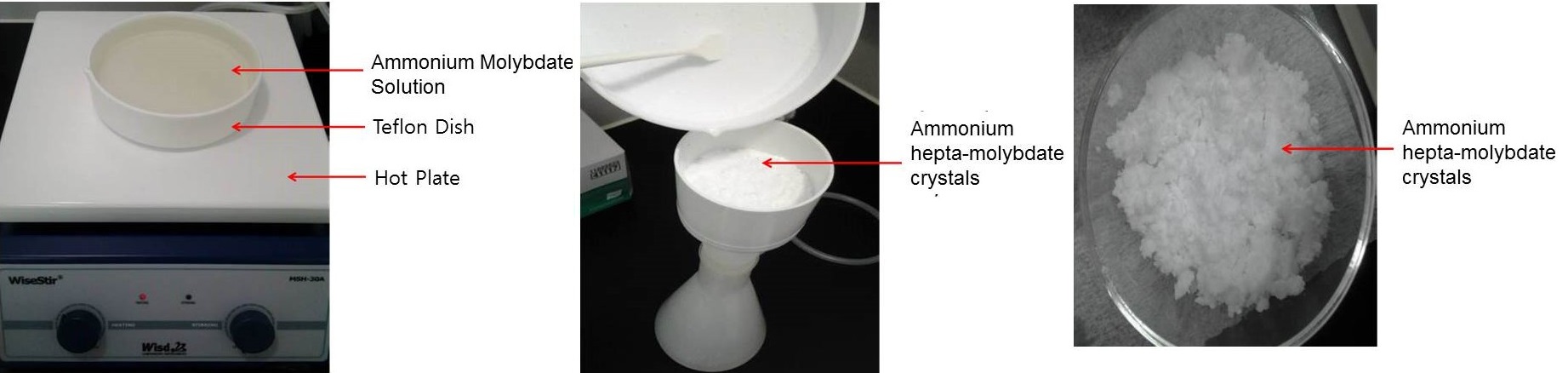}
\caption{The recrystallization method procedure.}
  \label{fig:fig6-5-7}
\end{figure}

In the recrystallization method, we have the highest purity product, but a low yield. The purified molybdenum is
in the form of ammonium hepta-molybdate $\mathrm{(NH_4)_6Mo_7O_{24}.4H_2O}$ in aqueous solution. In this process,
ammonium molybdate is continuously heated in a teflon evaporation dish to remove the aqueous media and purified
crystals are collected on filter paper (see Fig.~\ref{fig:fig6-5-7}). In this method, the chemical reaction is:
\[ \rm
(NH_4)_2MoO_4   \to   8NH_3\uparrow    + (NH_4)_6Mo_7O_{24}.4H_2O (Crystals).
\]

\begin{figure}
\includegraphics[width=\textwidth]{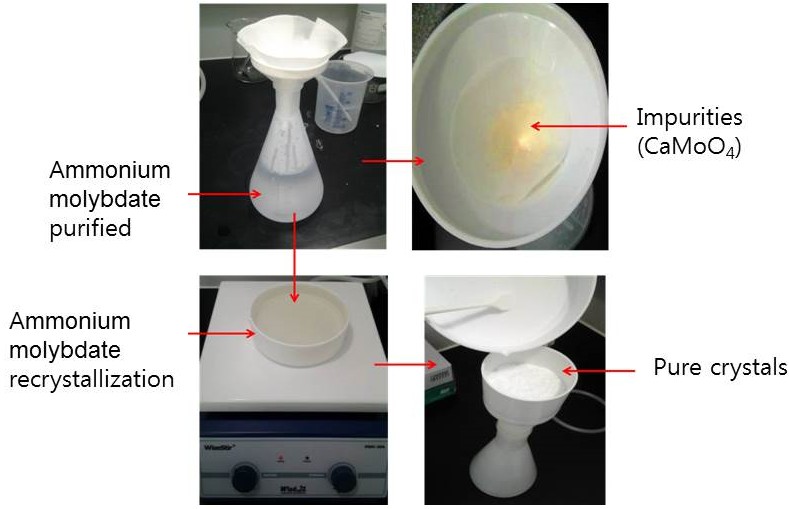}
\caption{The co-precipitation method procedure.}
  \label{fig:fig6-5-8}
\end{figure}

Co-precipitation is an impurity selective method. Impurities of radium are separated in the form of calcium molybdate
using calcium chloride ($\rm CaCl_2$) in aqueous solution. In this method, we have to prepare pure calcium chloride using
the reaction:
\[\rm CaCO_3 + HCl  \to   CaCl_2  +  H_2O  +  CO_2. \]
The main reaction in this method is 
\[\rm (NH_4)2MoO_4+ CaCl_2   \to   CaMoO_4  +  2NH_4Cl. \]
\noindent
About 1\% of the calcium molybdate crystals are collected on filter paper, while the filtrate solution contains purified
ammonium molybdate. The procedure for the co-precipitation method is presented in Fig.~\ref{fig:fig6-5-8}.

\begin{figure}\centering
 \includegraphics[width=\textwidth]{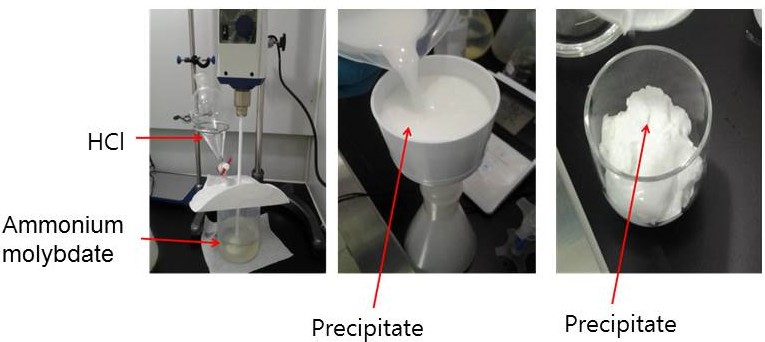}
\caption{The sedimentation method procedure.}
  \label{fig:fig6-5-9}
\end{figure}

In the sedimentation process we have a comparatively high product yield.  Molybdate is collected as ammonium
tetramolybdate $\mathrm{(NH_4)_2 \times Mo_4O_{13}}$ from an aqueous solution of ammonium molybdate. 
The theoretical yield for this process is  96\% to 97\%.  The procedure is (see Fig.~\ref{fig:fig6-5-9})
\begin{itemize}
\item  200~ml of ammonium molybdate is placed in a plastic beaker; 
\item  100~ml of DI water is mixed with 100~ml of HCl;
\item  this solution is added to ammonium molybdate solution drop wise until the pH of the solution reaches 1; 
\item  the precipitates are filtered off, collected on filter paper and dried.
\end{itemize}
The principal reactions for this method are:
\[\rm
4(NH_4)_6Mo_7O_{24}   +  6HCl  \to  6NH_4Cl  +  (NH_4)_2Mo_4O_{13}\\
(NH_4)_2Mo_4O_{13}     +     HCl  \to  MoO_3     +    H_2O+NH_4Cl.
\]

\begin{figure}\centering
 \includegraphics[height=\textwidth,angle=270]{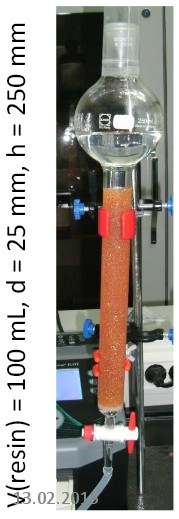}
\caption{The column chromatography method. The resin is inside the cylinder.}
  \label{fig:fig6-5-10}
\end{figure}

The column-chromatography method is based on the different behavior of components in their mixtures with two phases,
one stationary and the other mobile. Physical or chemical absorption takes place on the phase boundary. An initial
ammonium molybdate solution has anions, $\mathrm{Mo_8O_{24}}^{4-}$, $\mathrm{Mo_8O_{24}}^{4-}$ and $\mathrm{Mo_8O_4}^{2-}$,
and cations, $\mathrm{NH^{4+}}$, K, $\mathrm{Pb^{2+}}$, $\mathrm{(Th)^{n+}}$, $\mathrm{(U)^{n+}}$, etc. We use the
Amberlite IR-120 H-form cation exchange resin. A purified ammonium molybdate solution is produced. The column containing
the resin is shown in Fig.~\ref{fig:fig6-5-10}. In this study, the diameter and height of column are 25 mm and 250 mm,
respectively.

\section{Calcium and molybdenum recovery from \camo crystals}

In the procedure to recover calcium and molybdenum from $\mathrm{^{48depl}Ca^{100}MoO_4}$ crystals, we first have to make a
solution by means of a strong acid, either HCl or $\rm HNO_3$.

\subsection{Decomposition of \CMO crystals with 65\% $\mathrm{HNO_3}$}
The \CMO crystal is first crushed into small pieces using a pestle and mortar. Then the crushed \CMO is milled into a fine powder
using a pulverizing milling machine. After this, milled powder is sieved to select diameters between 0.10~mm and 0.25~mm using a
non-metallic sieve.

In the second step, about 300~g of the \CMO powder is placed inside a plastic bottle with 400~ml of 65\% $\rm HNO_3$.
The bottle is kept in a water bath and heated up to $\rm 70^{\circ} C- 80^\circ C$.
This mixture is constantly stirred with a high-speed mechanical stirrer for 6~hours. During this procedure a thick,
yellow-colored molybdic acid is formed via the reaction
\[\rm CaMoO_4   +   2HNO_3 \to    H2MoO_4 (solid) + Ca (NO_3)_2 (liq). \]
In this reaction, all of the ammonium molybdate is in the molybdic acid form and calcium nitrate is in the solution form.
In this way, we recover ammonium molybdate and calcium carbonate separately. The above solution is filtered under vacuum.
The filtered cake is washed with a dilute solution of nitric acid. After filtration we obtain molybdic acid as a residue
on the filter paper and calcium nitrate as the filtrate.

\begin{figure}\centering
\includegraphics[width=0.3\textwidth]{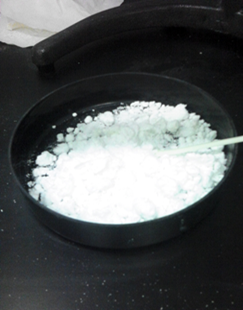}
\caption{Ammonium molybdate obtained from the recovery of \CMO crystals.}
\label{fig:fig6-3-11}
\end{figure}

In the third step, the molybdic acid is dissolved in a 25\% aqueous ammonium hydroxide solution until a transparent
solution is obtained.  The solution is then filtered under vacuum. The undissolved \CMO remains as a residue on filter
paper and is collected in a quartz crucible and the filtrate solution is poured into a clean beaker. The solution in
the beaker is neutralized with nitric acid up to pH=1.65--1.70. While this happens, a precipitate of fine 
polyammonium molybdate (PAM) crystals is formed. This precipitate solution is filtered under vacuum and the
crystals are washed with a dilute ammonium nitrate solution with pH=1.7. The PAM residue
on the filter paper is collected in a clean beaker and the filtrate solution is discarded. The PAM is again dissolved
in aqueous ammonia until a transparent solution is obtained. Some of this solution is evaporated, leaving crystals of
ammonium molybdate (AM).
These crystals are collected and the remaining solution is evaporated, producing more crystals (see Fig.~\ref{fig:fig6-3-11}).

\begin{figure}\centering
\includegraphics[width=\textwidth]{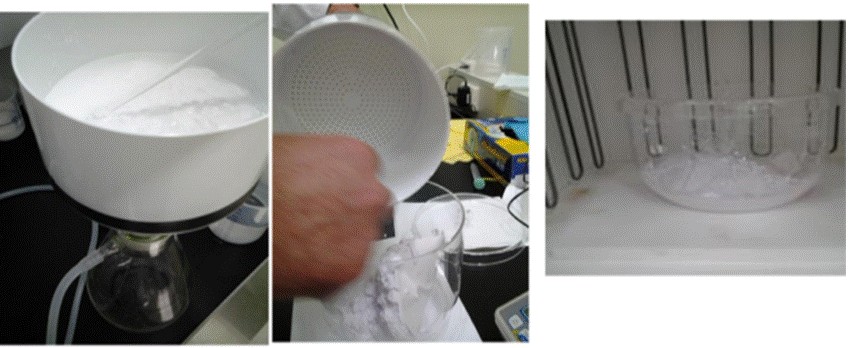}
\caption{Pure \caco on filter paper (left), the purified \caco is collected in the quartz crucible (middle) and
calcinated \caco in a furnace (right).}
  \label{fig:fig6-6-12}
\end{figure}

In the fourth step, the calcium nitrate solution is evaporated to remove excess nitric acid. After evaporation a wet salt
of calcium nitrate is obtained and the obtained salt is dissolved in water until a transparent solution is obtained. This
solution is neutralized with aqueous ammonia to pH=8. During this neutralization, a white precipitate of \camo is formed.
The precipitated solution is filtered under vacuum and the residue, which is collected on filter paper, is undissolved
\camo that is collected in a quartz crucible and the filtrate is collected in clean beaker. In the filtrate solution,
enough ammonium carbonate is added so that 5\% of the calcium carbonate precipitates out. 
During this procedure, aqueous ammonia is added to maintain pH=9. This solution is constantly stirred with a
mechanical stirrer. The above precipitate solution is filtered under vacuum. The precipitate on the filter paper is
calcium carbonate that contains a lot of impurities. This is discarded and the filtrate solution is collected in a
clean beaker. In the filtrate solution we again add ammonium carbonate to precipitate the remaining 95\% of the
calcium carbonate. During this procedure, liquid $\rm NH_3$ is added to maintain the pH at 9 and the solution is
constantly stirred by a mechanical stirrer. The precipitate solution is filtered under vacuum and here the residue
in the filter paper is purified \caco and the filtrate is waste. The purified \caco is collected in a quartz crucible
and calciniated at 4500C for two hours inside a closed furnace (see Fig.~\ref{fig:fig6-6-12}). 

\subsection{Decomposition of \camo crystal material with 36\% HCl}
In the first step of Mo recovery and purification, about 300 g of \camo powder is placed in a glass beaker,
and 775~ml of HCl and 250~ml of DI water are added. The mixture is heated for 30 minutes while constantly
being stirred by a magnetic stirrer until a transparent solution is obtained. The main reaction for this procedure is:
\[ \rm CaMoO_4 + 6HCl  \to   H_2MoO_2Cl_4 + CaCl_2 + 2H_2O. 
\]
\noindent
The $\rm CaCl_2$ precipitate is filtered off. The remaining solution is neutralized with liquid NH$_3$ to
pH=1--1.5. During the neutralization, a PAM precipitate is formed. The precipitated solution is then passed
through a vacuum and the PAM residue is collected in one beaker and the filtrate in another. The residue is
washed with a dilute ammonium chloride solution with pH=1.7.  From the PAM, we separate AM and obtain calcium
carbonate from the filtrated calcium chloride.

\begin{figure}\centering
\includegraphics[width=\textwidth]{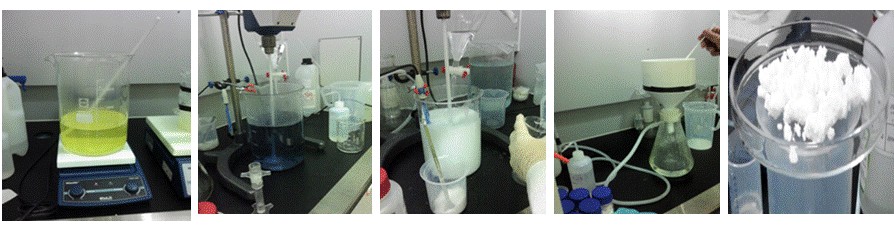}
\caption{The procedure to obtain crystals of molybdate; \CMO solution (left),
neutralization with liquid NH$_3$, vacuum filtration, and crystals of molybdate (right).}
\label{fig:fig6-6-13}
\end{figure}

In the second step, the PAM is dissolved in liquid ammonia until a transparent solution is obtained. The solution
is then vacuum filtered. The undissolved \CMO is collected as a residue on filter paper and collected in a quartz
crucible while the filtrate is collected in a clean beaker. The obtained solution is evaporated to obtain
ammonium molybdate crystals. The initially formed crystals are collected and the remaining solution is evaporated
to obtain the remaining, purified crystals. The procedure is illustrated in Fig.~\ref{fig:fig6-6-13}.

\begin{figure}
\includegraphics[width=\textwidth]{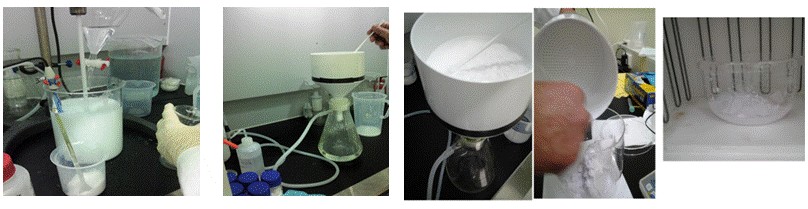}
\caption{Calcium carbonate production: neutralization with liquid NH$_3$ (left), filtration, and purified product (right).}
\label{fig:fig6-6-14}
\end{figure}

In the third step, we obtain calcium carbonate from the calcium chloride solution as shown in Fig.~\ref{fig:fig6-6-14}.
The neutralization of the calcium chloride solution is done with liquid $\rm NH_3$ at pH=8. A precipitate of dissolved
\CMO is formed.  The precipitate solution is vacuum filtered. The residue on the filter paper is undissolved \CMO that is
collected in the same beaker and the filtrate solution is collected in a clean beaker. In the filtrate solution, we add
enough ammonium carbonate so that 5\% of the calcium carbonate precipitates out. During this procedure aqueous ammonia
is added to maintain pH=9. This solution is constantly stirred by a mechanical stirrer. The above precipitate solution
is vacuum filtered. The precipitate on the filter paper is calcium carbonate that contains lots of impurities and is
discarded, while the filtrate solution is collected in a clean beaker. In the filtrate solution, we again add ammonium
carbonate to precipitate the remaining 95\% of the calcium carbonate. During this procedure, liquid NH$_3$ is added to
maintain pH=9 and the solution is constantly stirred by a mechanical stirrer. The precipitate solution is vacuum filtered.
The residue on the filter paper is purified \caco and the filtrate is discarded. The purified \caco is
collected in a quartz crucible and is calcinated at 450 $^\circ$C for two hours inside a closed furnace.

%% file: tables/table6-1.tex
\newcolumntype{Y}[1]{>{\centering}p{#1}}															
\begin{tabular}{|p{2.5cm}|Y{1.3cm}Y{1.3cm}Y{1.3cm}Y{1.3cm}Y{1.3cm}Y{1.3cm}|}															
\hline															
	Sample	&	Volume fraction \newline	&	\multicolumn{5}{c|}{Concentration [\ppb]}									\tabularnewline	
		&	[ml]	&	Fe	&	Sr	&	Ba	&	Th	&	U	\tabularnewline	\hline
	Initial solution 	&	-	&	16120	&	8129	&	67.2	&	0.73	&	20.4	\tabularnewline	
	Filtrate (1F)	&	840	&	122	&	1400	&	0.089	&	0.011	&	0.04	\tabularnewline	
	Filtrate (2F)	&	840	&	66	&	565	&	0.021	&	0.003	&	0.025	\tabularnewline	
	Filtrate (3F)	&	880	&	146	&	171	&	0.053	&	0.0004	&	0.017	\tabularnewline	
	Filtrate (4F)	&	1420	&	359	&	132	&	0.023	&	0.00006	&	0.039	\tabularnewline	
\hline															
\end{tabular}															

%% file: tables/table6-2.tex
\newcolumntype{Y}[1]{>{\centering}p{#1}}		
\begin{tabular}{|p{2.5cm}|Y{2cm}Y{1cm}Y{1cm}Y{1cm}Y{1cm}Y{1cm}|}
\hline															
	Sample	&	Volume fract.	&	\multicolumn{5}{c|}{DF}									\tabularnewline	
		&	[ml]	&	Fe	&	Sr	&	Ba	&	Th	&	U	\tabularnewline	\hline
	Filtrate(1F)	&	840	&	132	&	6	&	755	&	66	&	510	\tabularnewline	
	Filtrate(2F)	&	840	&	244	&	14	&	3200	&	243	&	816	\tabularnewline	
	Filtrate(3F)	&	880	&	110	&	48	&	1270	&	1825	&	1200	\tabularnewline	
	Filtrate(4F)	&	1420	&	45	&	62	&	2920	&	12170	&	523	\tabularnewline	
\hline															
\end{tabular}															

%% file: tables/table6-3.tex
\newcolumntype{Y}[1]{>{\centering}p{#1}}															
\begin{tabular}{|p{3.1cm}|Y{2cm}Y{1.2cm}Y{1.2cm}Y{1.2cm}Y{1.2cm}|}															
\hline															
			$\rm CaCO_{3}$ &	\multicolumn{5}{c|}{Concentration [\ppb]}									\tabularnewline	
				&	Fe	&	Sr	&	Ba	&	Th	&	U	\tabularnewline	\hline
			Initial powder	&	$\rm 1.3\times 10^{5}$	&	$\rm 6.5 \times 10^{4}$	&	538	&	5.8	&	163	\tabularnewline	
			Purified powder	&	$\rm 1.4\times 10^{4}$	&	$\rm 1.0 \times 10^{5}$	&	611	&	2.7	&	1.05	\tabularnewline	
			DF	&	9.3	&	0.63	&	0.88	&	2.1	&	155	\tabularnewline	
\hline															
\end{tabular}															

%% file: tables/table6-4.tex
\newcolumntype{Z}[1]{>{\centering$\rm}p{#1}<{$}}													
\newcolumntype{Y}[1]{>{\centering}p{#1}}													
\begin{tabular}{|Z{1.5cm}Z{1.5cm}Y{2.0cm}Y{2.0cm}Y{2.0cm}Y{2.0cm}|}													
\hline
		&		&	Peak Energy	&	Rate [counts/day]	&	Rate [counts/day]	&			\tabularnewline
	Chain	&	Element	&	[keV]	&	(Before)   	&	(After)	&	DF	\tabularnewline	\hline
	^{40}K	&^{40}K&	1460.83	&	164.9	&	9.31	&	17.61	\tabularnewline	
	^{238}U	&^{226}Ra	&	186.2	&	2246.9	&	6.97	&	322.36	\tabularnewline	
		&	^{214}Pb	&	295.22	&	4592	&	41.48	&	110.7	\tabularnewline	
		&		&	351.94	&	8387.9	&	133.99	&	62.59	\tabularnewline	
		&	^{214}Bi	&	609.31	&	6579.7	&	132.05	&	42.82	\tabularnewline	
		&		&	1120.28	&	1526.4	&	17.6	&	86.72	\tabularnewline	
		&		&	1764.51	&	1323.8	&	31.14	&	42.51	\tabularnewline	
		&		&		&		&	12.8	&	103.42	\tabularnewline	
		&		&		&		&	5.94	&	222.86	\tabularnewline	\hline
	^{232}Th	&	^{228}Ac	&	911.2	&	151.2	&	3.54	&	42.71	\tabularnewline	
		&		&	964.77	&	63.2	&	2.55	&	24.7	\tabularnewline	
		&		&	968.92	&	76.2	&	4.54	&	16.78	\tabularnewline	
		&	^{224}Ra	&	240.99	&		&	17.3	&		\tabularnewline	
		&	^{212}Pb	&	238.63	&	410.1	&	5.9	&	69.5	\tabularnewline	
		&	^{212}Bi	&	727.33	&	51.8	&	0.46	&	112.6	\tabularnewline	
		&	^{208}Tl	&	2614.53	&	45.4	&	4.51	&	10.06	\tabularnewline	
		&		&	583.14	&	180.1	&	2.74	&	65.72	\tabularnewline	
		&		&	510.77	&		&	2.06	&	87.42	\tabularnewline	\hline
\end{tabular}													

%% file: tex/offline_software.tex
\chapter{Offline software}

\begin{figure}\centering
\includegraphics[width=0.9\textwidth]{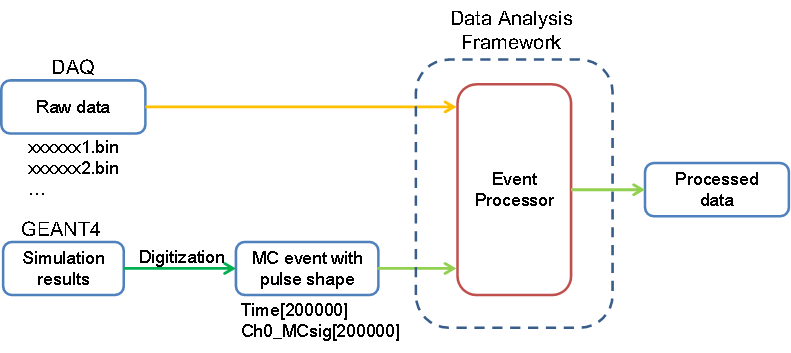}
\caption{Offline framework overview.}
\label{fig:fig9-1}
\end{figure}

The offline software will process measurement data recorded by the DAQ system  as well as Monte Carlo (MC) simulation results (Chapter~\ref{chap:Sim}) (see Fig.~\ref{fig:fig9-1}). The offline framework will support consistent and systematic data processing for different experimental configurations or data formats. It will be used to maintain and handle data sets. 

\begin{figure}\centering
\includegraphics[width=0.9\textwidth]{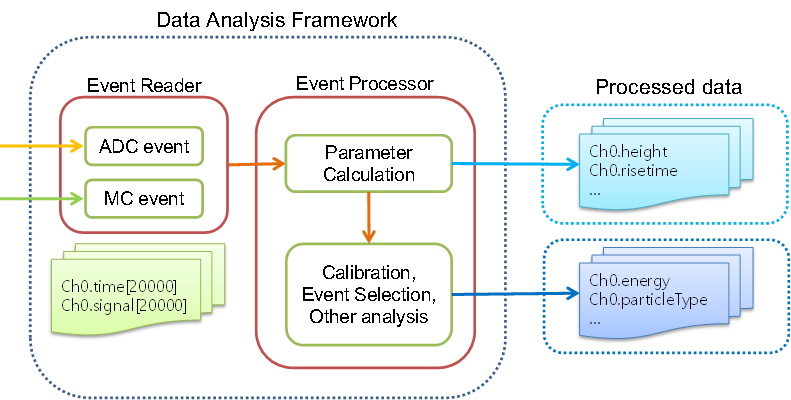}
\caption{Data analysis framework scheme.}
\label{fig:fig9-2}
\end{figure}

The raw data written by the DAQ system will be processed by the data analysis framework (Fig.~\ref{fig:fig9-2}), and processed measurements from triggered events will be stored in a file using a tree-structure called an ntuple. Similarly, MC simulation results will be processed by the same data analysis calculation routines, so that efficiencies of triggering and selection requirements can be determined.

An additional digitization processing of MC events will produce simulated pulse shapes, creating data records with the same structure as the real data for analysis by the same framework as the real data.  The current AMoRE-I simulation package outputs the energy that is deposited in crystals, but does not generate a time-domain pulse shape. Software for pulse-shape generation of MC events is in progress.

Design of data analysis framework is not finalized yet. Research and development on the framework is on-going. The basic structure of the current version of the framework is shown in Figure \ref{fig:fig9-1}.

\section{Environment}
The data analysis framework is being developmed on a linux server, Scientific Linux~6.3. The measured data and the simulation results are stored in the same system at the Center for Underground Physics, IBS. The data analysis framework software is written in C/C++ and uses ROOT libraries (Ver.~5.24). The framework software will be executed by ROOT scripts or binary executable files.

\section{Modules}
The main processor of the data analysis framework, EventProcessor, consists of two subclasses, EventReader and EventParameter. \\
\textrm{EventReader} is a routine to read events from files. When the DAQ data format is modified, the read-out routine in the framework should be updated accordingly. Depending on the data format version, a proper read-out routine can be used. Similarly, a read-out routine for MC simulation data will be added as a module.\\
\textrm{EventParameter} is a routine to calculate variables of each event such as, baseline, pulse height, rise time, etc. User-defined functions or calculation algorithm can be added as new modules. 

More analysis routines, such as optimal filtering, event selection requirements will be implemented in the future. 

%% file: tex/time_schedule.tex
\chapter{Time, Schedule }

\section{Schedules}

The AMoRE project will start with the preparation of the 1st phase experiment and R\&D for the 2nd phase. The 1st phase experiment will be performed also in two steps: first with 1.5 kg of \CMO crystals and later with 5 kg of the crystals. The 5 kg 1st phase will start at the end of 2016 and will run for about three years.

The R\&D for the 2nd phase includes cryogenic techniques, the development of light sensors, and the reduction of background levels. After three years of R\&D, the mass production of crystals and the design of the detector configuration for the 2nd phase will begin.  The setup for the 2nd phase will be ready by the end of the 1st phase data-taking.  The detailed schedule for each step is given in the Gantt plot below. The 2nd phase of the experiment will begin in early 2020 with 70 kg of \CMO, and will be upgraded to 200 kg in the following year. 

\begin{figure} \centering
  \includegraphics[width=\textwidth]{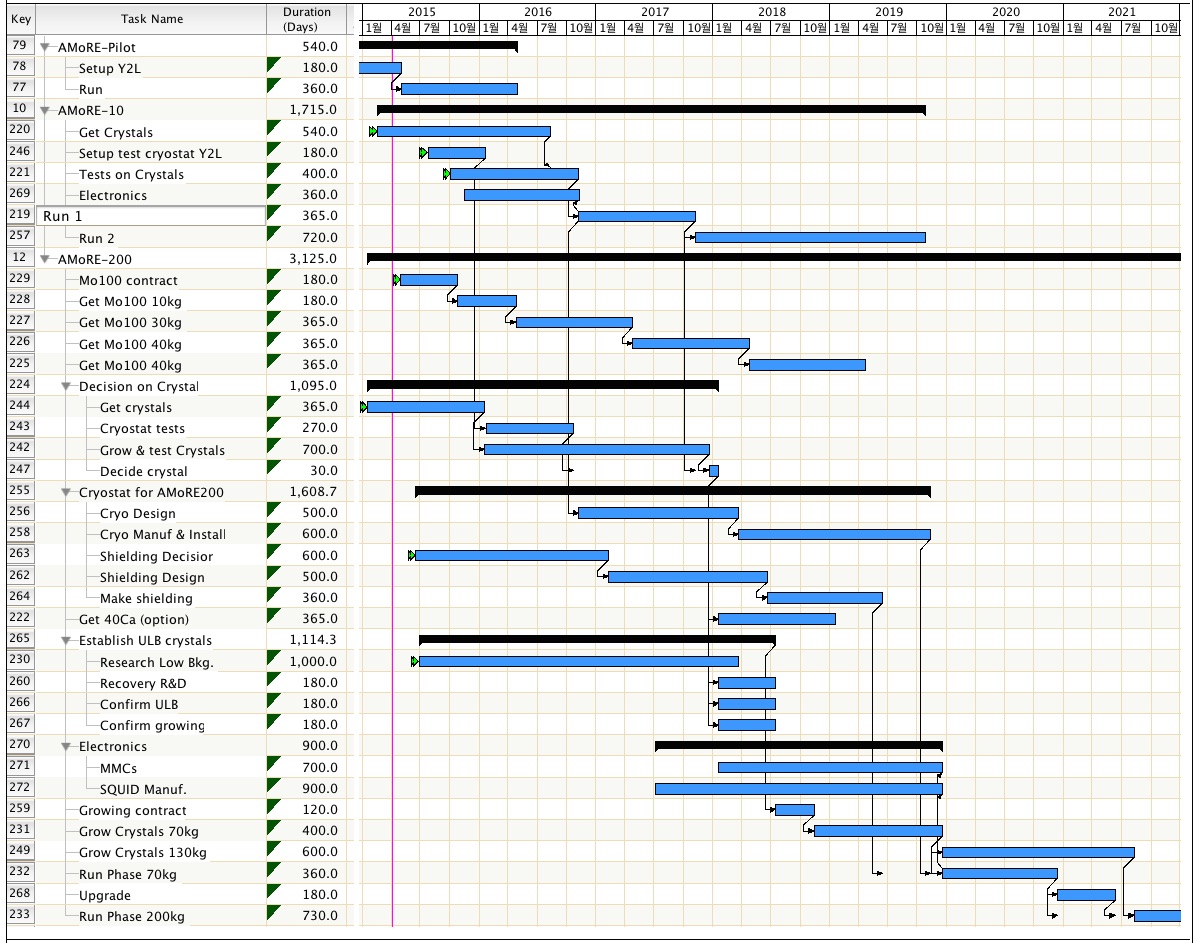}
\caption{The time schedule of AMoRE project.}
\label{fig:fig10-1}
\end{figure}